\documentclass[useAMS,usenatbib]{mn2e}

\usepackage{amsmath}
\usepackage{amssymb}
\usepackage{graphicx}
\usepackage{listings}
\usepackage{color}
\usepackage{booktabs}

\title[Formation and Composition of Super Earths]{On the Formation and Chemical Composition of Super Earths}
\author[Matthew Alessi, Ralph Pudritz and Alex Cridland]{Matthew Alessi$^{1}$\thanks{E-mail:
alessimj@mcmaster.ca (MA); pudritz@mcmaster.ca (REP);  cridlaaj@mcmaster.ca (AJC)}, 
Ralph E. Pudritz$^{1,2,3,4}$\footnotemark[1] and Alex J. Cridland$^{1}$\footnotemark[1]\\
$^{1}$Department of Physics and Astronomy, McMaster University, Hamilton, ON L8S 4M1, Canada\\
$^{2}$Origins Institute, McMaster University, Hamilton, ON L8S 4M1, Canada\\
$^{3}$Zentrum f\"{u}r Astronomie der Universit\"{a}t Heidelberg, Institut f\"{u}r Theoretische Astrophysik, Albert-Ueberle-Stra$\beta$e 2, 69120 Heidelberg, Germany \\
$^{4}$Max-Planck-Institut f\"{u}r Astrophysik, Karl-Schwarzschild-Stra$\beta$e 1, 85740 Garching bei M\"{u}nchen, Germany}
\begin{document}

\date{Accepted . Received ; in original form 2015 May 12}

\pagerange{\pageref{firstpage}--\pageref{lastpage}} \pubyear{2015}

\maketitle

\label{firstpage}

\begin{abstract}
Super Earths are the largest population of exoplanets and are seen to exhibit a rich diversity of compositions as inferred through their mean densities. Here we present a model that combines equilibrium chemistry in evolving disks with core accretion that tracks materials accreted onto planets during their formation. In doing so, we aim to explain why super Earths form so frequently and how they acquire such a diverse range of compositions. A key feature of our model is disk inhomogeneities, or planet traps, that act as barriers to rapid type-I migration. The traps we include are the dead zone, which can be caused by either cosmic ray or X-ray ionization, the ice line, and the heat transition. We find that in disks with sufficiently long lifetimes ($\gtrsim$ 4 Myr), all traps produce Jovian planets. In these disks, planet formation in the heat transition and X-ray dead zone produces hot Jupiters while the ice line and cosmic ray dead zones produce Jupiters at roughly 1 AU. Super Earth formation takes place within short-lived disks ($\lesssim$ 2 Myr), whereby the disks are photoevaporated while planets are in a slow phase of gas accretion. We find that super Earth compositions range from dry and rocky ($<$ 6 \% ice by mass) to those with substantial water contents ($>$ 30 \% ice by mass). The traps play a crucial role in our results, as they dictate where in the disk particular planets can accrete from, and what compositions they are able to acquire. 
\end{abstract}

\begin{keywords}
accretion, accretion discs, astrochemistry, planet-disk interactions, planets and satellites: composition, planets and satellites: formation, protoplanetary discs
\end{keywords}

\section{Introduction}

The growing sample of nearly 3000 observed exoplanets with over 2500 unconfirmed candidates has revealed new and unexpected populations of planets that do not share a Solar system analogue (Borucki et al. 2011, Mayor et al. 2011, Cassan et al. 2012, Rowe et al. 2014, Morton et al. 2016, see also exoplanets.org). The robustness of planet formation theories can be tested in their ability to reproduce these statistically significant populations of planets on the mass semi-major axis diagram, such as super Earths (1-10 M$_\oplus$), hot Neptunes ($\sim$ 10-30 M$_\oplus$), hot Jupiters, and 1 AU Jupiters \citep{IdaLin2004, IdaLin2008, Mayor2011, ChiangLaughlin2013, HP13}.

The distribution of observed planets on the mass-radius diagram adds another set of data to constrain models of planet formation. This distribution reveals a range of mean densities of planets that have similar masses \citep*{Fortney2007, Howard2013}, suggesting an interesting variety of chemical compositions among them. How, then, can planets that have similar masses and semi-major axes achieve such different compositions? This could arise for several reasons, such as variations in metallicities of their host stars, or the accretion of material at different locations in disks around stars with similar compositions. Our work studies the latter case, whereby planet compositions are intimately linked to their formation history.

In order to track materials accreted onto a planet throughout its formation, the physical and chemical conditions throughout the protoplanetary disk it is forming within must first be modelled.  One approach is to use equilibrium chemistry, whereby the Gibbs free energy of the system is minimized \citep*{White1958}. This technique is useful in determining chemical abundances throughout a complex system, and has been used in previous studies of disk chemistry \citep{Pasek2005, Pignatale2011}. Solid compositions are largely unaffected by non-equilibrium effects mainly due to their short equilibrium timescale \citep{Toppani2006}, allowing for equilibrium chemistry models to obtain good estimates of condensation sequences along the disk's midplane.  Due to the method's ability to model solid chemistry, this technique has largely been used to track compositions of terrestrial planets throughout their formation \citep*{Bond2010, Elser2012, Moriarty2014}. However, non-equilibrium effects such as photodissociation, grain-surface reactions, and ion driven chemistry are expected to be present within protoplanetary disks \citep*{VisserBergin2012, Cleeves2014} and will effect gas-phase chemistry. Gaseous abundances are therefore more reliably studied when taking non-equilibrium effects into consideration.

In this paper, we apply the technique to modelling planet compositions as they form in the core accretion model, a model of Jovian planet formation. We focus on the compositions of super Earths and hot Neptunes as their masses are mainly in solids, where our chemistry method is most applicable.

The core accretion model is a bottom-up process of planet formation whereby an initially small ($\sim 0.01 M_\oplus$) planetary embryo grows by accreting $\sim$ 1-10 km-sized planetesimals before becoming massive enough to accrete gas from the disk \citep*{Pollack1996, Hubickyj2005}. As was shown in \citet {HP12} and \citet{IdaLin2004}, this model predicts the formation of massive (1-10 Jupiter mass) gas giants in average to long-lived disks ($\gtrsim$ 2 Myr). If the process of photoevaporation is efficient enough to disperse the disk before the planetary core can accrete substantial amounts of gas, the core will be unable to continue growing. One can recognize the failed cores that result from these short-lived disks ($ \lesssim$ 2 Myr) as super Earths and hot Neptunes \citep{HP13}.  

The success of the core accretion model is seen in its ability to reproduce the observed distribution of planets on the mass-period diagram \citep{IdaLin2008, HP12, HP13, Mordasini2015}. One of the key processes shaping this distribution is planet migration. Througout its formation, the gravitational interaction between a planet and the surrounding disk results in an exchange of angular momentum \citep{GoldreichTremaine1980, MenouGoodman2004, HellaryNelson2012}. Properly accounting for migration throughout all stages of planet formation is critical to understand where in the disk a planet is forming and therefore what material it accretes.

Planet traps are a key feature of our core accretion model, and are used to model planet-disk interactions and radial migration throughout a large portion of planet formation. Planet traps arise from inhomogeneities in disks and act as barriers to rapid type-I migration \citep*{Masset2006, MP2007}. The inhomogeneities we study in our model are the outer edge of the dead zone (a transition from an MRI inactive to active region), the ice line (an opacity transition), and the heat transition (an entropy transition). Planet traps have been combined with a semi-analytic core accretion model in \citet{HP11, HP12, HP13} and have been shown to play a key role in reproducing the mass-period distribution of exoplanets. Here, our work builds upon these previous studies and attempts to determine how planet formation in traps affects their compositions.

The goal of this paper is to combine chemical models of protoplanetary disks with the core accretion model in order to account for the formation of different classes of planets as well as the chemical variety observed among super Earths. After obtaining planet masses and compositions at the end of their formation, we hope to provide initial conditions for modelling the interior structures of planets in the super Earth and hot Neptune population \citep*{Valencia2007}. By including the effects of trapped type-I migration, we will reveal what effect planet traps have on the compositions of planets formed in our models. 

As our work combines planet formation with migration, the materials a planet accretes change with both its position and time. Computing a time dependent chemical disk model offers an improvement on previous works which have limited their focus to the disk chemistry at a single time in the disk's evolution \citep{Bond2010}. The work we present here tracks planet formation from oligarchic growth (core formation) through the end of runaway gas accretion, and can be considered a global model of planet formation as it covers physical processes over a wide range of planet masses. This builds on previous works that have studied one aspect of Jovian planet formation in detail, such as gas accretion, migration, or oligarchic growth (\citet{Lissauer2009, Kley1999, HellaryNelson2012}, respectively). 

We first outline our model of the physical and chemical conditions in sections 2.1 and 2.2, respectively. With a disk model in hand, the locations of different planet traps can be calculated, and are discussed in section 2.3. The planet formation model we use is then outlined in section 2.4. In section 3, we present individual planet formation tracks and resulting compositions while varying important parameters in our model such as disk mass and lifetime. We focus primarily on conditions giving rise to super Earths and hot Neptunes, and what range of compositions among these planets our model predicts. We leave a complete statistical treatment of this to our next paper (Alessi, Pudritz, \& Cridland 2016, in prep.). Finally, in section 4 we discuss key implications and conclusions of our work. 

\section[]{Model}

\subsection{Accretion Disk Model}

The core accretion model predicts that Jovian planet formation occurs on a timescale of a few million years \citep{Pollack1996}. This timescale is comparable to the viscous timescale for protoplanetary disks. Therefore, the disk that a Jovian planet is forming within will evolve substantially throughout its formation as accretion onto the host star takes place \citep{Chambers2009}. Due to this, we require a dynamic and evolving disk in our model to account for the changes in disk properties over the course of a planet's formation. This leads to disk chemistry being inherently time dependent, as the governing temperatures and pressures throughout the disk are decreasing. Disk evolution is crucial in our model, as time-dependent physics and chemistry throughout the disk lead to planet traps sweeping through the disk, allowing planets forming within them to encounter regions with different materials available for accretion. 

The analytic, 1+1D disk model presented in \citet{Chambers2009} will be used throughout this paper. An analytic disk model is advantageous for our work as it allows us to efficiently calculate the conditions throughout the disk while modelling disk chemistry and planet formation. The self-similar approach adopted by \citet{Chambers2009} simultaneously models viscously heated, active inner regions of the disk and the outer regions that are passively heated by direct irradiation of the host star. In doing so, it merges the viscous disk models that are used for planet formation \citep{HP12, IdaLin2004} with models that are aimed at reproducing observed spectra of disks \citep{DAlessio1998, DAlessio1999} that only consider heating by radiation. The \citet{Chambers2009} model also has one of the three traps we are interested in tracking, the heat transition, built into the mathematical framework. However, it does not include the effects of an ice line or dead zone, which is a drawback of the model. To track these two traps, we use our disk chemistry (see section 2.2) and ionization (see section 2.3) models. We note that while we are able to use the disk structure to calculate the location of the ice line and dead zone, the back-reaction of these effects on the disk structure is not included.

The disk model in \citet{Chambers2009} gives an analytic solution to the viscous evolution equation describing the surface density profile $\Sigma(r, t)$ of a circumstellar disk in polar coordinates, 
\begin{equation} \frac{\partial \Sigma}{\partial t} = \frac{3}{r}\frac{\partial}{\partial r}\left[r^{1/2}\frac{\partial}{\partial r}\left(r^{1/2} \nu \Sigma\right) \right]\, , \end{equation}
where $\nu$ is the disk's viscosity. As shown in \citet{LBP1974}, self-similar solutions to this equation can be obtained for alpha disk models where the viscosity in the disk is taken to be proportional to the sound speed $c_s$ and disk scale height $H$ \citep{SS1973},
\begin{equation} \nu = \alpha c_s H\;, \end{equation}
where $\alpha$ is the effective viscosity coefficient. 

We expect there to be multiple sources of angular momentum transport in disks, such as by torques exerted by MHD disk winds as well as MRI generated turbulence (the latter, ourside the dead zone). The disk's effective $\alpha$ can then be written as a sum of individual $\alpha_i$ parameters characterizing particular angular momentum transport mechanisms. For example, in the case where disk angular momentum is transported through a combination of disk winds and MRI turbulence,
\begin{equation} \alpha = \alpha_{\textrm{wind}} + \alpha_{\textrm{turb}} \; . \label{sumalpha} \end{equation}
While our model uses $\alpha$ values corresponding to these two means of angular momentum transport, other possible mechanisms such as the hydrodynamic zombie vortex instability \citep*{Mohanty2013, Marcus2015} can fit within this framework by adding subsequent $\alpha$ parameters to equation \ref{sumalpha}.

The activity of the MRI instability depends on the ionization rates throughout the disk, discussed in detail in section 2.3. In the MRI active regions of the disk, $\alpha_{\textrm{turb}} \sim10^{-3} - 10^{-2}$, whereas it is $\sim 10^{-5}$ in the MRI inactive regions (referred to as the dead zone). It has been shown recently in \citet{Gressel2015} and \citet{Gressel2015b} that disk winds can maintain disk accretion rates through the MRI dead zones in disks. We therefore make the assumption that the disk's effective $\alpha$ is a constant throughout the disk with a particular value of $\alpha = 10^{-3}$ used.

The \citet{Chambers2009} model describes disk evolution under the influence of only viscous processes. An additional source of disk evolution is expected to be caused by high energy radiation from the protostar slowly dispersing the disk material, known as photoevaporation \citep*{Pascucci2009, Owen2011}. As was shown in \citet{HP13}, viscous evolution alone cannot reproduce the mass-period distribution of observed exoplanets, and results in low-mass planets being formed too far from their host stars. Photoevaporation's gradual removal of material acts as a means to accelerate disk evolution in the viscous framework. This allows planets to migrate inwards on a shorter timescale, forming planet populations consistent with exoplanet data, such as super Earths and hot Jupiters \citep{HP13}. Motivated by these results, we make the following modification to the disk accretion rate presented in \citet{Chambers2009},
\begin{equation} \dot{M}(t) = \frac{\dot{M}_0}{(1 + t/\tau_{vis})^{19/16}} \exp\left(-\frac{t - \tau_{int}}{\tau_{dep}}\right)\;, \label{ViscousAccretion}\end{equation}
which includes an exponentially decaying factor which models photoevaporation's effect on the disk's viscous evolution. In the above equation, $\dot{M}_0$ is the disk accretion rate at initial time $\tau_{int} = 10^5$ years, $\tau_{dep} = 10^6$ years is the depletion timescale, and $\tau_{vis}$ is the viscous timescale,
\begin{equation} \tau_{vis} = \frac{3 M_0}{16 \dot{M}_0}\;, \end{equation}
where $M_0$ is the disk's mass at time $t=0$. 

\begin{table}
\caption[Constants used in the Chambers Model]{Constants used in the Chambers disk model.}
\begin{center}
\begin{tabular}{| c | c | c |}

\hline
Constant & $T_{vis} > T_{rad}$ & $T_{vis} < T_{rad}$ \\
\hline
$\Sigma_{vis}$ & $\frac{7 M_0}{10 \pi s_0^2}$ & $\Sigma_{rad}\left(\frac{T_{rad}}{T_{vis}}\right)^{4/5}$ \\
$\Sigma_{rad}$ & $\Sigma_{vis}\left(\frac{T_{vis}}{T_{rad}}\right)$ & $\frac{13 M_0}{28\pi s_0^2}\left[1 - \frac{33}{98}\left(\frac{T_{vis}}{T_{rad}}\right)^{52/33}\right]^{-1}$ \\
$\Sigma_0$ & $\Sigma_{vis}$ & $\Sigma_{rad} $\\
$T_0$ & $T_{vis}$ & $T_{rad}$\\
\hline

\end{tabular}
\end{center}
\label{Chambers_constants}
\end{table}

\begin{table*}
\caption[Surface Density and Temperature Formulas in the Chambers Model]{Results of the \citet{Chambers2009} disk model. The surface density $\Sigma$ and temperature $T$ in each of the three zones are given below. In all cases, both $\Sigma$ and $T$ are found to have power law dependences on the radius in the disk, and the disk's accretion rate (causing them to be functions of time as $\dot M = \dot M(t)$).}
\centering
\begin{tabular} {|c|c|c|c|}
\hline
& $r<r_e$ &$r_e < r < r_t$ & $r > r_t$ \\
\hline
$\Sigma(r,t)$ & $ \Sigma_{evap}\left(\frac{\dot{M}}{\dot{M}_0}\right)^{17/19}\left(\frac{r}{s_0}\right)^{-24/19}$ & $\Sigma_{vis}\left(\frac{\dot{M}}{\dot{M}_0}\right)^{3/5}\left(\frac{r}{s_0}\right)^{-3/5} $ & $\Sigma_{rad}\left(\frac{\dot{M}}{\dot{M}_0}\right)\left(\frac{r}{s_0}\right)^{-15/14}$ \\
$T(r,t)$ & $\frac{T_0 \Sigma_0}{\Sigma_{evap}}\left(\frac{\dot{M}}{\dot{M}_0}\right)^{2/19}\left(\frac{r}{s_0}\right)^{-9/38}$ & $\frac{T_0 \Sigma_0}{\Sigma_{vis}}\left(\frac{\dot{M}}{\dot{M}_0}\right)^{2/5}\left(\frac{r}{s_0}\right)^{-9/10}$ & $\frac{T_0 \Sigma_0}{\Sigma_{rad}}\left(\frac{r}{s_0}\right)^{-3/7}$ \\
\hline
\end{tabular}
\label{ChambersSigmaT}
\end{table*}

The lifetimes of disks are dictated by the efficiency of the photoevaporation process. The disk lifetime, $t_{LT}$, is a key parameter in our model, as it sets an upper limit on the timescales that disk evolution, disk chemistry, and planet formation have to take place \citep{Pascucci2009, Owen2011}. A fiducial value for the disk lifetime that we adopt in this paper is 3 Myr, although a range of lifetimes as short as 0.5 Myr and up to 10 Myr for the longest lived disks are considered reasonable in our model, as they match with disk lifetimes inferred through disk observations in young star clusters \citep{Hernandez2007}. While calculating our disk models, we use equation \ref{ViscousAccretion} to compute accretion rates for all times $t \leq t_{LT}$. At $t=t_{LT}$, we assume the disk is rapidly dispersed in less than $10^4$ years due to photoevaporation dominating disk evolution. Thus, at this time we set the disk accretion rate and mass to zero, halting all subsequent disk evolution, planet formation, and planet-disk interactions.

Throughout the entire disk, the opacity is assumed to be a constant value of 3 g cm$^{-2}$. This assumption is simplistic, as condensation fronts will play a role in changing the opacity. However, previous works which have used complicated piecewise opacity power laws obtain surface densities and midplane temperatures that are weakly sensitive to the disk's opacity \citep{Stepinski1998}. Moreover, these models have neglected variations in opacity due to time dependent dust compositions and size distributions. Since we are not employing sophisticated models of dust growth and composition, our assumption of constant opacity simplifies the problem and allows for analytic disk models to be used. 

Within the innermost region of the disk, the temperature is so high that dust grains are evaporated. Thus, within the \emph{evaporative radius}, $r_e$, will the opacity drop below the assumed constant value due to a reduced dust content. The evaporative radius can be calculated using, 
\begin{equation} r_e = s_0\left(\frac{\Sigma_{evap}}{\Sigma_{vis}}\right)^{95/63}\left(\frac{\dot{M}}{\dot{M}_0}\right)^{4/9}\;, \label{Evaporative_Radius} \end{equation}
where $s_0$ is the initial disk radius, and,
\begin{equation} \Sigma_{evap} = \Sigma_0 \left(\frac{T_0}{T_{vis}}\right)^{4/19}\left(\frac{T_0}{1380 \;\textrm{K}}\right)^{14/19}\;, \end{equation}
is the surface density constant in the evaporative zone. The viscous heating temperature constant, $T_{vis}$, is defined in equation \ref{Tvis}, while the constants $\Sigma_{vis}$, $\Sigma_0$, and $T_0$ can be found in table \ref{Chambers_constants}. Values of $r_e$ for disk masses in the range 0.01-0.05 $M_\odot$ after 1 Myr of evolution are generally $\sim 0.1$ AU, in agreement with observations \citep{Eisner2005}. Thus this inner region with reduced opacity comprises a small fraction of the disk. The opacity in this region takes on a temperature power law of the form \citep{Stepinski1998},
\begin{equation} \kappa = 3\; \textrm{g cm}^{-2}\left(\frac{T}{1380 \, \textrm{K}}\right)^{-14}\; , \end{equation}
and this only applies when $T > T_e \equiv 1380$ K.

We now summarize the formulation of our disk model. All of the remaining equations presented in this section are taken from \citet{Chambers2009}. The input parameters to the model are the viscosity parameter $\alpha$, the initial disk mass $M_0$, initial disk radius $s_0$, as well as the mass, radius, and temperature of the protostar ($M_*$, $R_*$, $T_*$). The output of the calculations gives the disk accretion rate as a function of time, as well as time-dependent radial profiles of surface density, and midplane temperature. 

When starting the calculation, it must first be determined whether or not an irradiation-dominated region is present by comparing initial temperatures caused by viscous heating and irradiation at the outer edge of the disk. The initial temperature at the outermost point of the disk, $s_0$, caused by viscous heating is,
\begin{equation} T_{vis} = \left(\frac{27\kappa_0}{64\sigma}\right)^{1/3}\left(\frac{\alpha \gamma k}{\mu m_H}\right)^{1/3}\left(\frac{7 M_0}{10 \pi s_0^2}\right)^{2/3}\left(\frac{G M_*}{s_0^3}\right)^{1/6} \;, \label{Tvis} \end{equation}
where $\sigma$ is the Stefan-Boltzmann constant, $\gamma$ is the adiabatic index ($\simeq 1.4$), $\mu$ is the mean molecular weight ($\simeq 2.4$), $k$ is Boltzmann's constant, $m_H$ is the mass of the hydrogen atom, and $G$ is Newton's gravitational constant. The initial outer temperature caused by irradiation from the central protostar is, 
\begin{equation} T_{rad} = \left(\frac{4}{7}\right)^{1/4}\left(\frac{T_*}{T_c}\right)^{1/7}\left(\frac{R_*}{s_0}\right)^{3/7}T_*\;, \end{equation}
where,
\begin{equation} T_c = \frac{G M_* \mu m_H}{k R_*} . \end{equation}

\begin{figure*}
\includegraphics[width = 3in]{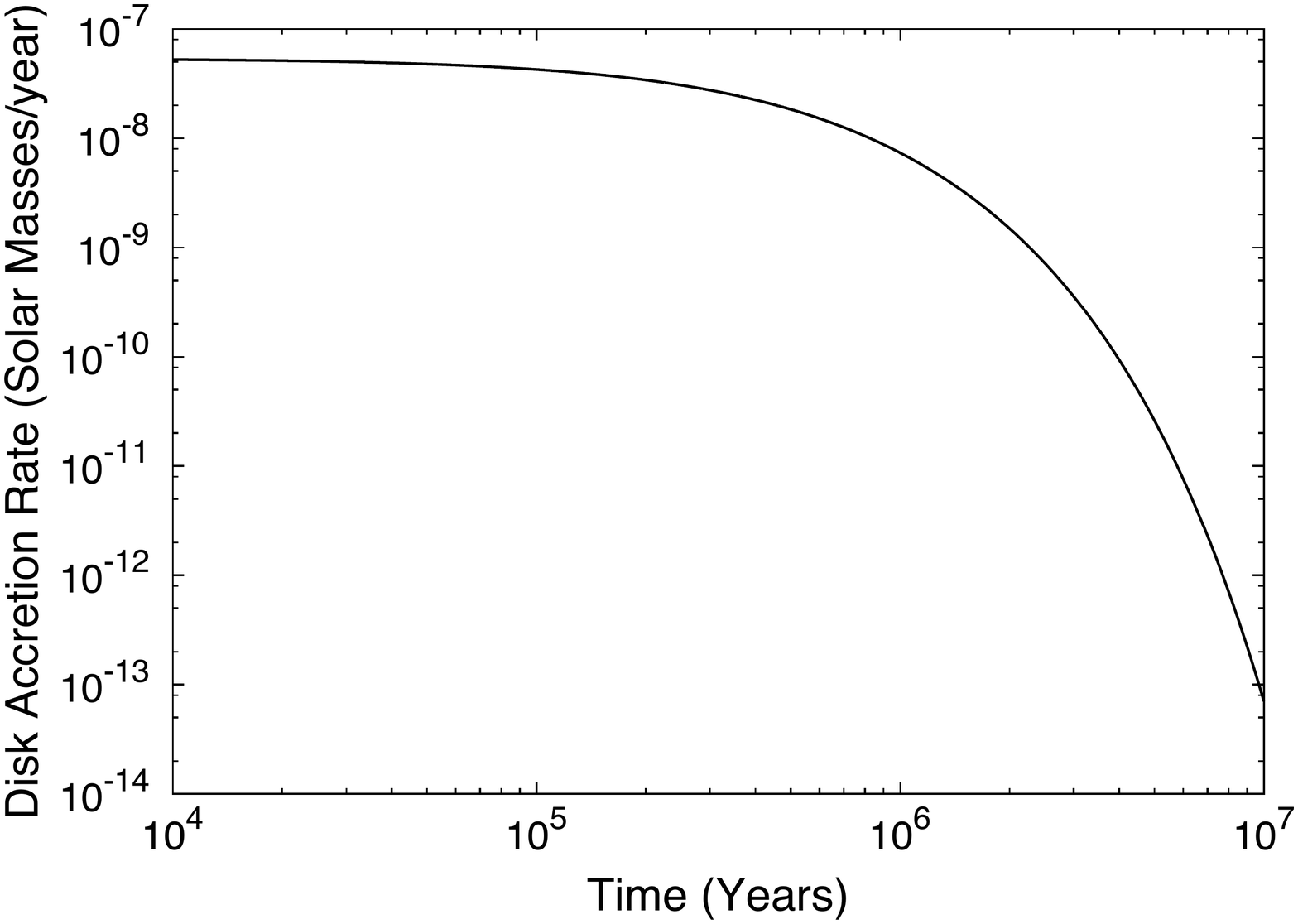}
\includegraphics[width = 3 in]{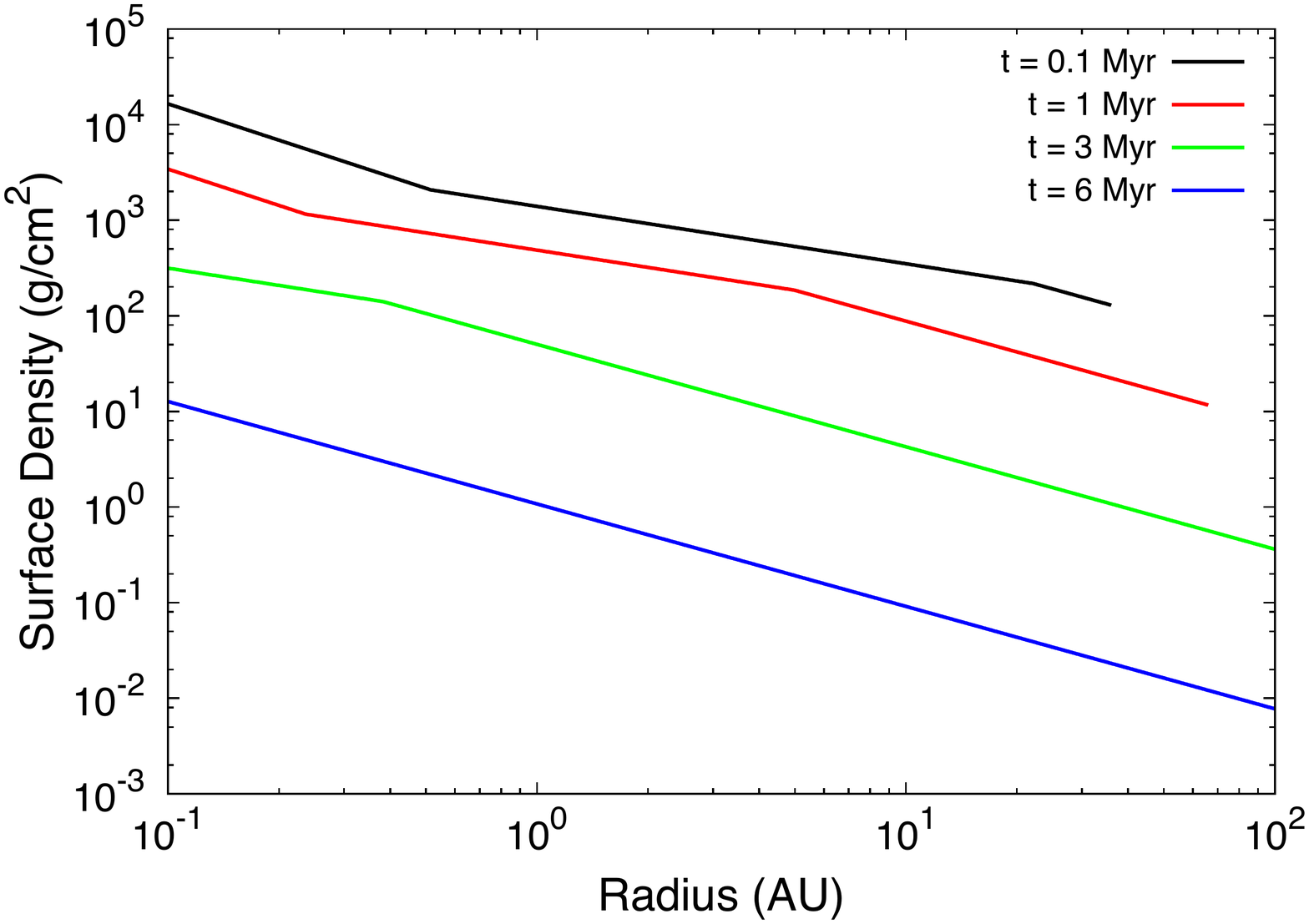}
\\ \includegraphics[width = 3 in]{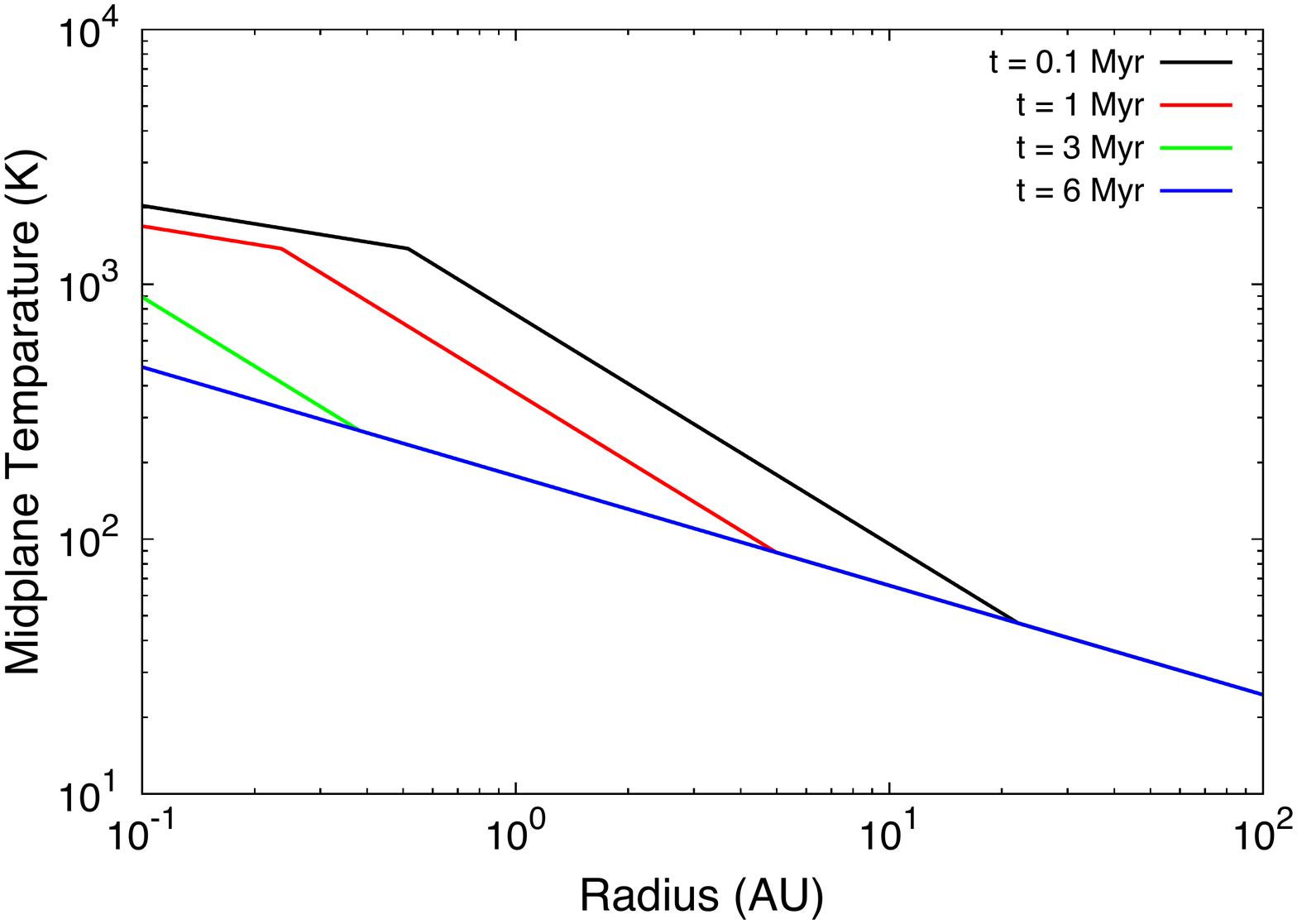}
\includegraphics[width = 3 in]{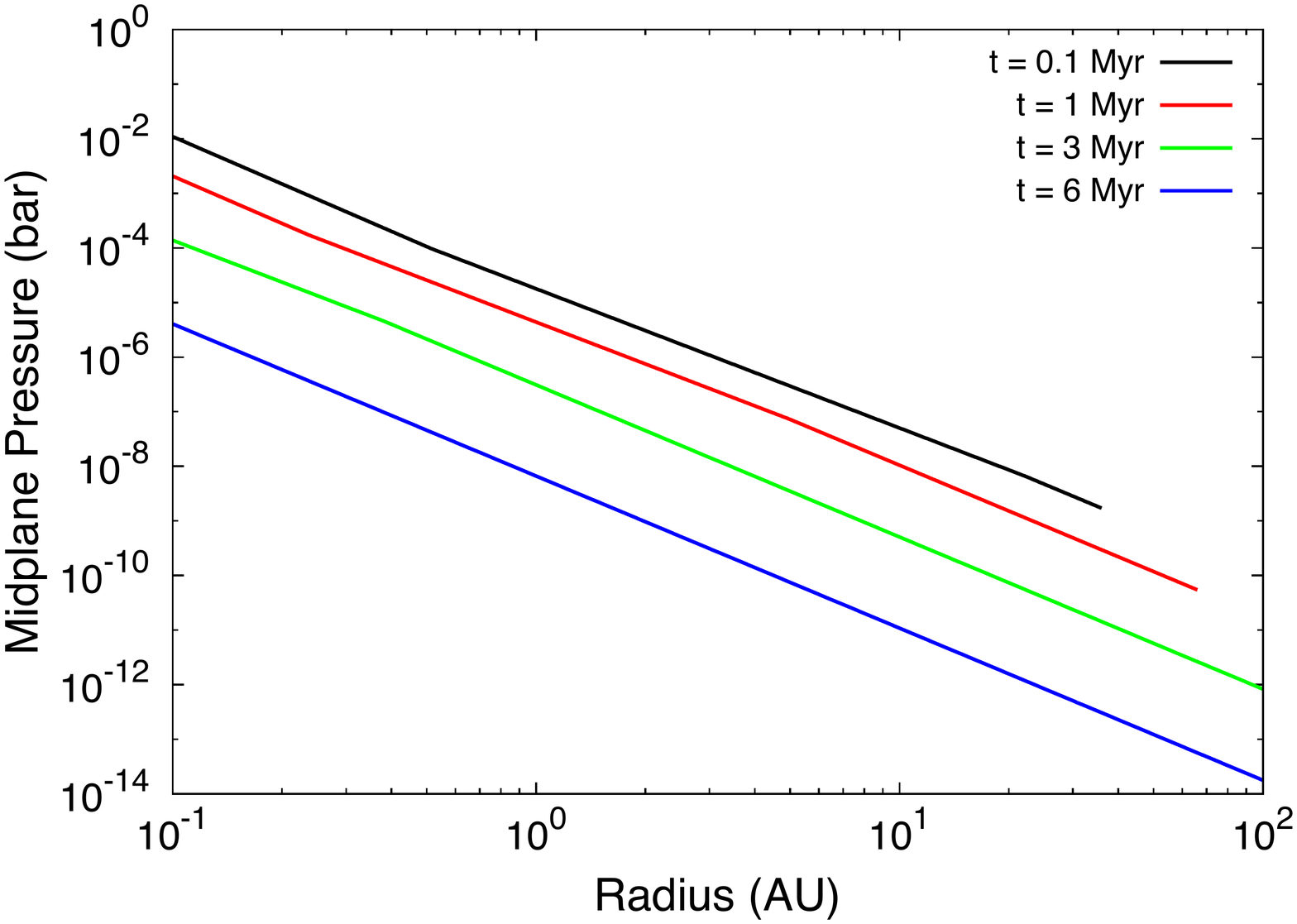}
\caption{The time-evolution of the disk accretion rate along with radial profiles of surface density, midplane temperature and pressure at various times throughout the disk's evolution. These plots pertain to a fiducial disk model whose parameters are outlined in the text. The outer radius of the disk is seen to increase with time in the radial profiles as a consequence of angular momentum conservation as the disk evolves.}
\label{DiskPlot}
\end{figure*}

After comparing the two initial outer temperatures, several constants are set as shown in table \ref{Chambers_constants}.  If the outer radius is initially in the viscous regime, the input time $t$ needs to be compared with the time at which an irradiated regime is first present at the outer edge of the disk, $t_1$. This time is determined by first calculating how much the disk radius needs to expand for the two heating mechanisms to produce the same temperature at the outer edge. The time at which the disk expands to this radius is,
\begin{equation} t_1 = \tau_{vis}\left[\left(\frac{T_{vis}}{T_{rad}}\right)^{112/73} -1\right]. \end{equation}
We note that $t_1$ is set to zero for disks that initially have outer regions dominated by irradiation. Equation \ref{ViscousAccretion} can be used to determine the accretion rates whenever $t < t_1$, as these disks are entirely viscous. Alternatively, if there is an irradiated region present ($t> t_1$), then the accretion rate at time $t_1$, defined as $\dot{M}_1$, first needs to be found using equation \ref{ViscousAccretion},
\begin{equation} \dot{M}_1 \equiv \dot{M}(t_1) = \frac{\dot{M}}{(1 + t_1/\tau_{vis})^{19/16}} \exp\left(-\frac{t_1 - \tau_{int}}{\tau_{dep}}\right)\;.\end{equation}
Then, the accretion rate for time $t > t_1$ is,
\begin{equation} \dot{M}(t) = \frac{\dot{M}_1}{(1 + (t-t_1)/\tau_{rad})^{20/13}} \exp \left(-\frac{t-t_1}{\tau_{dep}}\right), \end{equation}
where, 
\begin{equation} \tau_{rad} = \frac{7M_1}{13\dot{M}_1}.\end{equation}

The disk is divided into three regions; the innermost one being the small ($\sim$ 0.1 AU) evaporative zone previously discussed. The remaining two are defined by regions where the two mechanisms that heat the disk, viscous dissipation and irradiation from the central star, dominate. The model assumes a flared profile in the disk's vertical direction, allowing the outermost regions of this disk to intercept radiation from the star most efficiently. This arises because stellar radiation is the primary source of heating in outer regions of the disk. In the inner regions, the disk's surface density is highest, allowing for viscous dissipation to dominate heating in the inner disk. The radius separating the innermost evaporative zone and the viscously heated zone is given in equation \ref{Evaporative_Radius}. The heat transition, $r_t$, separates the viscous and irradiated regions. It is calculated by determining the radius where the two heating mechanisms result in the same midplane temperature. In the \citet{Chambers2009} model, the heat transition's location is given by,
\begin{equation} r_t = s_0 \left(\frac{\Sigma_{rad}}{\Sigma_{vis}}\right)^{70/33}\left(\frac{\dot{M}}{\dot{M}_0}\right)^{28/33}. \label{Heat_transition} \end{equation}
Note that both $r_e$ and $r_t$ move inwards with time due to their dependencies on accretion rate. The input radius $r$ is compared with these two radii to deduce what region of the disk is being considered before calculating surface density and midplane temperature. In table \ref{ChambersSigmaT} we present the surface density and midplane temperature profiles within each of the three regions of the disk.

Motivated by defining the external parameters (temperature and pressure) of a chemical system, we include a calculation of the disk's midplane pressure. In order to obtain a midplane pressure from surface density and midplane temperature, the ideal gas equation of state is used,
\begin{equation} P(r) = \frac{\rho_M(r)kT(r)}{\mu m_H}\;, \end{equation}
where $\rho_M$ represents the density at the midplane,
\begin{equation} \rho_M(r) = \frac{\Sigma(r)}{2\pi H(r)}\;,\end{equation}
and the scale height, $H$ is given by,
\begin{equation} H(r) = \sqrt{\frac{kT(r)r^3}{\mu m_H G M_*}} \;.\end{equation}
We assume that the disk is isothermal in the vertical direction $z$. In the viscous regime of the disk, the disk's effective temperature $T_{\textrm{eff}}$ differs from the midplane temperature by a factor of $(\kappa \Sigma /2)^{1/4}$. We find that this is a factor of order unity using our disk opacity and surface density values within the viscous region. While this vertical temperature gradient is important to consider when calculating disk chemistry away from the midplane, we feel that the assumption of a vertically isothermal disk is justified for our purpose of calculating the midplane pressure. Under this assumption, the density $\rho$ at height $z$ is,
\begin{equation} \rho(r,z) = \rho_M \exp\left(-\frac{z^2}{2H^2(r)}\right)\;,\end{equation}
Using the definition of surface density the midplane density can be solved for and input into the ideal gas equation, resulting in,
\begin{equation} P(r) = \Sigma(r) \sqrt{\frac{GM_*kT(r)}{2\pi \mu m_H r^3}}\;. \end{equation}

Figure \ref{DiskPlot} shows the disk accretion rate as a function of time, along with radial profiles of surface density, midplane temperature and pressure at several times throughout disk evolution for a fiducial set of model parameters,
\begin{equation} \begin{aligned} M_0 &= 0.1\; \textrm{M}_\odot \,,\;  \alpha = 10^{-3} \;,\; s_0 = 33\; \textrm{AU} \,,
\\ M_* &= 1 \; \textrm{M}_\odot\,,\; T_* = 4200\; \textrm{K} \,,\; R_* = 3 \textrm{R}_\odot\,. \end{aligned} \label{FiducialParameters} \end{equation}
Our choice of stellar parameters models a pre main sequence Solar type star \citep*{Siess2000}, while our initial disk mass is chosen such that disk evolution results in disk masses similar to the observed MMSN after 3 Myr \citep{Cieza2015}. We find that our model produces surface density and midplane temperature profiles that compare reasonably well (within a factor of 2 over all disk radii) to those found in \citet*{DAlessio2001} and \citet{Hueso2005} when using the same initial conditions and disk accretion rate. The kinks present in the radial profiles in figure \ref{DiskPlot} occur at boundaries between the three zones of the disk. We emphasize their presence, predominantly the heat transition, as they are locations of planet traps. The temperature profiles can be seen to all converge to a final profile in this figure. This is due to the assumed constancy of the irradiating stellar flux over the disk's lifetime. This differs from viscous heating as it does not depend on the disk accretion rate (see table \ref{ChambersSigmaT}). At late times in the disk's evolution, a decreasing surface density causes the viscous regime to shrink, eventually disappearing altogether. Thus, the entire disk becomes radiation dominated, resulting in a passive, or time-independent, temperature structure.

\subsection{Equilibrium Disk Chemistry}

In order to track accreted materials throughout planet formation simulations, and to constrain the dust to gas ratio within the disk, chemistry has been integrated into our accretion disk model. Here, we assume that the materials present in circumstellar disks are formed in situ rather than being accreted directly from their pre-stellar cores.

The question of ``reset" (in-situ formation) vs. ``inheritance" (direct transport from the stellar core) is debated as both are plausible mechanisms for chemical evolution of disks \citep{Pontoppidan2014}. While a combination of both mechanisms is likely responsible for the chemical structures of disks, there is evidence that the short chemical timescales lead to the ``reset" scenario dominating the chemical evolution in the inner disk regions \citep{Oberg2011, Pontoppidan2014}. Conversely, direct inheritance likely has a dominant effect in the outer disk \citep{Aikawa1999}.

In our core accretion model, planets accrete materials within 10 AU in the majority of cases (see sections 2.3 \& 2.4). While tracking planet compositions, we only consider the in-situ formation of materials to simplify disk chemistry, and assume that the ``reset" scenario has the most significant effects on our planetary compositions. We note, however, that the effects of direct inheritance from the stellar core on disk chemistry will likely be important (especially for planets that accrete solids at large disk radii), but are not considered here.

\begin{table}
\caption[Elemental Abundances in Chemistry Model]{Elemental abundances used in equilibrium chemistry calculations. Taken from \citep{Pignatale2011}.}
\begin{center}
\begin{tabular}{|c|c|}
\hline
Element & Abundance (kmol) \\
\hline
H & 91\\
He & 8.89\\
O & 4.46 $\times$ 10$^{-2}$ \\
C & 2.23 $\times$ 10$^{-2}$ \\
Ne & 1.09 $\times$ 10$^{-2}$\\
N & 7.57 $\times$ 10$^{-3}$ \\
Mg & 3.46 $\times$ 10$^{-3}$\\
Si & 3.30 $\times$ 10$^{-3}$ \\
Fe & 2.88 $\times$ 10$^{-3}$\\
S & 1.44 $\times$ 10$^{-3}$\\
Al & 2.81 $\times$ 10$^{-4}$\\
Ar & 2.29 $\times$ 10$^{-4}$\\
Ca & 2.04 $\times$ 10$^{-4}$\\
Na & 1.90 $\times$ 10$^{-4}$\\
Ni & 1.62 $\times$ 10$^{-4}$\\
\hline
\end{tabular}
\end{center}
\label{Abundances}
\end{table}

The time-dependent midplane temperature and pressure define the local conditions for a chemical system at each radius within the disk. Equilibrium abundances of gases, ices, and refractories are calculated by determining the set of abundance values that minimizes the disk's total Gibbs free energy. The Gibbs free energy of a chemical system is defined as,
\begin{equation} G = H - TS\;, \end{equation}
where $H$ is the enthalpy, $T$ is the system's temperature, and $S$ is the entropy. For a system being composed of $N$ species, the total Gibbs free energy is,
\begin{equation} \label{Gibbs} G_T = \sum_{i=1}^{N}X_i G_i = \sum_{i=1}^{N} X_i(G_i^0 + RT\ln X_i)\;,\end{equation}
where $X_i$, $G_i$, and $G_i^0$ are the mole fraction, Gibbs free energy, and Gibbs free energy of formation of species $i$, respectively.

\begin{table*}
\centering
\caption[Substances] {A list of species present in the chemistry model. Solids that are present in figure \ref{Solids} have their common names bracketed following their chemical formulae.}
\begin{tabular}{|l l l |l l l |}
\hline
\multicolumn{3}{ |l| }{Gas Phase} & \multicolumn{3}{ |l| }{Solid Phase} \\
\hline
Al& H& NO$_2$& & &\\
Ar& H$_2$& Na&Al$_2$O$_3$ & Fe$_2$O$_3$ (Hematite)& Na$_2$SiO$_3$ \\
C& H$_2$O& Ne & CaAl$_2$SiO$_6$&Fe$_3$O$_4$ (Magnetite) &SiO$_2$ \\
C$_2$H$_2$& HCN & Ni&CaMgSi$_2$O$_6$ (Diopside) & FeSiO$_3$ (Ferrosilite) &FeS (Troilite)\\
CH$_2$O& HS& O& CaO &Fe$_2$SiO$_4$ (Fayalite) & NiS\\
CH$_4$& H$_2$S& O$_2$& CaAl$_{12}$O$_{19}$ (Hibonite)& H$_2$O & Ni$_3$S$_2$\\
CO& He& OH&CaAl$_2$Si$_2$O$_8$ &MgO & Al \\
CO$_2$& Mg& S&Ca$_2$Al$_2$SiO$_7$ (Gehlenite) &MgAl$_2$O$_4$ & C\\
Ca& N& Si&Ca$_2$MgSi$_2$O$_7$ & MgSiO$_3$ (Enstatite) & Fe\\
CaO& N$_2$& SiO&FeAl$_2$O$_4$ (Hercynite) &Mg$_2$SiO$_4$ (Forsterite) & Ni\\
Fe& NH$_3$& SiO$_2$ & FeO& NaAlSi$_3$O$_8$ & Si\\
FeO & NO &SiS & & & \\
\hline
\end{tabular}
\label{Substances}
\end{table*} 

In order to determine the equilibrium state, the set of $X_i$ which minimize equation \ref{Gibbs} for a chemical system defined by temperature $T$ and pressure $P$ must be calculated. An additional constraint based on mass considerations is,
\begin{equation} \sum_{i=1}^N a_{ij} x_i = b_j \;\;\;\; (j = 1, 2, \ldots , m) , \end{equation} 
where $m$ is the number of elements in the chemical system, $x_i$ is the total number of moles of species $i$, $a_{ij}$ is the number of atoms of element $j$ contained in species $i$, and $b_j$ is the total number of moles of element $j$. The total number of moles, $x_i$, and the mole fraction, $X_i$ of species $i$ are related by $x_i = X_i \times 100\,\textrm{kmol}$.

We adopt the HSC Chemistry software package to perform equilibrium chemistry calculations (HSC website: http://www.outotec.com/en/Products--services/HSC-Chemistry/). It includes thermodynamic data, such as enthalpies, entropies, and heat capacities for all chemical species we consider in our model. The Gibbs free energy minimization technique with HSC software has been previously used in astrophysical contexts for chemical modelling of accretion disks \citep{Pasek2005, Pignatale2011}, as well as for tracking abundances of terrestrial planets during N-body simulations \citep{Bond2010, Elser2012, Moriarty2014}. 

Elemental abundances must be specified as initial conditions for equilibrium chemistry calculations and were taken from the Solar photosphere, scaled up to a total of 100 kmol \citep{Pasek2005}. In order to reduce computation time, only the fifteen most abundant elements have been included. The remaining ones have abundances $b_j < 10^{-4}$ kmol in the 100 kmol system, and are considered negligible for the calculations. The abundances of the fifteen elements considered in the 100 kmol system are listed in table \ref{Abundances}.

HSC has thermodynamic data on an extensive list of roughly 100 gaseous and 50 solid phase species that can form from the fifteen elements considered. Ideally, the calculation could be done with each of these having a possibility of forming in the chemical system. However, having such a large number of species to track in a calculation is computationally expensive, so a low resolution trial was first performed to determine the species that are not expected to be present within the protoplanetary disk. The low resolution trial was performed over a temperature range of 50-1850 K with a large temperature step of $\Delta T \simeq 100$ K and at pressures 10$^{-11}$, 10$^{-10}$, $\ldots$ ,10$^{-1}$ bar. These limits were chosen to cover the range of temperatures and pressures calculated with the disk model between 0.1-100 AU and 10$^5$ - 10$^7$ years. All species that did not form in this low resolution trial were omitted from future calculations to reduce computation time. Among the species that did form in the low resolution trial were 36 gases and 30 solids recorded in table \ref{Substances}. This reduced list of 66 species was used in all subsequent high resolution simulations as a set of possible species that could form in equilibrium chemistry calculations. 

Using this reduced list, we then performed a high resolution equilibrium chemistry calculation within the same limits outlined above. We used 200 temperature points in the temperature range of $50-1850$ K, resulting in a temperature spacing of $\Delta T = 9$ K. We calculated abundances of all species in table \ref{Substances} at each of these 200 temperatures for 2000 pressures that were equally spaced logarithmically within the range of $10^{-11} - 10^{-1}$ bar. The high resolution calculations allowed us to compute 200$\times$2000 arrays of equilibrium abundances for each substance in table \ref{Substances}, with each value in the array corresponding to a particular temperature and pressure. Abundances at $T$ and $P$ values within these grid points are calculated using linear interpolation, which is justified due to the high resolution of the grid. We note that abundances of each individual species are often much more sensitive to temperature than they are to pressure. However, since the pressures that are of interest span several orders of magnitude along the disk's midplane, pressure's effect on the abundances must be taken into account. 

Using the disk model, we are able to calculate the temperature and pressure throughout the disk and map our abundances to a location within the disk at a particular time. We emphasize that the abundances throughout the disk are time dependent due to the evolving temperatures and pressures within the disk. This results in time-dependent radial abundance profiles for each substance in our chemistry simulation. 

\begin{figure*}
\begin{center}
\includegraphics[width = 2.25 in]{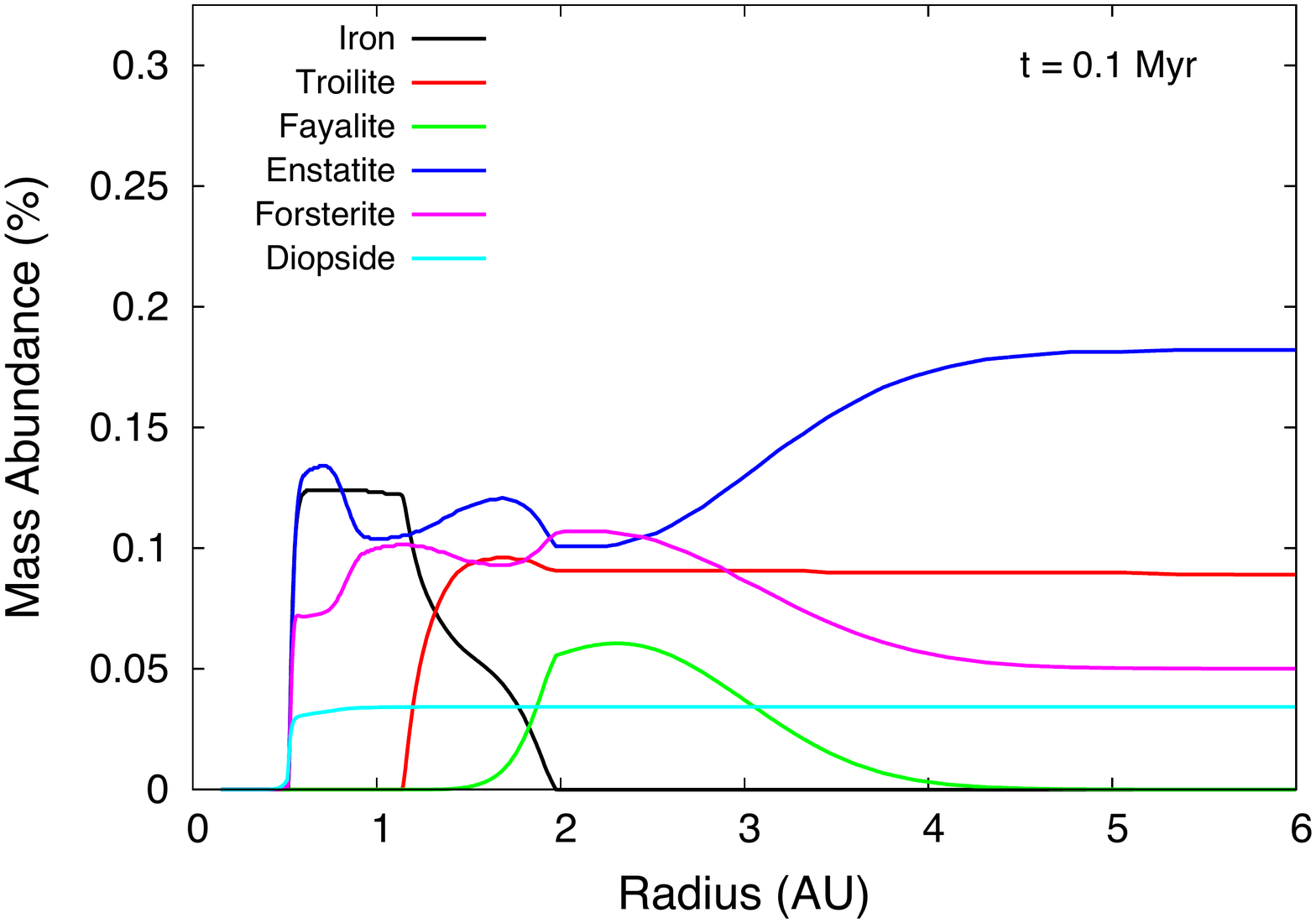} \includegraphics[width = 2.25 in]{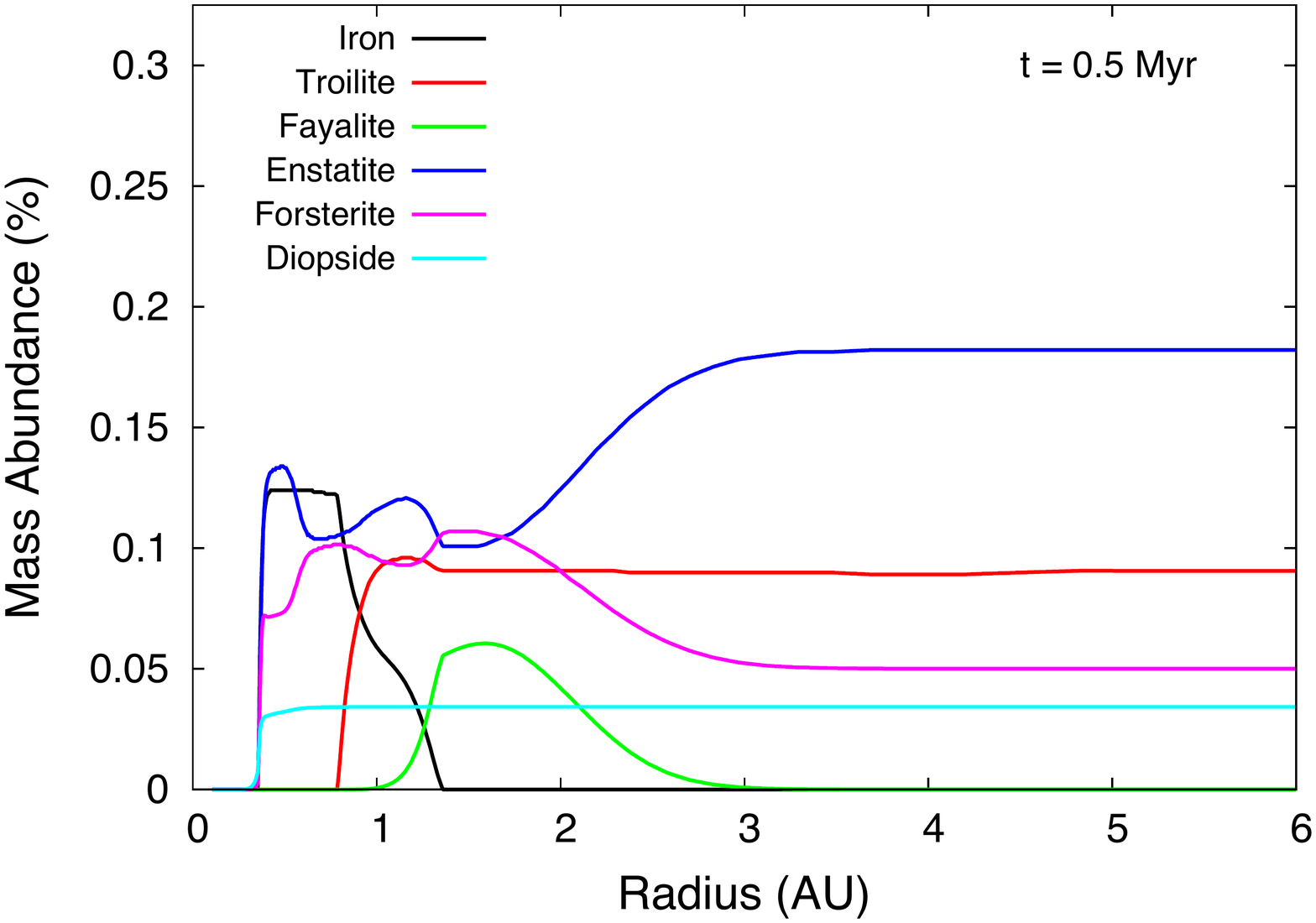}
\includegraphics[width = 2.25in]{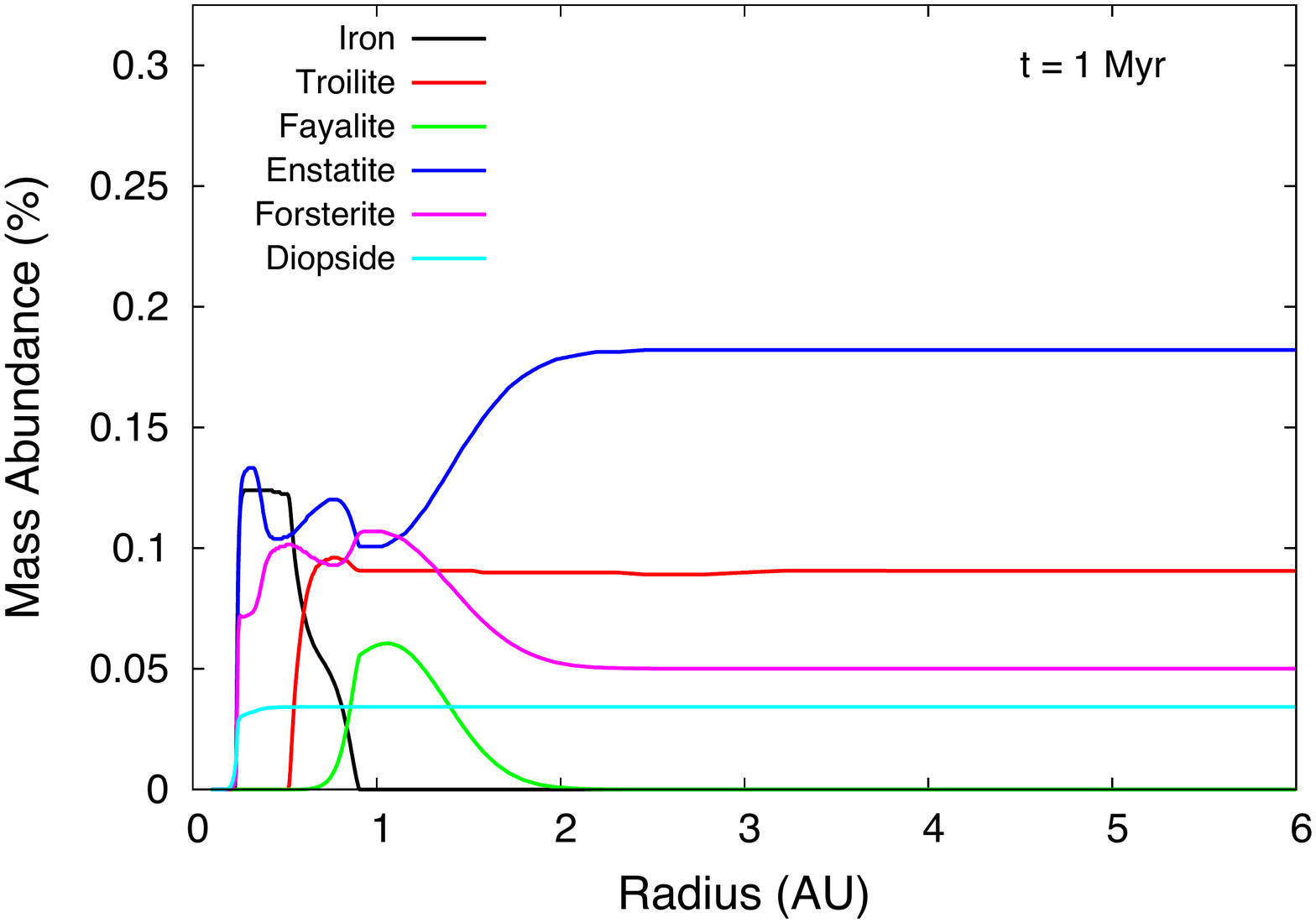}
\\ \includegraphics[width = 2.25 in]{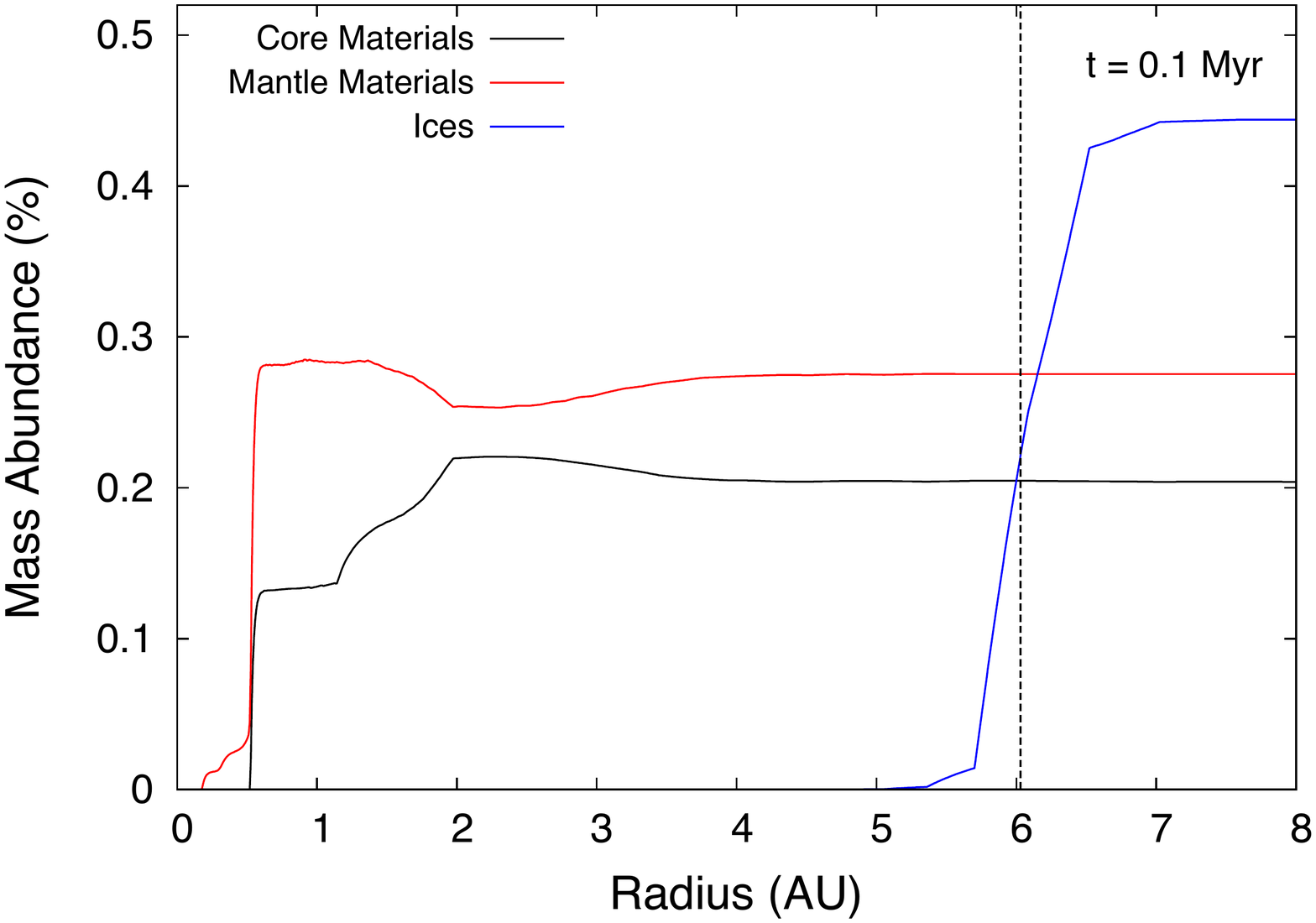} \includegraphics[width = 2.25 in]{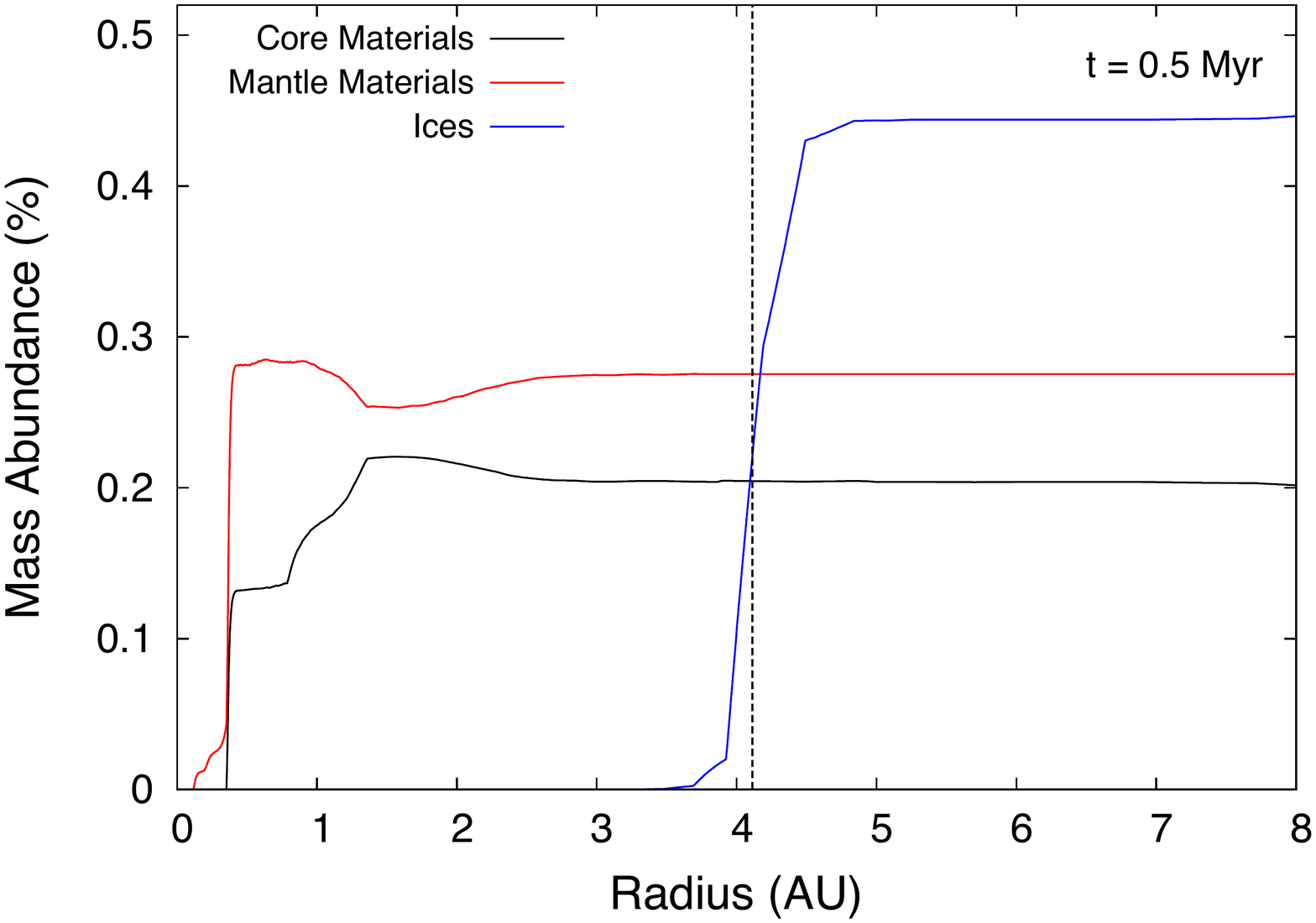}
\includegraphics[width = 2.25in]{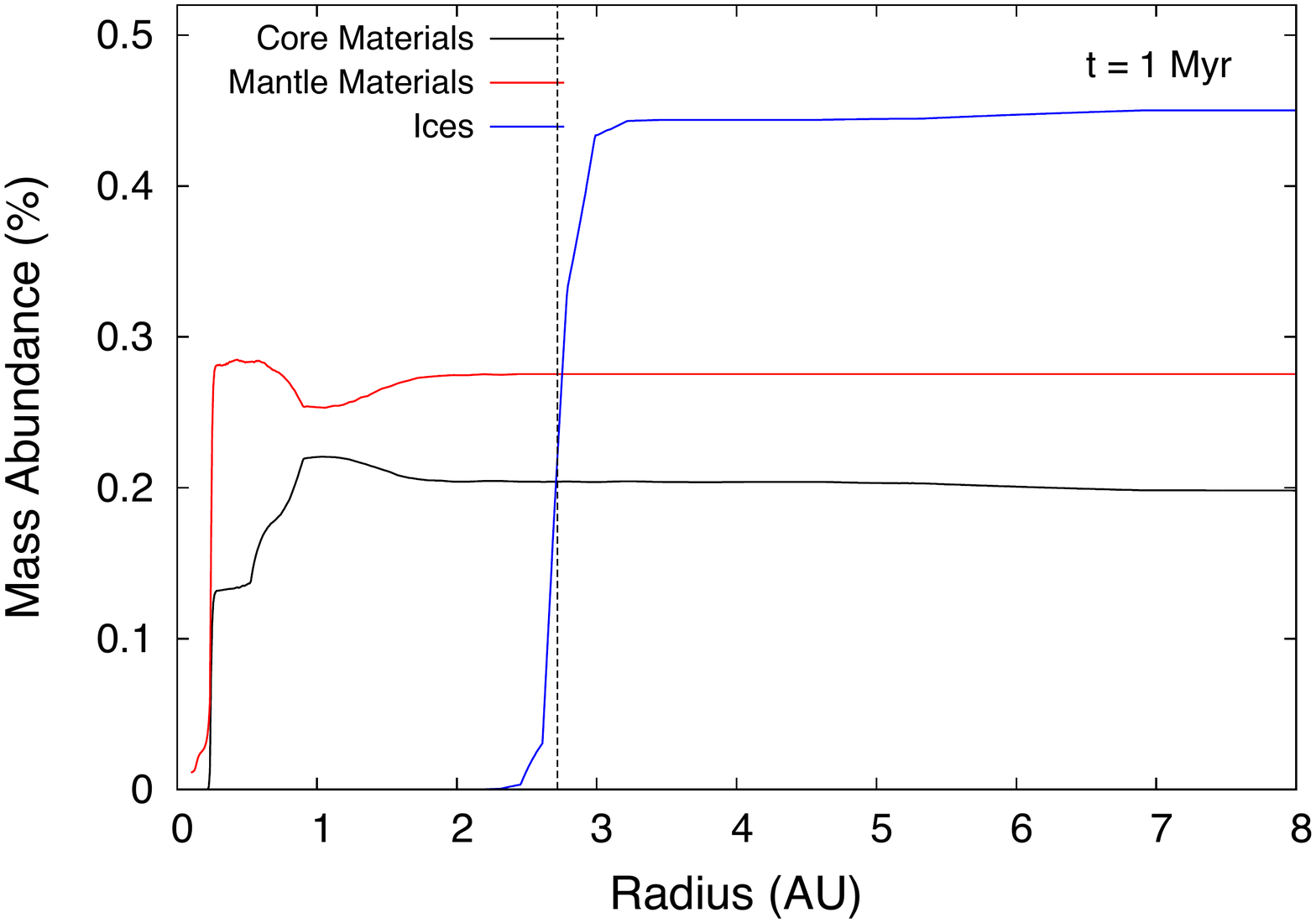}
\caption{\textbf{Upper Panels}: Midplane abundance profiles of several important solids at 0.1 Myr (left), 0.5 Myr (center), and 1 Myr (right) into the disk's evolution for a fiducial disk (equation \ref{FiducialParameters}). Notice that key features in the abundance profile move inwards with time as the disk viscously evolves. \textbf{Lower Panels}: Midplane abundance profiles of the three summed solid components at the same times. The water ice line is marked with a black vertical line. }
\label{Solids}
\end{center} \end{figure*}

We note that while we do compute abundance profiles of solids throughout the disk, we do not consider their effect on the disk's structure through changing the disk's opacity. While this is a simplification, we note that the disk's midplane temperature has a weak dependence on opacity of $T \sim \kappa ^{1/3}$ (see equation \ref{Tvis}). Therefore, even opacity changes by a factor of 10 will lead to corrections of order unity on our overall disk structure. We have confirmed this by comparing our disk model to the one presented in \citet{Stepinski1998} who used a detailed disk opacity structure, including the ice line's effect on opacity, and finding that our overall surface density and midplane temperature profiles were similar even though our model assumes a constant opacity.

In figure \ref{Solids}, top panels, we show several snapshots of the abundance profiles of several prominent minerals along the disk's midplane. Features in the radial abundance profiles of these minerals are seen to shift inwards with time as the disk viscously evolves. We note that while graphite is listed as a solid material that can form in our chemistry model, we do not produce an appreciable amount anywhere in the disk using Solar abundances as our initial condition. The midplane solid abundances we obtain are quantitatively similar with those shown in \citet{Bond2010} \& \citet{Elser2012} who also performed equilibrium chemistry calculations on a disk of Solar abundance.

Interior structure models of super Earth-mass planets typically are not interested in abundances of specific minerals. Rather, the abundances of broad groups of solids that characterize where they will end up within the planet's interior after differentiation is of importance \citep{Valencia2007}. Motivated by this, we categorize the solids in our chemical data into three groups: 
\begin{itemize}
\item \textbf{Core Materials : }Iron and nickel based materials, which will build up the core of a differentiated planet. This subset contains eleven of the thirty solids present in the chemistry simulation. The most abundant solids in this subset are iron (Fe), troilite (FeS), fayalite (Fe$_2$SiO$_4$), and ferrosilite (FeSiO$_3$).
\item \textbf{Mantle Materials : }Magnesium, aluminum, and silicate materials, which will build up the mantle of a differentiated planet. This subset contains eighteen of the thirty solids in the chemistry simulation. The most abundant solids in this subset are enstatite (MgSiO$_3$), forsterite (Mg$_2$SiO$_4$), diopside (CaMgSi$_2$O$_6$), gehlenite (Ca$_2$Al$_2$SiO$_7$), and hibonite (CaAl$_{12}$O$_{19}$).
\item \textbf{Ices} which will lie on the planet's solid surface. This subset only contains water. The omission of CO ices, among others is a limitation of our model, and is discussed in section 2.3.2.
\end{itemize}

Radial abundance profiles of these three summed components can be seen in the bottom panels of figure \ref{Solids}. We find that the ratio between the abundances of mantle materials and core materials throughout the disk is roughly constant, with mantle materials being slightly more abundant. The abundance profile of ice displays a step function profile, with its abundance increasing from zero to its maximum amount of 0.45\% in less than an AU. In this sense, the ice line is quite well defined, and we mark its location with a vertical dashed line in figure \ref{Solids}. The ice line, along with all other chemical signatures, is seen to shift inwards with time as the disk evolves viscously. The time-dependence of the ice line will be further discussed in section 2.3, as it is one of the planet traps in our model. Lastly, figure \ref{Solids} shows that virtually no solids are present within 0.1 AU as this is the evaporative region of the disk discussed in section 2.1, where the chemistry simulation confirms that the disk temperature is too high for any solids to exist at this location.

\begin{figure*}
\begin{center}
\includegraphics[width = 2.25 in]{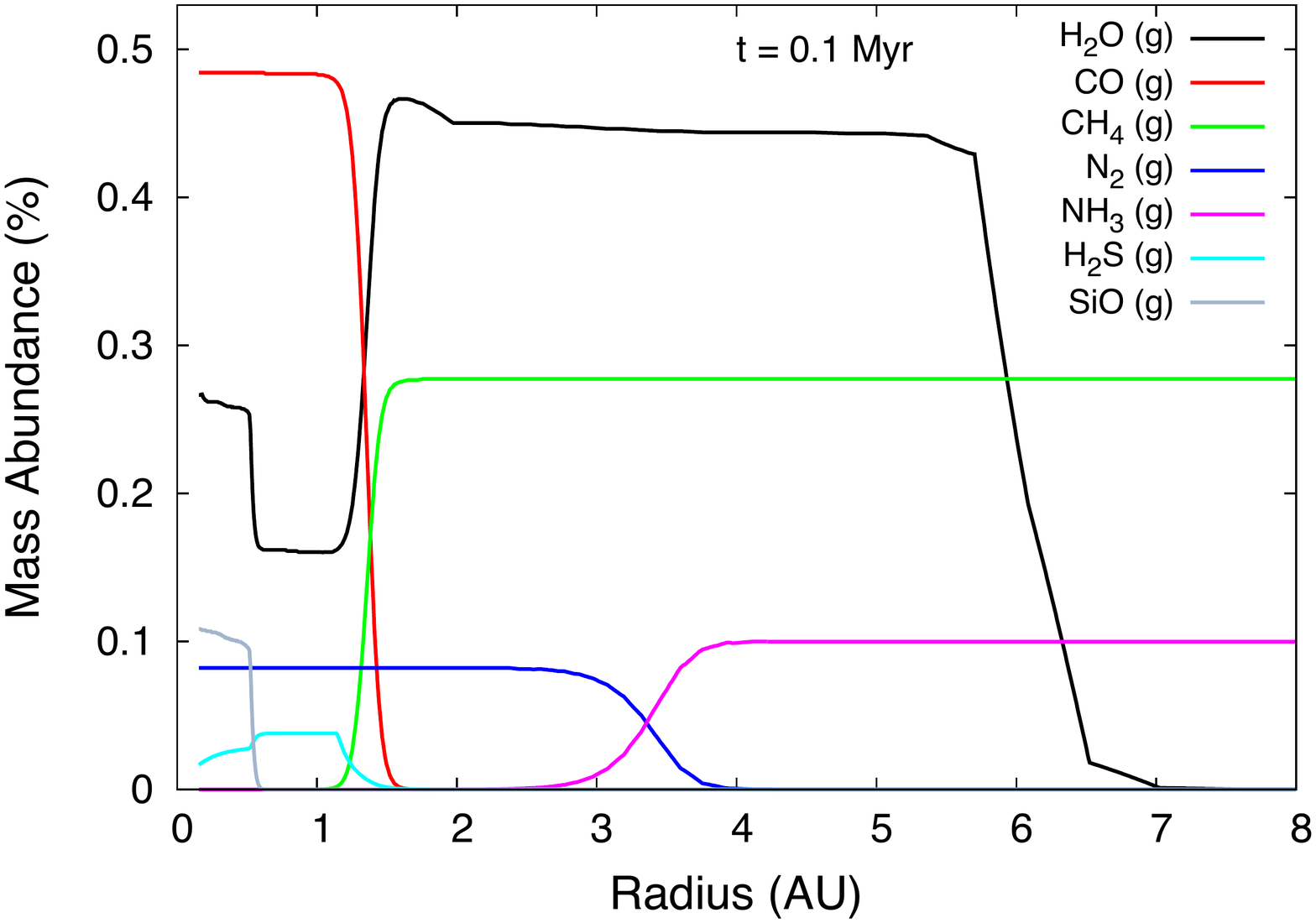} \includegraphics[width = 2.25in]{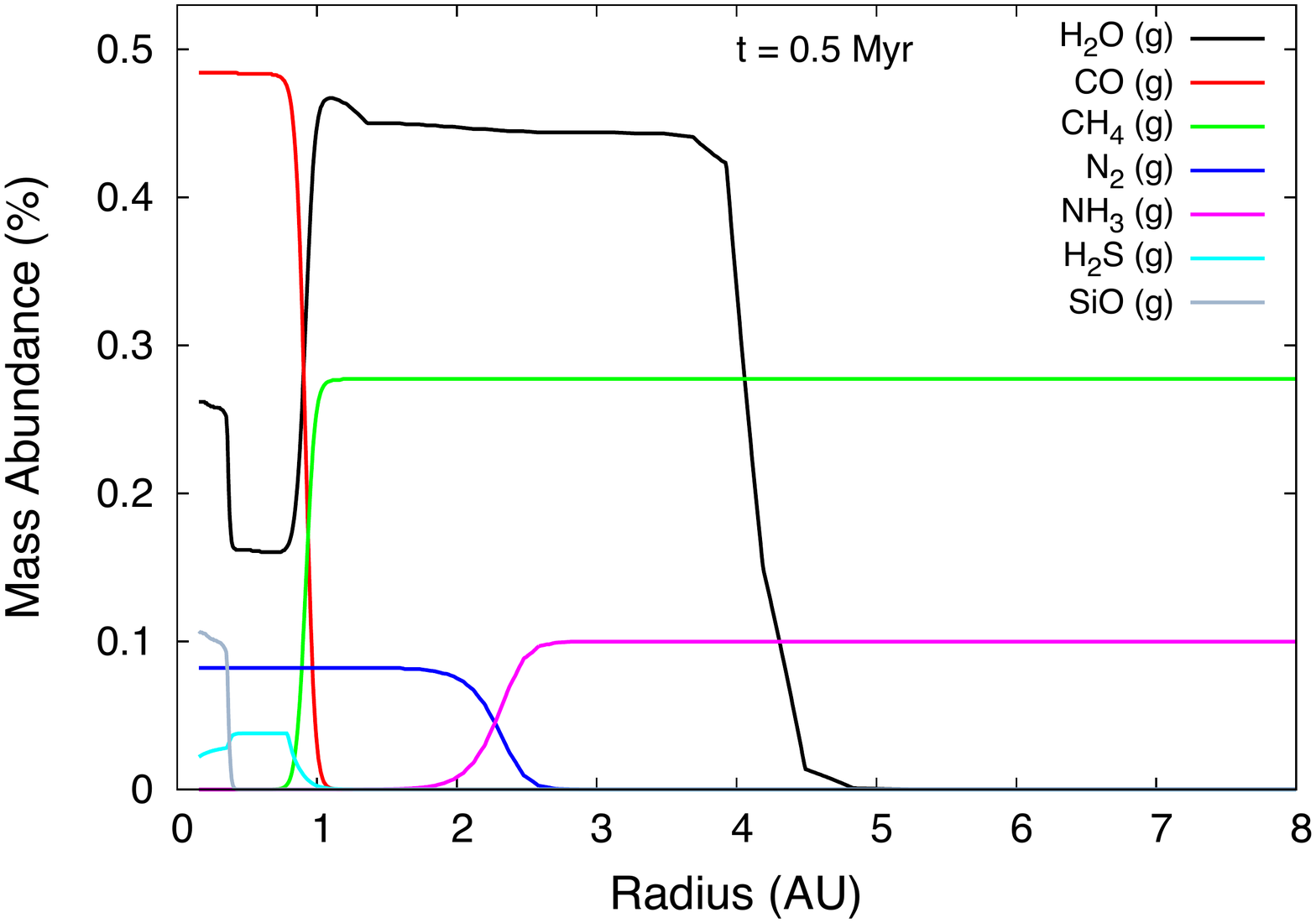}
\includegraphics[width = 2.25 in]{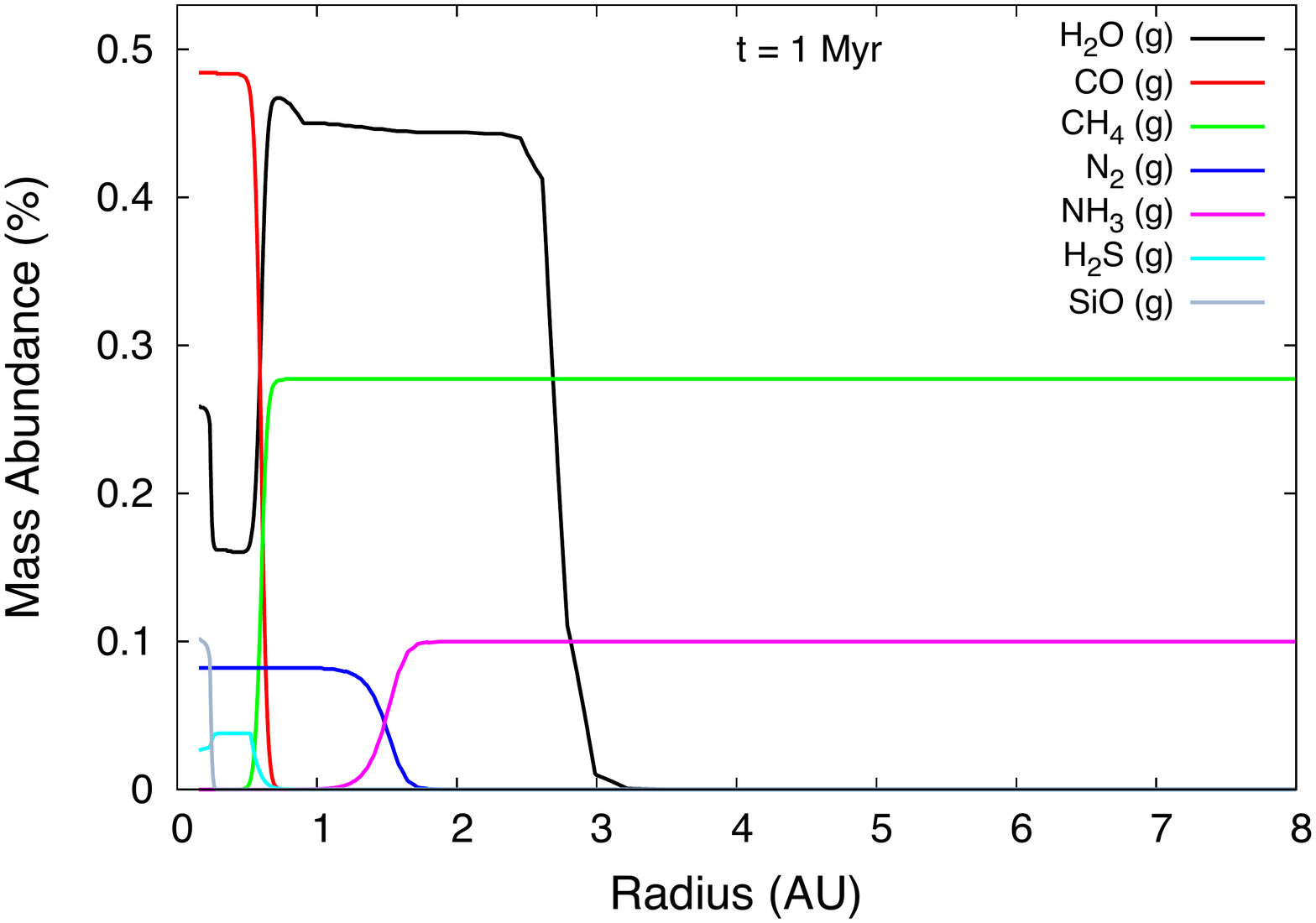}
\caption{Abundance of several prominent gases at 0.1 Myr (left), 0.5 Myr (center), and 1 Myr (right) into the disk's evolution for a fiducial disk (equation \ref{FiducialParameters}). Crossovers in abundances are shown to exist between carbon monoxide and methane, as well as molecular nitrogen and ammonia. This shows in what molecules it is energetically favourable for carbon and nitrogen, respectively, to exist at different regions of the disk governed by different temperatures. As was the case in figure \ref{Solids}, these features are seen to shift inwards with time as the disk evolves.}
\label{Gases}
\end{center} \end{figure*}

In order for equilibrium chemistry to be accurate, the timescale for chemical equilibrium must be shorter than the viscous timescale in the disk, which is $\sim$1 Myr. If this were not the case, the local temperature and pressure governing the chemistry would change faster than the material could find itself in chemical equilibrium. As was found in the experimental work presented in \citet{Toppani2006}, solids condense out of nebular gas on short timescales ($\sim$ 1 hour, on average). Thus, they are well represented by an equilibrium approach \citep{Pignatale2011}. Gases on the other hand have equilibrium timescales that are comparable to or longer than 1 Myr, and thus the equilibrium approach in inadequate for this subset of our chemical system. Examples of important non-equilibrium effects are grain surface reactions and UV dissociation \citep{VisserBergin2012}. 

In figure \ref{Gases}, we include abundance profiles of several prominent gases within the disk for completeness. We note that the abundances of molecular hydrogen and helium are by far the most abundant substances in the chemical system. The gases present in figure \ref{Gases} are the gases which have the highest abundances aside from these dominating gases.

Figure \ref{Gases} shows two interesting chemical features among these secondary gases. The first of which occurs at roughly 1.3 AU at 0.1 Myr. This feature displays a crossover in abundances of carbon monoxide and methane, along with an increase in abundance of water vapour, and takes place at a temperature of 1000 K \citep{Molliere2015}. At this location, as the midplane temperature and pressure decrease, it becomes chemically favourable for carbon to exist in methane as opposed to carbon monoxide. The leftover oxygen then combine with the molecular hydrogen, which is extremely abundant throughout the disk, to form more water vapour. This transition between CO and CH$_4$ is quite abrupt, spanning only a few tenths of an AU. 

The second interesting chemical feature shown in figure \ref{Gases} is a crossover between the abundances of molecular nitrogen and ammonia at roughly 3.3 AU at 0.1 Myr. This transition (along with the CO - CH$_4$ transition) provides a means for explaining the abundances of nitrogen in Terrestrial planet atmospheres in the Solar System, and the amounts of methane and ammonia in the Solar System's Jovians. Here, as the temperature decreases, it becomes more chemically favourable for nitrogen to exist within NH$_3$ as opposed to N$_2$. This crossover is much less abrupt, spanning several AU. We note that these distinct transitions in abundances of gaseous molecules is a feature of the equilibrium chemistry model. Such a sharp transition is not observed when photon driven chemistry and other non-equilibrium effects are taken into account, such as in \citet{Cleeves2013} and \citet*{Cridland2016}.

\subsection{Planet Traps}

As a planet forms within its natal disk, the mutual gravitational forces cause an exchange of angular momentum between the planet and the disk, leading to planet migration. The two torques that must be accounted for to track a planet's migration through the disk are the Lindblad torque and the corotation torque. As the planet forms, it excites spiral density waves throughout the disk at Lindblad resonances. The Lindblad torque is the summed interaction of the planet with disk material in these spiral waves. For most disk surface density and temperature structures, the Lindblad torque leads to low mass planetary cores losing angular momentum rapidly \citep{GoldreichTremaine1980}. This mechanism of transferring angular momentum from the planet to the disk leads to the planet migrating into its host star on a timescale of roughly $10^5$ years. This is problematic, as the core accretion model predicts planet formation to complete on timescales of at least $10^6$ years \citep{Pollack1996}. If only the Lindblad torque was operating, then this timescale argument would predict that gas giants cannot form without being tidally disrupted by their host stars. This problem is known as the type-I migration problem.

\begin{figure*}
\begin{center}
\includegraphics[width = 2.25 in]{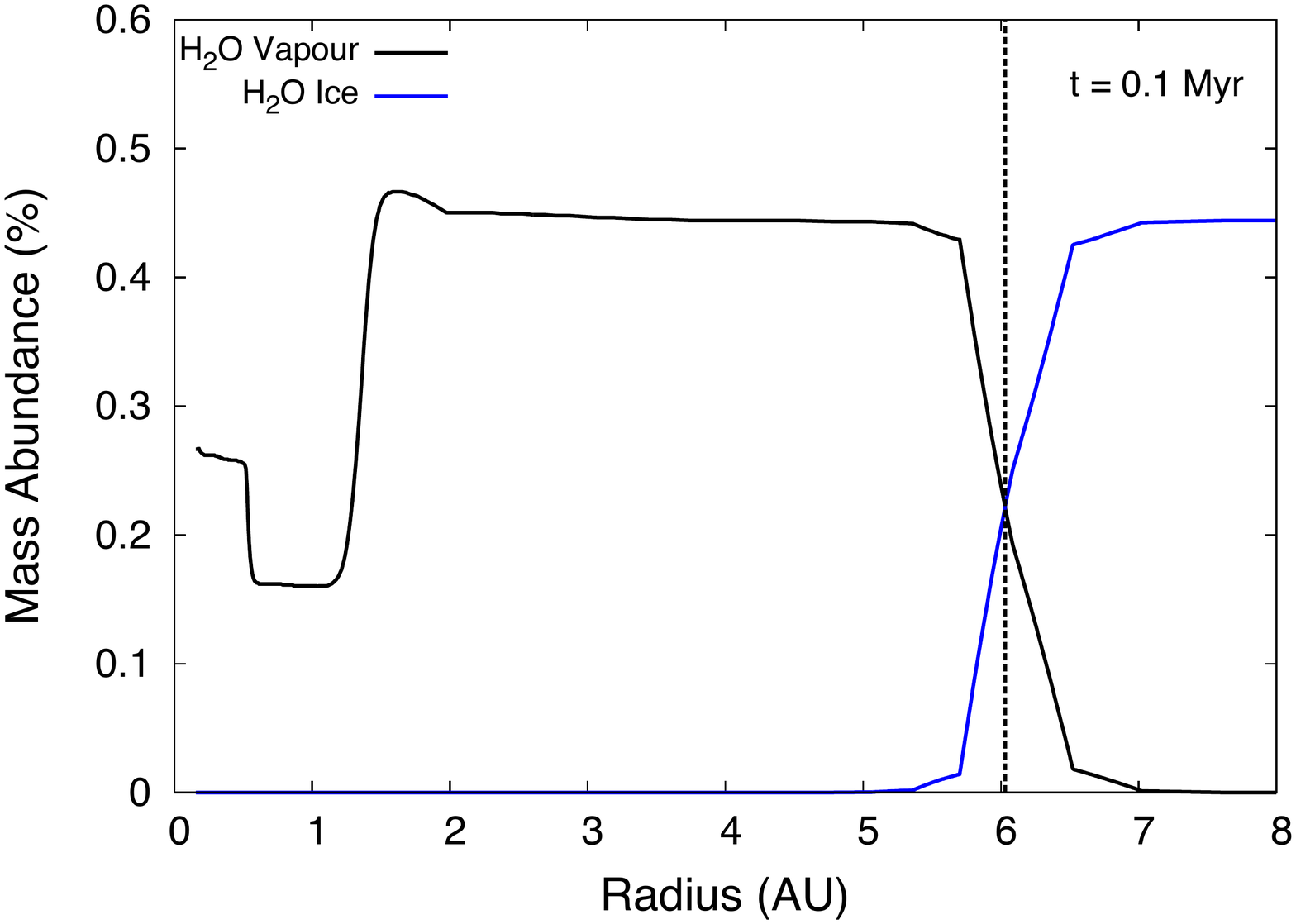} \includegraphics[width = 2.25 in]{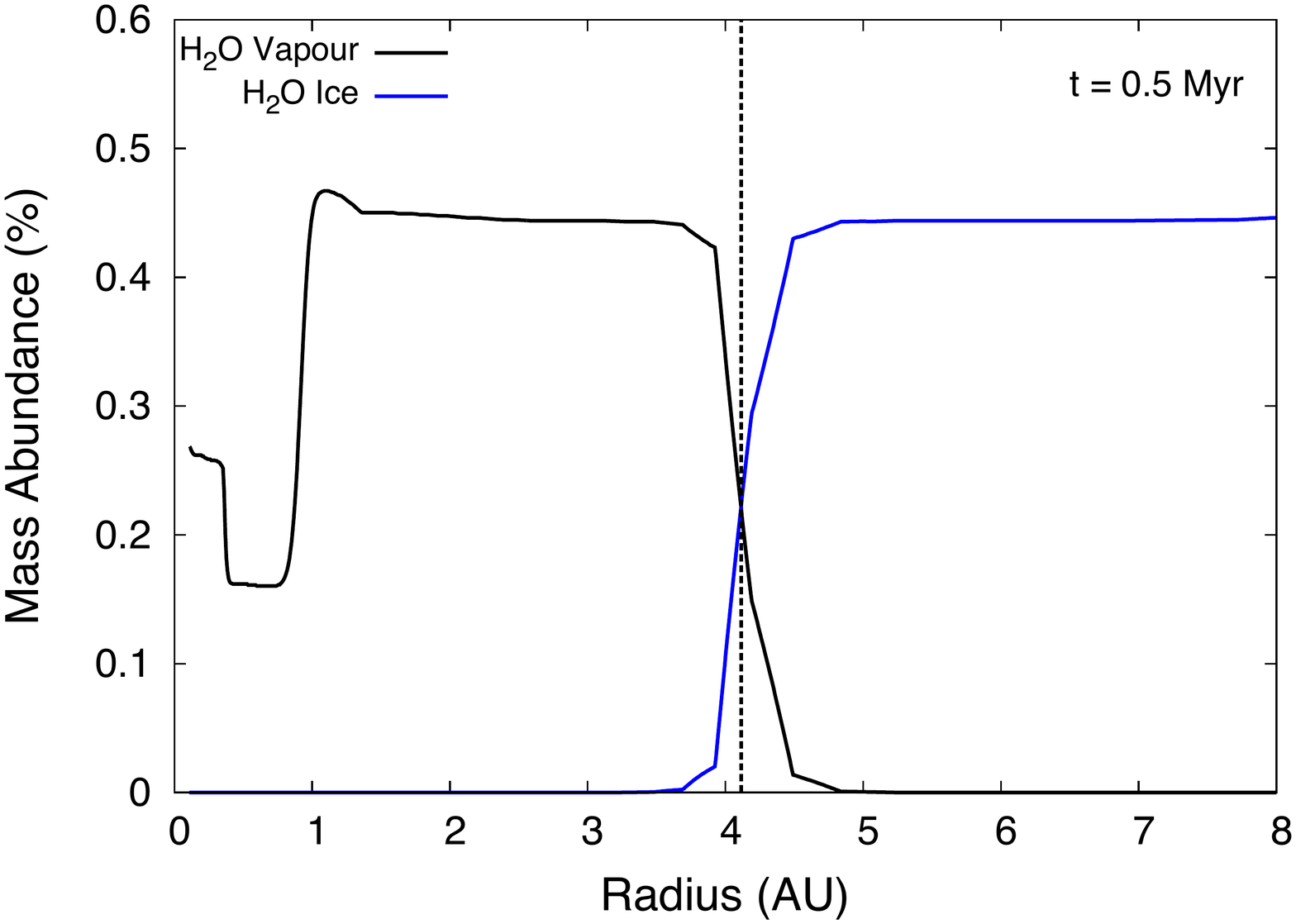} \includegraphics[width = 2.25in]{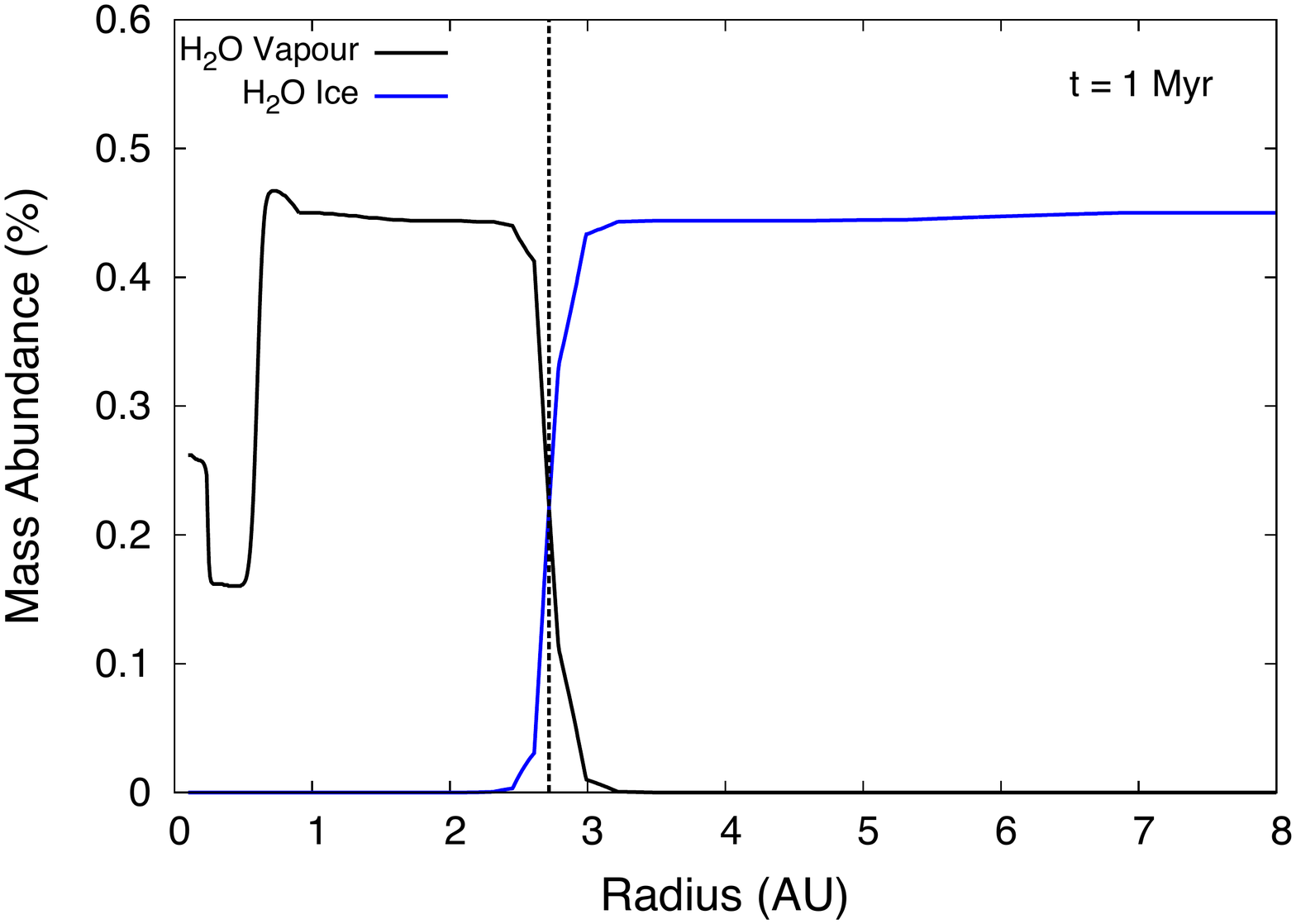}
\caption{Abundance profiles of gaseous and solid water are displayed at 0.1 Myr (left), 0.5 Myr (center), and 1 Myr (right) in a fiducial disk. The vertical lines depict the water ice line in our disk, defined as the radius where the abundance profiles intersect.}
\label{Water}
\end{center} \end{figure*}

As a possible mechanism to increase the planet's migration timescale, the corotation torque must also be considered. The corotation torque arises due to gravitational interactions between the planet and disk material orbitting the host star with a similar orbital frequency as the planet. This disk material undergoes horseshoe orbits transitioning from slightly lower orbits than the planet to slightly higher orbits on the libration timescale \citep{Masset2001, Masset2002}. If the disk material on horseshoe orbits does not exchange heat with surrounding fluid, there will be no net angular momentum transfer with the planet, and the corotation torque is said to be saturated. In this scenario, the corotation torque cannot act to slow down planet migration, and we are left with the same type-I migration problem outlined above. On the other hand, if the disk material on horseshoe orbits does exchange heat with surrounding disk material, the corotation torque is said to be unsaturated, and acts as a means to transfer angular momentum to the planet \citep{Masset2001, Masset2002}. The corotation torque is unsaturated as long as the libration timescale of horseshoe orbits is longer than the disk's viscous timescale. The operation of the corotation torque can act as a means to increase the planet's inward migration timescale to more than $10^6$ years as it exerts on outward torque on the planet. This gives planets enough time to form in the core accretion model, and is a solution to the type-I migration problem.

As was shown in \citet*{Lyra2010} and \citet{HP11}, disks with inhomogeneities in their temperature and surface density structures have unsaturated corotation torques near the inhomeneities. Planets that migrate into these disk inhomogeneities experience zero net torque due to planet-disk interactions. Thus, these inhomogeneities are appropriately named planet traps \citep{Masset2006}. A type-I migrating planet core that migrates to a radius coinciding with a trap will have its inward migration halted, and will grow within the trap. As is discussed in detail in \citet{HP11, HP12, HP13}, planet traps play a key role in preventing rapid inward migration of forming jovian planets and can reproduce the mass-semimajor axis distribution of exoplanets. The traps themselves migrate inwards on the disk's viscous timescale of roughly 1 Myr, which sets the timescale for the planet's formation. This migration timescale gives the planet enough time to build its core and accrete gases until it becomes massive enough to open up an annular gap in the disk, and liberate itself from the trap.

In this work, we only consider two main migration regimes: trapped type-I migration, and type-II migration following gap formation. Other works, such as \citet{HellaryNelson2012} and  \citet{Dittkrist2014} consider several type-I migration regimes (one of which is the trapped regime), which depend on the viscous, libration, u-turn, and cooling timescales. These works find that low mass cores (up to $\sim 4 M_\oplus$) are not trapped, but are rather in a locally isothermal migration regime. Additionally, they find that after the planet is trapped, the corotation torque can saturate for many disk configurations prior to the planet opening a gap in the disk.

We calculate these timescales using our model parameters and find that low mass cores are governed by the locally isothermal migration regime until the reach a mass of $\sim 3-5 M_\oplus$, depending on the particular trap used. We do not include the effects of this migration regime in this work, and rather force the low-mass cores to be in the trapped regime. Additionally, we find that the corotation torque does not saturate in our planet formation runs until after the planets have opened an annular gap in the disk. Therefore, corotation torque saturation does not affect planets forming in our model. We present this calculation in Appendix A.

The traps that are present within our disk are the heat transition, ice line, and outer edge of the dead zone. Since planets will be forming within the traps, the materials they have available for accretion are dictated by the location of the traps. Therefore, in order to track the materials a planet accretes throughout its formation, it is necessary to have a detailed understanding of the traps' locations in the disk. Below we discuss a summary of the physical origin of each of the three traps in our model, and how they are computed.

\subsubsection {Heat Transition}

The heat transition exists at the boundary between regions of the disk heated by different mechanisms. As discussed in section 2.1, the inner region of the disk with high surface density is heated predominantly by viscous dissipation, while the outer disk is heated by radiation from the host star due to the disk's flared profile. In order to calculate the location of the boundary throughout the disk's evolution, a heating model must be used. Since this is built into the \citet{Chambers2009} disk model, we can use equation \ref{Heat_transition} to track its location as the disk accretion rate evolves. The heat transition's location is shown to have a power law relationship with $\dot{M}$,
\begin{equation} r_t \propto \dot{M}^{28/33}\; . \end{equation}
At the location of the heat transition, the disk's surface density and temperature profile exhibit kinks, or inhomogeneities. Physically, this originates due to an entropy transition across the trap \citep{HP11}.

\subsubsection {Ice Line}

At an ice line, also known as a condensation front, the disk's opacity changes due to an increased amount of solid grains. While opacity change is not built into our disk model, having a sharp increase in opacity at the ice line would not result in globally different temperature or surface density profiles. However, at the location of the ice line, the local surface density and temperature profiles would change as a result of the opacity transition \citep{Stepinski1998}, giving rise to the conditions necessary for a trap \citep{MenouGoodman2004}. It is a drawback of the \citet{Chambers2009} model that these effects are omitted. We can still include this trap in the model by a slight modification discussed below.

To first-order, the location of the water ice line can be calculated by tracking the midplane location in the disk that has the condensation temperature of water, 170 K \citep{JangCondellSasselov2004}. However, this misses the second-order effects that pressure gradients throughout the disk can have on the ice line's location. Here, we use the equilibrium chemistry code to directly calculate the ice line's location. 

Figure \ref{Water} shows radial abundance profiles of gaseous and solid water along the disk's midplane. We define the location of the ice line, $r_{il}$, as the point of intersection of the two abundance profiles,
\begin{equation} X_{H_2O\textrm{ gas}}(r_{il}) = X_{H_2O\textrm{ solid}}(r_{il}) \; .\end{equation}
The ice line in figure \ref{Water} is denoted by a vertical dashed line. As the disk viscously evolves, the water ice line shifts inwards with decreasing disk accretion rate,
\begin{equation} r_{il} \propto \dot{M}^{4/9} \; ,\end{equation}
which is the same scaling obtained in \citet{HP11}.

Our definition of the ice line pinpoints one exact radius at each time as the transition between the two phases of water. Figure \ref{Water} shows that this transition takes place over roughly a few tenths of an AU, which is a small, but non-zero range of radii in the disk. Our model predicts that the disk opacity will be transitioning over this small region, and our definition of the ice line characterizes the average radius where a trapped planet will reside.

The abrupt phase transition of water near the ice line (spanning at most 0.3 AU) may be a result of our simplified 1D model which assumes a constant opacity. The model presented in \citet{Min2011} used a more detailed 2D disk opacity structure while performing radiative transfer calculations. At high accretion rates ($10^{-7}-10^{-6}$ M$_\odot$/yr), their model found that water undergoes a phase transition along the midplane spanning a larger range of up to $\sim 1.5$ AU. At lower accretion rates more comparable with typical disk accretion rates ($\dot{M}\sim10^{-8}$) in our model, however, the \citet{Min2011} model found that the water phase transition spans no more than 0.5 AU along the midplane, which is comparable to the results found in this work.

Our equilibrium chemistry code cannot compute a carbon monoxide ice line, which has been observed around other stars \citep{Qi2011}. Our equilibrium chemistry calculations have resulted in CO having a negligible abundance outside the CO-CH$_4$ abundance transition, taking place at roughly 1 AU. Given this result, our model does not predict any CO gas in the outer disk for a phase transition to take place. Photon-driven chemistry can cause dissociation of larger molecules, producing CO at intermediate and large radii. Our equilibrium chemistry model does not have the capability to include photon-driven effects. Therefore non-equilibrium chemistry models that include radiation effects, such as those presented in \citet{Cleeves2013, Cleeves2014} are best suited to track the structure and location of the CO ice line \citep*{Cridland2016}.

We note that we omit the ice line's effect on the disk opacity and resulting temperature and surface density structure in our model. We expect there to be an increase in surface density at the ice line that leads to the dynamic effect of a planet trap, but that is an unnecessary detail for our model as we do not directly compute the lindblad and corotation torques during the trapped type-I phase. Recently, \citet{Coleman2016} have shown that condensation fronts are the location of mass-independent planet traps, which further motivates our assumption of trapped migration throughout the type-I migration regime at the ice line.

\subsubsection {Dead Zone}

The dead zone is a region in the disk where the ionization fraction is insufficient for the magnetorotational instability (MRI) to be actively generating turbulence. Within the dead zone, rapid dust settling takes place due to a lack of turbulence. The outer edge of the dead zone separates the MRI active and inactive regions, and turbulence at this location gives rise to a wall of dust whose radiation heats the dead zone, leading to a thermal barrier on planet migration \citep{HP10}. This section will discuss our method of calculating the location of the dead zone's outer edge, which is a planet trap in our model.

\citet{HP11} incorporated a dead zone into their model using a piecewise function for the $\alpha$ parameter governing MRI viscosity. Other models that focus on detailed calculations of ionization rates throughout disks and resulting $\alpha$ values utilize 3D MHD simulations that include the non-ideal effects of ohmic dissipation, ambipolar diffusion, and the Hall effect \citep{Gressel2015}. These works result in $\alpha$ values that vary continuously throughout the disk, resulting in disk accretion rates that are both radially and time dependent. The choice of a constant $\alpha \sim 10^{-3}$ in our analytic model is an average value of these 3D simulations.

Chemical networks are also extremely useful for calculating ionization rates throughout disks. Non-equilbrium chemistry networks are particularly useful as they are able to account for photochemistry and ionization chemistry effects. Inclusion of these important effects allow these models to track ionization and recombination events \citep*{Cridland2016}, leading to detailed estimations of ionization fractions throughout the disk and the dead zone's location. The equilibrium chemistry model used throughout this work is limited as it cannot take into account these non-equilibrium effects necessary to track ionizations from first principles. We therefore employ an analytic ionization model as an alternative.

Our calculation of the dead zone follows the analytic model presented in \citet{MP2003}. By using an analytic model we are able to efficiently calculate ionization rates and dead zone locations over Myr of disk evolution while capturing the main results of detailed 3D simulations. The complete damping of MRI driven turbulence can be estimated analytically by balancing the MRI growth timescale with the ohmic diffusion timescale for all scales smaller than the disk's pressure scale height \citep{Gammie1996}. The resulting condition for a dead zone is then expressed via the magnetic Reynolds number \citep*{Fleming2000, MP2005},
\begin{equation} \textrm{Re}_M = \frac{V_A H}{\eta} \lesssim 100 \;, \label{DeadZoneCriterion} \end{equation}
where $V_A$ is the Alfv\'en speed, which is given by,
\begin{equation} V_A = \frac{B}{(4\pi \rho)^{1/2}} \simeq \alpha_{\textrm{turb}}^{1/2}c_s = \sqrt{\frac{\alpha_{\textrm{turb}} k T}{\mu m_H}} \,, \label{AlfvenSpeed} \end{equation}
and $\eta$ is the diffusivity of the magnetic field \citep{Blaes1994},
\begin{equation} \eta = \frac{234}{x_e}T^{1/2} \textrm{cm}^2\,\textrm{s}^{-1}. \end{equation}
It is through the magnetic diffusivity that the magnetic Reynolds number depends on the electron fraction, $x_e$. It is clear that in regions with sufficiently small electron fractions, $\eta$ will be large and the Reynolds number will be small, such that the condition for a dead zone (equation \ref{DeadZoneCriterion}) is satisfied. 

Recent works that use 3D MHD simulations use the magnetic Elsasser number as a measure MRI activity \citep{Blaes1994, Simon2013},
\begin{equation} \Lambda_0 = \frac{V_A^2}{\eta \Omega_K} \lesssim 1\; . \label{Elsasser} \end{equation}
The magnetic Reynolds number and the magnetic Elsasser numbers have the same physical origin of a ratio between MRI dissipation and growth, but have slightly different definitions based on the Elsasser number's inclusion of non-ideal MHD effects. Using equations \ref{DeadZoneCriterion}, \ref{AlfvenSpeed}, and \ref{Elsasser} we see that the magnetic Reynolds number and Elsasser numbers are related by,
\begin{equation} \Lambda_0 = \alpha_{\textrm{turb}} ^{1/2} \textrm{Re}_M \; .\end{equation} 
Since the $\alpha$ parameter in our disk model is 10$^{-3}$, the critical magnetic Reynolds number of 100 is consistent within a factor of order unity with a critical Elsasser number of 1. Therefore, our definition of the MRI active regions are consistent with current estimates using the Elsasser number, which take into account 3D MHD effects.

The electron fraction can be calculated in our ionization model as a solution to the following third-degree polynomial \citep{Oppenheimer1974},
\begin{equation} x_e^3 + \frac{\beta_t}{\beta_d}x_Mx_e^2 - \frac{\zeta}{\beta_d n}x_e - \frac{\zeta \beta_t}{\beta_d \beta_r n} x_M = 0 , \label{polynomial} \end{equation}
where $x_M = 0.0011$ is the metal fraction taken from the initial conditions to our chemistry model (see table \ref{Abundances}), and $n$ is the local number density of material in the disk. The ionization rate, $\zeta$, takes into account ionization from X-rays ($\zeta = \zeta_X$) or cosmic rays ($\zeta = \zeta_{CR}$).

There are three $\beta$ terms representing different recombination rate coefficients in equation \ref{polynomial}: the dissociative recombination rate coefficient for electrons with molecular ions ($\beta_d = 2\times 10^{-6} T^{-1/2}\,\textrm{cm}^3 \,\textrm{s}^{-1}$), the radiative recombination coefficient for electrons with metal ions ($\beta_r = 3\times10^{-11} T^{-1/2}\,\textrm{cm}^3\,\textrm{s}^{-1}$), and the rate coefficient of charge transfer from molecular ions to metal ions ($\beta_t = 3\times10^{-9} \textrm{cm}^3 \,\textrm{s}^{-1}$) \citep{MP2003}.

The ionization rate from X-ray sources is given by \citet{MP2003},
\begin{equation} \zeta_X = \left[\left(\frac{L_X}{kT_X4\pi d^2}\right)\sigma(kT_X)\right]\left(\frac{kT_X}{\Delta \epsilon}\right)J(\tau,x_0)\;, \end{equation}
where $L_X \simeq 10^{30}$ ergs s$^{-1}$ is the X-ray luminosity of the protostar, and $\sigma(kT_X)$ and $\tau(kT_X)$ are the absorption cross section and optical depth at the energy $kT_X = 4$ keV, which we choose to be an average X-ray energy. The distance between the X-ray source (taken to be 12 $R_\odot$ above the midplane at r= $12 R_\odot$, to represent magnetospheric accretion onto the protostar) and some point on the disk surface is denoted by $d$, and the energy to make an ion pair is $\Delta \epsilon \simeq 13.6$ eV. The first factor in the above equation in square brackets represents primary ionizations, assuming the same energy $E=kT_X$ for all primary electrons. The second term $kT_X/\Delta \epsilon$ represents secondary electrons produced by a photoelectron with energy $kT_X$. The last factor $J(\tau,x_e)$  represents attenuation of X-rays. The dimensionless energy parameter is defined as $x = E/kT_X$, and the attenuation factor $J$ is written as,
\begin{equation} J(\tau, x_0) = \int_{x_0}^\infty x^{-n}\exp(-x-\tau(kT_X)x^{-n}) dx \,. \label{RadiativeTransfer}\end{equation}
The optical depth $\tau(kT_X)$ is given by,
\begin{equation} \tau(kT_X) = N_H\sigma(kT_X)\;,\end{equation}
and the absorption cross section is,
\begin{equation} \sigma(KT_X) = 8.5\times10^{-23} \,\textrm{cm}^2\left(\frac{kT_X}{\textrm{keV}}\right)^{-n}, \end{equation}
where n = 2.81 \citep*{Glassgold1997}. The surface number density $N_H$ is measured along the ray path from the X-ray source. If $\alpha'$ is the angle between the ray path and the radial axis, then the surface number density is,
\begin{equation} N_H = \frac{\int_z^\infty n(a, z') dz'}{\sin \alpha'} \end{equation}
where $a$ is the disk radius. The integral in equation \ref{RadiativeTransfer} was numerically evaluated by setting the lower limit $x_0 = 1$ and an upper limit of $x = 100$ \citep{MP2003}. 

\begin{figure} \begin{center}
\includegraphics[width = 3.5 in]{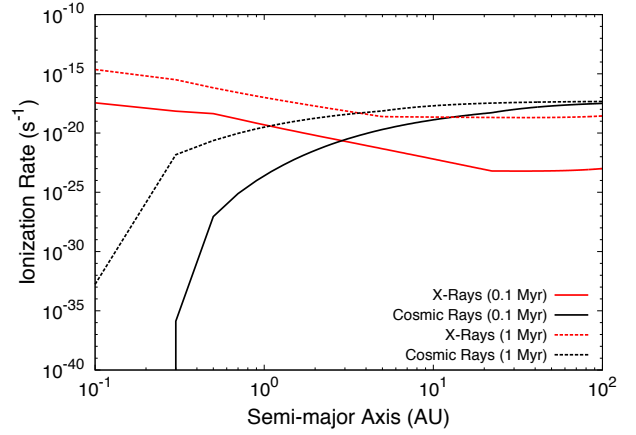}
\caption{Here we picture ionization rates throughout the disk midplane that result from each ionization source. X-rays dominate ionization at small radii, close to the X-ray source, while cosmic rays dominate ionization in the outer disk due to low surface densities. At later times, ionization rates increase due to lower surface densities resulting from disk evolution.}
\label{Ionization}
\end{center} \end{figure}

The ionization rate by cosmic rays is estimated to be $10^{-17}$ s$^{-1}$ \citep{Spitzer1968}. Using this with an attenuation length for cosmic rays of 96 g cm$^{-2}$ \citep{Umebayashi1981}, we calculate the ionization rate due to cosmic rays at the disk midplane using \citep{Sano2000},
\begin{equation} \zeta_{CR}(a) = \frac{10^{-17}\,\textrm{s}^{-1}}{2}\exp \left(-\frac{\Sigma(a)}{96\,\textrm{g}\;\textrm{cm}^{-2}}\right). \end{equation}

It is currently unclear if X-rays or cosmic rays dominate ionization in protostellar disks. Some models suggest that protostellar winds can attenuate cosmic rays prior to them reaching the disk, resulting in cosmic ray ionization rates 1 to 2 orders of magnitude lower than the assumed value of $10^{-17}$ s$^{-1}$ \citep{Cleeves2014}. Conversely, recent observations of young protostellar systems have suggested a higher ionization rate throughout molecular clouds \citep{Ceccarelli2014} than these models would predict. These observations have been attributed to the presence of ionizing cosmic rays which are generated in protostellar jets in the model presented in \citet{Padovani2015}. 

Due to the uncertainty in the importance of cosmic ray ionization in disks, the ionization model we present here takes the conservative approach of including both X-rays and cosmic rays individually. By separately considering the two ionization effects, we can discern differences in the resulting dead zone locations and time evolutions. Additionally, we can determine the types of planets that form as a result of the dead zone traps caused by the two ionization sources. In a future work, we will use a population approach to determine if the dead zone resulting from each of the two ionization sources can form a mass-period distribution of planets consistent with exoplanetary data (Alessi et al. 2016, in prep.).

In Figure \ref{Ionization} we plot ionization rates throughout the disk midplane caused by X-rays and cosmic rays. The ionization rates we obtain agree reasonably well with those presented in \citet{Gressel2015}, which were obtained using 3D MHD simulations. A key difference between the two ionization sources in our model is that X-rays originate at the protostar, thus having a diminishing flux at large radii in the disk, while cosmic rays shine down on the disk from an external source and have a constant flux across all radii. We find that X-rays dominate disk ionization in the inner disk, as these regions are closer to the X-ray source, and experience a much higher X-ray flux then outer regions. Additionally, the higher surface densities in the inner disk heavily attenuate cosmic rays, causing the cosmic ray ionization in these regions to be small. In the outer disk, the surface density is smaller, allowing cosmic rays to dominate disk ionization in this region. As the outer regions of the disk are farther from the X-ray source region, more of the X-rays are attenuated by the time they reach the outer disk resulting in a low X-ray ionization rate. Including the effects of X-ray scattering would cause the X-ray ionization rate to be larger in the outer disk than our model predicts, since a portion of the X-rays would be scattered to the outer disk instead of being attenuated. Additionally, figure \ref{Ionization} shows that at later times, the ionization rates throughout the disk due to both X-rays and cosmic rays increase. This is due to the disk surface density decreasing as evolution takes place, resulting in lower attenuation rates for both sources.

The X-ray dead zone is the trap that exists at the largest semimajor axes in our model. Planet formation in this trap will lead to planets accreting material within a region of the disk whose chemistry is strongly affected by inheritance from the stellar core (\citet{Pontoppidan2014}, see discussion at start of section 2.3), that we do not account for in our chemistry model. We note, however, that the X-ray dead zone trap quickly evolves towards the inner regions of the disk, within several $10^5$ years, where the chemistry is dominated by in-situ formation of material that our model considers. Therefore, we do not expect the process of inheritance of chemical materials from the stellar core to have a strong impact on our resulting planet compositions.

After the ionization rates throughout the disk have been determined, the electron fraction $x_e$ as a function of radius can be obtained by numerically solving equation \ref{polynomial} at each disk radius. Lastly, the particular radius that gives an $x_e$ satisfying equation \ref{DeadZoneCriterion} will be the location of the outer edge of the dead zone.

\begin{figure} \begin{center}
\includegraphics[width = 3.5 in]{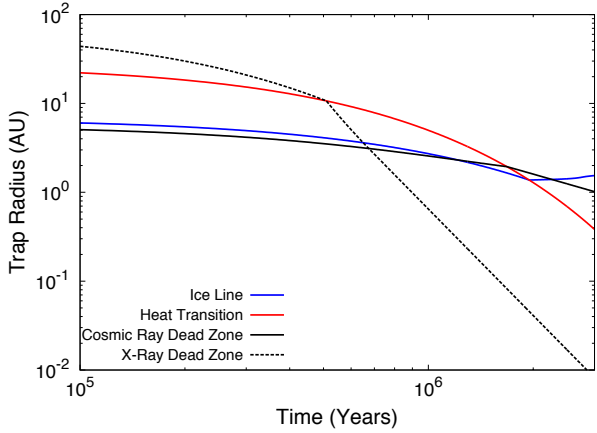}
\caption{Time evolution of planet traps throughout the disk with fiducial parameters (see section 2.1). We compute dead zone locations by considering both X-ray and cosmic ray ionization individually.}
\label{traps}
\end{center} \end{figure}

Figure \ref{traps} shows the location of the planet traps in the Chambers disk with fiducial parameters (equation \ref{FiducialParameters}) evolving with time. In this figure, we calculate the dead zone's location by considering X-ray and cosmic ray ionization rates individually. We note that the ionization source does not affect the location of the heat transition or ice line. For the majority of a typical disk's lifetime, the cosmic ray dead zone lies interior to the ice line while the heat transition lies outside. The X-ray dead zone is seen to lie exterior to the ice line only at the earliest times in figure \ref{traps}. After intersecting the heat transition at several $10^5$ years, the X-ray dead zone quickly migrates to the innermost regions of the disk, and is the only planet trap in our model to migrate interior to 0.1 AU for a fiducial disk. This behaviour shows that the evolution of the X-ray dead zone is sensitive to the local surface density and temperature profiles. Within the viscous regime, the X-ray dead zone evolves drastically, whereas in the irradiated regime its evolution is much slower. 

Throughout the disk's lifetime, the traps intersect, and planets forming within these traps will have a non-negligible dynamical interaction. Dynamical interactions between planets forming within different traps has been considered in \citet*{IdaLin2010, HellaryNelson2012, Ida2013, Alibert2013} \& \citet{Coleman2014}. Dynamic effects between multiple forming planets in one disk is not accounted for in our work, as we assume that individual planets form in isolation. Including dynamics between forming planets in our model will be the subject of future work.

\subsection{Core Accretion Model}

We follow the formalism presented in \citet{HP12} to calculate accretion and migration rates throughout a planet's formation, which is based upon the model developed in \citet{IdaLin2004}. In this model, there are several critical masses that act as boundaries between various migration and accretion regimes that a forming planet must surpass as it accretes material, building up a Jovian mass planet. We discuss below in detail these various critcial masses and timescales while summarizing in table \ref{MassRegimes}. 

We begin our planet formation calculations at $10^5$ years into the disk's lifetime with a 0.01 $M_\oplus$ core situated at a semimajor axis that coincides with a particular trap in the disk. While it is unlikely that a planetary core will initialize at the exact location of a planet trap, the type-I migration timescale is short enough that the core will rapidly migrate inwards until it encounters a trap. Low mass cores accrete solids via the oligarchic growth process in our model. During this phase, we increase the disk's solid surface density to $0.1\Sigma$. This is an order of magnitude beyond the dust-to-gas ratio predicted by the chemistry model. This increase was necessary in order for our model to produce gas-accreting cores on timescales smaller than a fiducial disk lifetime of 3 Myr that lead to Jovian planets forming. Physically, this enhancement of solids can be caused by the effects of dust trapping \citep{Lyra2009}, but currently the value of the solid increase is a free parameter in our model. The planet's accretion timescale in this regime is \citep{KokuboIda2002}, 
\begin{equation} \begin{aligned} \tau_{c,acc}  \simeq &1.2\times10^5\;\textrm{yr}\;  \left(\frac{\Sigma_d}{10\;\textrm{g cm}^{-2}}\right)^{-1}
\\ & \times \left(\frac{r}{r_0}\right)^{1/2}\left(\frac{M_p}{M_\oplus}\right)^{1/3}\left(\frac{M_*}{M_\odot}\right)^{-1/6} 
\\ & \times\left[\left(\frac{b}{10}\right)^{-1/5}\left(\frac{\Sigma_g}{2.4\times10^3\;\textrm{g cm}^{-2}}\right)^{-1/5} \right.
\\ & \left. \times \left(\frac{r}{r_0}\right)^{1/20}\left(\frac{m}{10^{18} \;\textrm{g}}\right)\right]^2\;,\;\label{Solid_Accretion} \end{aligned}\end{equation}
where $\Sigma_d = 0.1\Sigma$ is the surface density of solids, $M_p$ is the mass of the core, $b \simeq 10$ is a parameter used to define the feeding zone of the core, $\Sigma_g = 0.9\Sigma$ is the surface density of gas, and $m \simeq 10^{18}$ g is the mass of planetesimals being accreted. Using this timescale, the growth of cores is given by,
\begin{equation} \frac{dM_p}{dt} = \frac{M_p}{\tau_{c,acc}} \propto M_p^{2/3} \;.\label{Solid_Accretion2}\end{equation}
This stage of core formation, where the planet is accreting solids from a planetesimal disk prior to gas accretion will be referred to as stage I of planet formation.

\begin{table} 
\caption{A summary of accretion and migration for various mass regimes throughout our core accretion model.}
\begin{center}
\begin{tabular}{|c|c|c|}
\hline
\textbf{Mass Range} & \textbf{Migration} & \textbf{Accretion} \\
\hline
$M<M_{c,crit}$ & Trapped type-I& Planetesimals \\
\hline
$M_{c,crit} < M < M_{\textrm{GAP}}$ & Trapped type-I & Gas \& Dust \\
\hline
$M_{\textrm{GAP}} < M < M_{\textrm{crit}}$ & Type-II & Gas \& Dust \\
\hline
$M_{\textrm{crit}} < M < M_{\textrm{MAX}} $ & Slowed type-II & Gas \& Dust \\
\hline
$M > M_{\textrm{MAX}}$ & Slowed type-II & Terminated \\
\hline
\end{tabular} 
\end{center}
\label{MassRegimes}
\end{table}

During oligarchic growth, a small gaseous envelope surrounding the planetary core will be in hydrostatic balance with pressure provided by the energy released by accreted planetesimals. This hydrostatic balance prevents the planet from accreting any appreciable amount of gas. As found in \citet*{Ikoma2000}, the envelope is no longer in hydrostatic balance when the mass of the core exceeds,
\begin{equation} M_{c,crit} \simeq 2\,M_\oplus\left(\frac{1}{10^{-6} M_\oplus \;\textrm{yr}^{-1}}\frac{dM_p}{dt}\right)^{1/4} \;, \label{Critical_Core_Mass} \end{equation}
where we have not included the dependence of $M_{c,crit}$ on the envelope opacity. This chosen parameterization is not unique, but rather corresponds to a low envelope opacity of $10^{-4} - 10^{-3}$ cm$^2$ g$^{-1}$. When the planet's mass exceeds $M_{c,crit}$, it is able to start accreting appreciable amounts of gas from the disk \citep{IdaLin2004}. The planet continues to accrete planetesimals, albeit at a reduced rate, as its inward migration continues to replenish its feeding zone \citep{Alibert2005}.

We assume that the availability of solids is reduced after the oligarchic growth stage takes place, and change the solid surface density to be that which coincides with the dust to gas ratio from the chemistry calculation, $\Sigma_d = 0.01\Sigma$. Solid accretion is still governed by the timescale in equation \ref{Solid_Accretion}, albeit at a reduced rate due to the lower dust to gas ratio. Growth of the planet is now dominated by accretion of gases, governed by the Kelvin-Helmholtz timescale, given by,
\begin{equation} \tau_{KH} \simeq 10^c \, \textrm{yr}\left(\frac{M_p}{M_\oplus}\right)^{-d}\;.\label{Gas_Accretion}\end{equation}
where $c=9$ and $d=3$ are parameters that depend on the planet's envelope opacity \citep{Ikoma2000}. We note that, in contrast to equation \ref{Critical_Core_Mass}, the Kelvin-Helmholtz parameters chosen correspond to larger envelope opacity values of 0.1-1 cm$^2$ g$^{-1}$. While the envelope opacity, which itself is uncertain, links the parameterization of the critical core mass and Kelvin-Helmholtz timescale, previous works have treated these as independent parameters \citep{IdaLin2008, HP12, HP13}, similar to the model presented here.\footnote{In a future population synthesis paper, we will restrict our parameterizations of $M_{c,crit}$ and $\tau_{KH}$ to be self-consistent in terms of envelope opacity, which will reduce our model's parameter set by two. We note that gas accretion timescales will have small effects on our super Earth masses and compositions, and thus will not affect the main conclusions of this work.} The gas accretion rate is then given by,
\begin{equation} \frac{dM_p}{dt} \simeq \frac{M_p}{\tau_{KH}}\,. \label{Gas_Accretion2} \end{equation}

We note that our model does not consider the enrichment of the planet's atmosphere due to impacting planetesimals, which is expected to be an important process when considering the atmospheric composition of super Earths or Neptunes \citep{Fortney2013}. Instead, our model assumes gas accretion is solely due to direct accretion from the disk. It is unclear, however, how large of an effect impacting planetesimals will have on gas abundances in Jovian planets' atmospheres.

Initially, when the planet's mass has just increased beyond $M_{c,crit}$, the timescale for gas accretion is long ($\sim 10^6$ years). This stage of slow gas accretion will be referred to as stage II. As the mass of the planet increases, it eventually will become large enough that it will be accreting gas at a fast enough rate such that its atmosphere will no longer be pressure supported, giving rise to an instability. When this occurs, the atmosphere collapses and the planet rapidly accretes its atmosphere. Quantitatively, this takes place when $\tau_{KH} < 10^5$ years. This segment of the formation process is referred to as runaway growth, and will be denoted as stage III.

Throughout the early phases of slow gas accretion, the planet remains in the trapped type-I migration regime (see discussion in section 2.3 and Appendix A). As the planet increases its mass, it exerts an increasingly large torque on the disk, eventually leading to the formation of an annular gap. Gap formation liberates the planet from the trap it was forming within.  To estimate the mass at which a planet opens up a gap, two arguments can be used. The first is that the planet's torque on the disk must be greater than the torque that disk viscosity can provide. Otherwise, the disk's viscosity will suppress gap formation. The second argument is that the planet's Hill sphere must be larger than the disk's pressure scale height, or else disk pressure will prevent a gap from opening. This critical mass is referred to as the gap-opening mass, and is given by \citep{MP2006},
\begin{equation} M_{\textrm{GAP}} = M_*\;\textrm{min}\left[3h^3(r_p), \sqrt{40\alpha h^5(r_p)}\right]\,, \label{GapOpeningMass} \end{equation}
where $r_p$ is the radius of the planet, and $h(r_p) = H_p/r_p$. During the phase where the planet is forming within a gap in the disk, the planet's migration is referred to as type-II migration. We note that our gap-opening criteria predicts the planet to be in the type-II migration regime once it overcomes the gap-suppressing effect of either disk thermal pressure or viscosity, considered independently of one-another. This causes our predicted $M_{GAP}$ values to be smaller than the model shown in \citet*{Crida2006}, which considers both gap-suppressing effects simultaneously.
 
Once a planet opens a gap in its natal disk, the migration rate of the planet is governed by the accretion timescale of the disk material onto the star ($\sim10^6$ years). In this regime, the planet migrates inwards with velocity,
\begin{equation} v_{\textrm{mig,II}} \simeq -\frac{\nu}{r}\,. \end{equation}
When the planet reaches a critical mass \citep*{Ivanov1999},
\begin{equation} M_{\textrm{crit}} = \pi r_p^2 \Sigma_g(r_p)\;,\end{equation}
it will be massive enough that its inertia will resist inward migration occurring with the evolution of the disk \citep{HP12, HellaryNelson2012}. In this regime, the migration velocity is,
\begin{equation} v_{\textrm{mig,slowII}} \simeq -\frac{\nu}{r(1 + M_p/M_{\textrm{crit}})} \;.\end{equation}
Typically, type-II migration applies to planets midway through stage II of their formation, while slowed type-II migration applies to planets in the late phases of stage II and throughout stage III in our models.

The last critical mass in our core accretion model is one that acts as an upper limit to how massive a planet will become. We scale a planet's maximum mass with its gap opening mass as follows \citep{HP13},
\begin{equation} M_{\textrm{MAX}} = f_{\textrm{MAX}}M_{\textrm{GAP}}\;, \label{MaximumMass} \end{equation}
with $f_{\textrm{MAX}}$ being the parameter that expresses the ratio between a planet's final mass and the mass at which it opened a gap. Previous works have shown that accretion onto a planet slows and eventually terminates after a planet opens a gap in the disk \citep{Lissauer2009}. However, flow onto the planet does not terminate immediately when a gap is opened in the disk. Numerical works have shown that a substantial amount of disk material can flow through the gap and be accreted by the planet \citep*{Lubow1999, Lubow2006}. The parameterization shown in equation \ref{MaximumMass} acknowledges gap opening as a key stage in terminating the accretion onto a planet, linking the planet's final mass with that at which it opens a gap.

Motivated by this, we use the parameterization given equation \ref{GapOpeningMass} to estimate the mass reservoir that planets can accrete from post-gap formation in our model. Typically $f_{\textrm{MAX}}$ is in the range of 10 to 100, with an $f_{\textrm{MAX}}$ of 10 producing Jovian planets of mass comparable to Jupiter \citep{HP13}. Planets with $f_{\textrm{MAX}}$ outside this range are also possible, as an $f_{\textrm{MAX}} \simeq 1$ would produce a planet whose accretion is sharply truncated when it opens a gap. Alternatively, an $f_{\textrm{MAX}}$ of several hundred would represent a planet whose accretion is driven long after it opens up a gap. This scenario has been shown to be possible if the disk possesses sufficient viscosity \citep{Kley1999} or if the planet can excite spiral density waves giving rise to an eccentric disk \citep{KleyDirksen2006}. 

Other works, such as \citet{Machida2010}, \citet{Dittkrist2014}, and \citet{Bitsch2015} use a disk-limited accretion phase to model the growth of planets in the mass range of $\gtrsim$ 30 M$_\oplus$. In these models gas accretion onto a planet post-gap formation is limited by the local supply of material from the disk. With such a model, the accretion rate of gas onto a planet decreases with time after gap formation occurs, in agreement with results found in hydrodynamic simulations such as \citet{Lubow1999}.

Conversely, our work only considers the Kelvin-Helmholtz timescale for gas accretion at all planet masses (from stage II onward) until the planet reaches its maximum mass given by equation \ref{MaximumMass}. This approach is limited as it produces accretion rates that increase with planet mass even after gap formation has taken place, contrary to results of hydrodynamic simulations. While our step-function model of accretion onto high mass planets is a simplistic treatment of a continuous process controlled by planet and disk properties, it has been shown in \citet{HP13} to produce planet populations in agreement with observations. Additionally, both the Kelvin-Helmholtz and disk-limited accretion methods, accretion onto high mass planets is sensitive to the planets' envelope opacities \citep{HP14, Mordasini2014}. Depending on the particular envelope opacity that is used, both methods can produce similar mass-period and core mass-envelope mass distributions.

\begin{figure}
\includegraphics[width = 3.5 in]{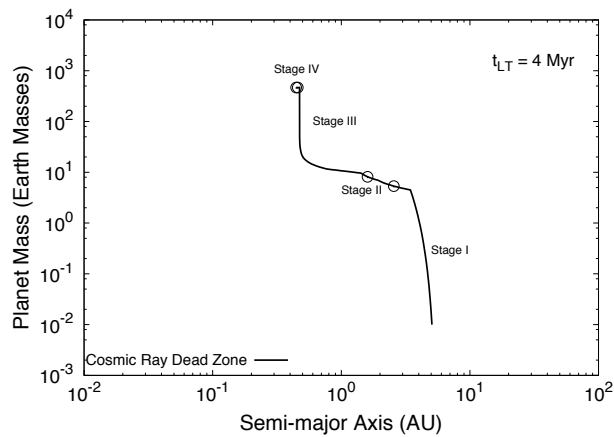}
\caption{An example formation track for a planet forming in our model within the cosmic ray dead zone trap in a fiducial disk (equation \ref{FiducialParameters}). Open circles along the track represent 1 Myr time markers. Oligarchic growth (stage I) takes place in $\sim 10^6$ years, which is shorter than the migration timescale causing the planet to move nearly vertically in this diagram. Stage II of the planets formation takes $\simeq 2$ Myr for this planet, causing it to move more horizontally as slow gas accretion takes place. Runaway growth (stage III) takes place in $< 10^5$ years until gas accretion is terminated as the planet reaches its maximum mass. The planet undergoes slow type-II migration in stage IV until the disk photoevaporates at 4 Myr, giving a final planet mass of 1.56 Jupiter masses and semimajor axis of 0.65 AU.}
\label{ExampleTrack}
\end{figure}

Further study of the late stages of planet formation in our model will be the subject of future work (Alessi \& Pudritz 2016, in preparation). Here, we adopt a fiducial value of $f_{\textrm{MAX}} = 50$, as this value results in Jovian planet masses that give an average fit to masses of giant exoplanets. After the planet has reached its maximum, or final mass, accretion is terminated. From this time onward, the planet will undergo slowed type-II migration until the disk photoevaporates at $t = t_{LT}$. We refer to this final stage of terminated accretion as stage IV.

In Figure \ref{ExampleTrack} we show the resulting formation track for a planet forming within the dead zone trap caused by cosmic ray ionization in a disk with initial mass 0.1M$_\odot$. The figure outlines the four stages, and by plotting the planet's mass as a function of its semimajor axis throughout formation, accretion and migration timescales can easily be compared. In oligarchic growth (stage I) the planet builds up its solid core in a short timescale of $\lesssim 10^6$ years, accreting a few M$_\oplus$ of solids while its trapped inward migration allows it to only move radially roughly 1 AU. The timescale to build the core is significantly shorter than the disk lifetime, which is 4 Myr in this case. 

As the core mass reaches a few M$_\oplus$ the solids in the planet's feeding zone have been depleted and its main accretion source becomes the gas and dust in the disk. Initially, gas accretion takes place slowly in stage II, and the evolution of the cosmic ray dead zone trap causes the planet to migrate appreciably as it accretes its atmosphere. Midway through stage II, the planet's mass exceeds the gap opening criterion, whereby the planet is no longer trapped and begins to undergo type-II migration. The timescale for stage II is roughly 2 Myr in this case, which is significantly longer than the oligarchic growth timescale. We emphasize that the timescale for slow gas accretion is comparable to disk lifetimes for planets forming in our model. As the planet enters stage III, runaway growth proceeds, whereby the planet rapidly accretes gas and reaches its maximum mass in less than $10^5$ years. Runaway growth allows the planet to satisfy the mass criterion for slowed type II migration. This allows it to migrate inwards on a timescale longer than the disk's viscous timescale during stage IV after accretion has been terminated. Thus, the planet does not migrate inwards appreciably for the remaining 1-2 Myr of the disk material being present, prior to photoevaporation taking place. The end of the disk's lifetime marks the final mass and semimajor axis of the planet.

We emphasize that the disk lifetime sets an upper limit to the time that the planet formation process can take. Planets that have a formation time which is less than the disk lifetime are able to reach their maximum mass defined in equation \ref{MaximumMass}. For the alternate scenario, planet accretion and migration ceases at the disk lifetime as the disk material is no longer present. Depending on the timing of disk dispersal, planets can be stranded during stages I, II or III of their formation. Since the timescale for stage II to take place is much longer than stages I or III, it is much more probable that a planet will be stranded in stage II than other stages of formation. Comparing with figure \ref{ExampleTrack}, stranding a planet during stage II would result in a planet with mass consistent with a super Earth or mini Neptune.

During our planet formation runs, we use the disk's abundance at the planet's current location to characterize the abundance of the material accreted onto the planet. In doing so, we assume that planets are sampling the disk's local abundance throughout their formation. We present a detailed algorithm describing our process of tracking planets' compositions throughout their formation in Appendix B.

\begin{figure}
\includegraphics[width = 3.5 in]{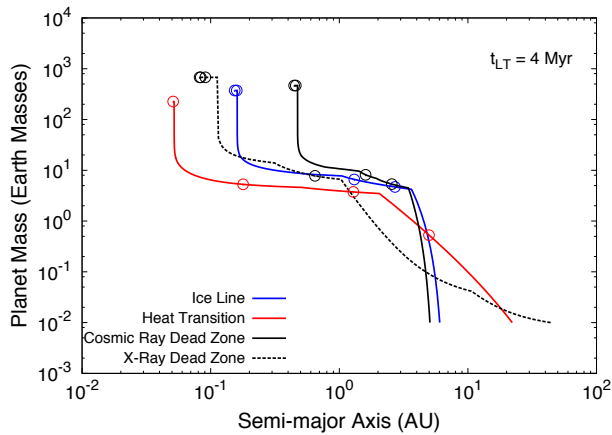}
\caption{Formation tracks for planets forming within each of the traps in a fiducial disk with a 4 Myr lifetime. Open circles along the tracks represent time stamps at 1 Myr intervals. The disk lifetime is sufficiently long for all four planets to complete stage III of their formation, resulting in four Jovian planets with distinct semimajor axes.}
\label{4Jupiters}
\end{figure}

\section{Results}

\begin{figure}
\includegraphics[width = 3.5 in]{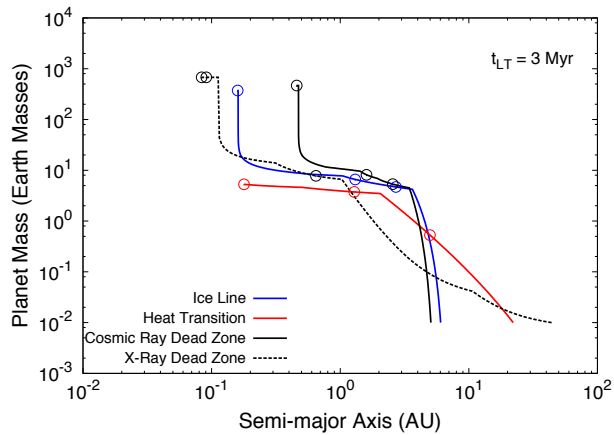}
\includegraphics[width = 3.5 in]{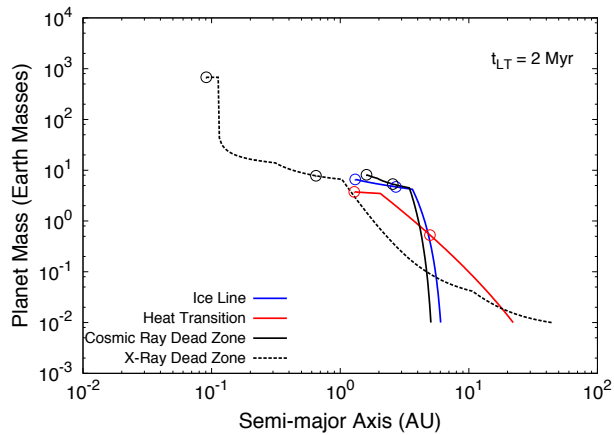}
\caption{Here we plot planet formation tracks in each of the four traps in a fiducial disk (0.1 M$_\odot$ initial mass) with reduced disk lifetimes. In the upper panel, $t_{LT}$ = 3 Myr, and this set up results in the heat transition trap producing a super Earth. In the lower panel, $t_{LT}$ = 2 Myr, and planets forming ice line, cosmic ray dead zone, and heat transition traps are all super Earths. Only the planet forming in the X-ray dead zone trap, which has the shortest formation timescale, produces a Jupiter-mass planet in both cases.}
\label{DecreasedLifetime}
\end{figure}

\subsection{Dependence of Planet Evolutionary Tracks on Disk Lifetime}

Figure \ref{4Jupiters} shows evolutionary tracks for planets forming within each of the traps in our model, in a 0.1 M$_\odot$ disk with a 4 Myr lifetime. This disk is sufficiently long-lived for Jovian planets to result from planet formation in all of the traps. The cosmic ray dead zone planet formation track is the same track that was presented in figure \ref{ExampleTrack}. We now compare accretion and migration timescales for planets forming in all four of the traps by contrasting the shapes of the evolutionary tracks on this diagram.

The formation tracks pertaining to the ice line and cosmic ray dead zone traps appear similar due to the traps themselves occupying nearby regions of the disk. The resulting planets, however, have substantial differences in their semimajor axes, with the cosmic ray dead zone producing a 0.45 AU gas giant while the ice line gives rise to a warm gas giant at 0.15 AU in this case. The difference in the final semimajor axes of these two planets is caused by the ice line's inward migration occurring on a shorter timescale than the inward migration of the cosmic ray dead zone.

The heat transition is shown to produce a hot Jupiter in the 4 Myr-lived disk, as is shown in figure \ref{4Jupiters}. The trap itself migrates inwards the fastest out of the three traps. Also, due to the trap being the farthest out in the disk, the planet forming within this trap is in a region with the smallest surface density of solids among these three planets. This causes the planet forming in the heat transition to have a small accretion rate during stage I, so the resulting timescale for stage I of formation is $\sim$ 2 Myr. These factors cause the planet to have a significantly lower mass at the beginning of gas accretion compared to the other tracks, causing the timescale for slow gas accretion to be longest ($\gtrsim$ 1.5 Myr) in the heat transition. This long timescale causes the planet to migrate inwards past 0.1 AU prior to runaway growth taking place, and the planet resulting from formation within the heat transition is a hot Jupiter.

Lastly, the planet forming within the X-ray dead zone trap starts the farthest out in the disk. However, due to the X-ray dead zone's rapid inward migration (see figure \ref{traps}), the planet migrates within an AU prior to 1 Myr into the disk's lifetime. This allows the planet to build its solid core in a region with a large amount of solids, completing stage I of its formation in $\lesssim 10^6$ years. With such a high accretion rate of planetesimals, its mass at the beginning of stage II is the largest among any of the four planets shown in figure \ref{4Jupiters}. Due to this, gas accretion takes place quickly when compared with the other formed planets. The planet reaches its maximum mass prior to the 2 Myr mark, showing that the X-ray dead zone lends itself to forming planets the fastest out of all the traps in our model. 

Figure \ref{DecreasedLifetime} shows one of the key issues this paper addresses; namely, how do super Earths form? Specifically, we show the effects of decreasing the disk lifetime by plotting the same tracks as shown in figure \ref{4Jupiters}, but in disks that get photoevaporated after 3 Myr (top) or 2 Myr (bottom). In the case of a 3 Myr disk, the planets forming in the ice line and both dead zone traps remain unaffected as their formation is complete prior to the disk lifetime. However, the planet forming within the heat transition is still in its slow gas accretion phase at the point of disk dispersal and gets stranded with a mass of 5.4 M$_\oplus$ at roughly 0.2 AU, resulting in a super Earth. 

In the case of a disk with a lifetime of 2 Myr (figure \ref{DecreasedLifetime}, lower panel), we find that planets forming in the ice line, cosmic ray dead zone, and heat transition traps become stranded in their slow gas accretion phase of formation at the time of disk dispersal. Planet formation in each of these three traps in this short-lived disk result in failed cores at roughly 1 AU. The heat transition produces a super Earth with a mass of 4 M$_\oplus$, and the ice line and cosmic ray dead zone traps produce planets with masses of roughly 10 M$_\oplus$. Conversely, the planet forming in the X-ray dead zone has a formation timescale of less than 2 Myr, so its formation completes prior to disk dispersal, and is again unaffected by the shorter disk lifetime. The X-ray dead zone is the only trap in our model that produces a Jovian planet in a 2 Myr-lived disk.

\subsection{Dependence of Planet Evolutionary Tracks on Disk Mass}

Up to this point, we have focused only on the variation of the disk's lifetime, and how small values of $t_{LT}$ can lead to super Earth and hot Neptune formation. The disk's mass is another key parameter in our model that has been shown by \citet{IdaLin2004}, \citet{Mordasini2012a}, \& \citet{HP13} to play a key role in shaping the mass period relation of exoplanets. 

\begin{figure}
\includegraphics[width = 3.5 in]{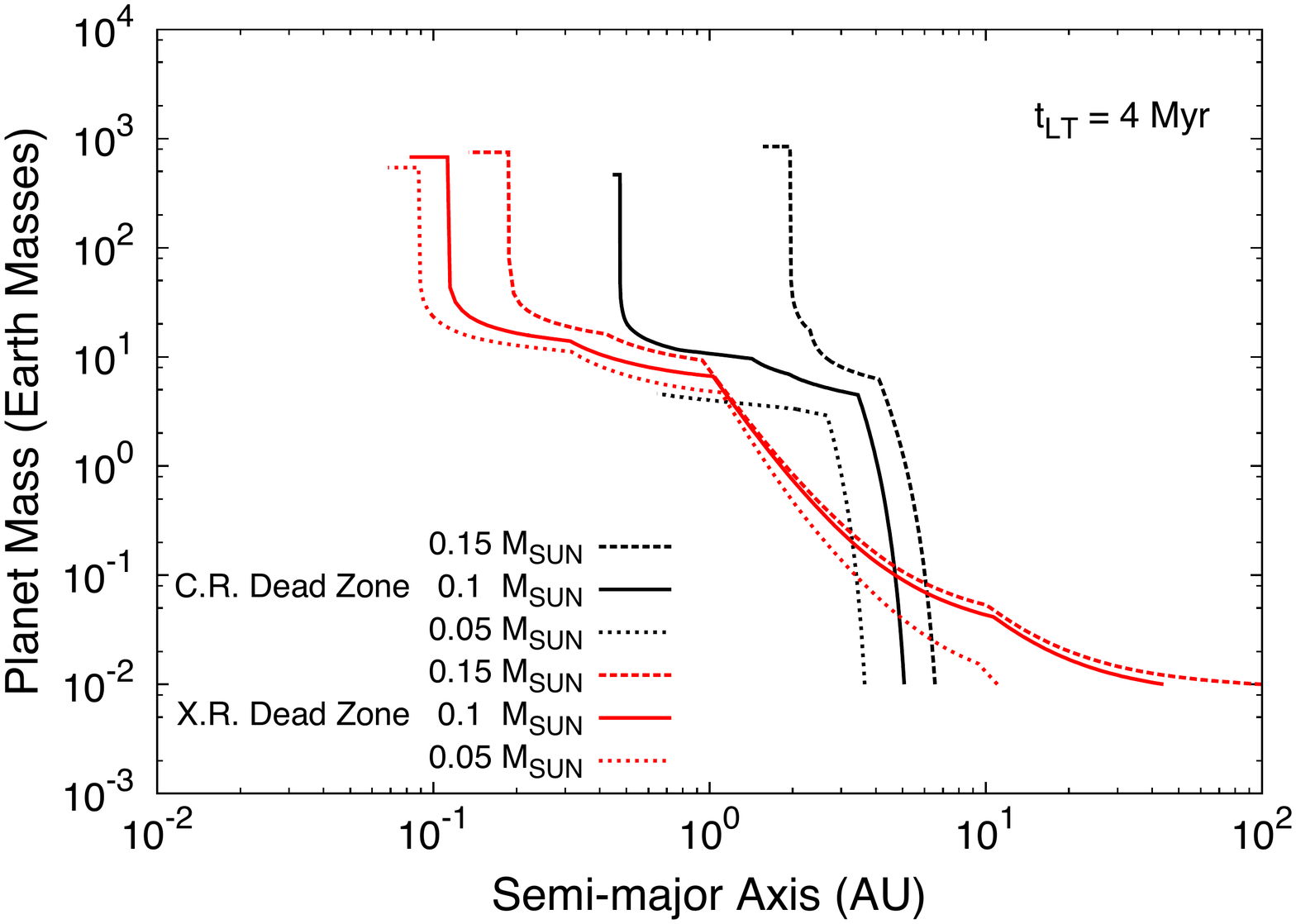}
\includegraphics[width = 3.5 in]{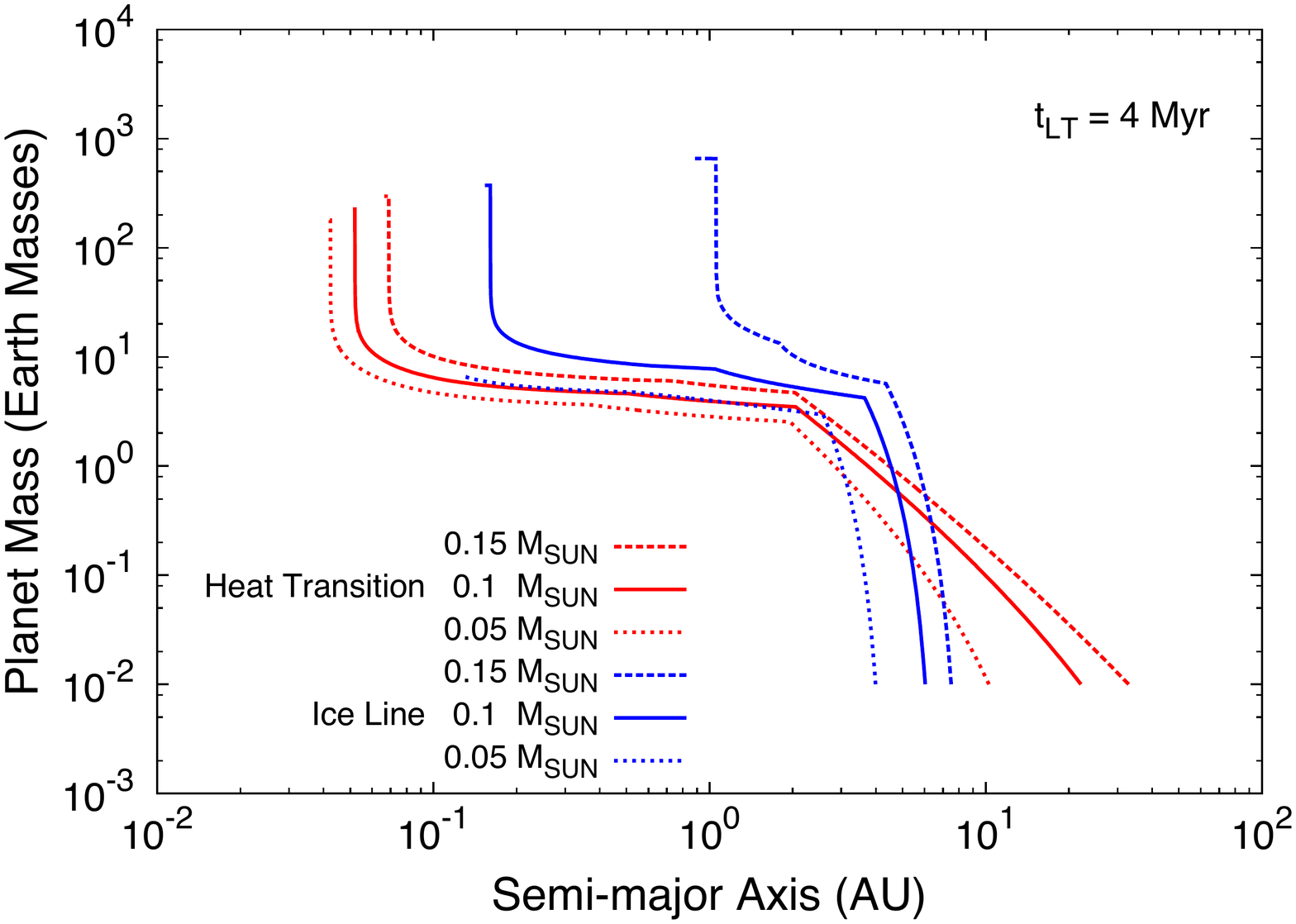}
\caption{Planet formation tracks are shown within each of the four traps while the initial disk masses are varied. Initial masses considered are 0.05 M$_\odot$, 0.1M$_\odot$, and 0.15 M$_\odot$. We find that the heat transition and X-ray dead zone traps are the least sensitive to the disk's mass, producing planets of similar masses and final locations in all three cases. Conversely, the cosmic ray dead zone and ice line traps are quite sensitive to this parameter. Both traps produce super Earths in the case of the 0.05 M$_\odot$ disk.}
\label{MassParameter}
\end{figure}

We chose an initial disk mass of 0.1 M$_\odot$ as a fiducial value, and we now vary this initial disk mass in order to determine the effect on resulting planet masses and final locations. In figure \ref{MassParameter}, we plot planet tracks from each of the four traps in our model that are computed in disks with initial masses of 0.05 M$_\odot$, 0.1 M$_\odot$, and 0.15 M$_\odot$. We hold the disk lifetime at a constant value of 4 Myr as this value resulted in all four traps producing a gas giant in a fiducial mass disk (see figure \ref{4Jupiters}).

In figure \ref{MassParameter} we see that traps move out to larger radii as the disk mass increases. This causes the planets to begin their formation farther from their host stars. In all four traps, planet formation in a more massive disk (0.15 M$_\odot$) takes place on a shorter timescale, and results in more massive planets orbiting at larger semimajor axes than the fiducial disk mass produces. Conversely, smaller disk masses result in lower mass planets forming at smaller separations from their host stars. Additionally, planet formation takes longer as the disk mass decreases. This is shown in figure \ref{MassParameter}, as the ice line and cosmic ray dead zone traps produce super Earths in the 0.05 M$_\odot$ disk mass case, showing that the planet formation timescale increased beyond the 4 Myr disk lifetime in both traps. The results are consistent with those presented in \citet{HP11, HP12}.

In figure \ref{MassParameter}, we find that the ice line and cosmic ray dead zone traps are the most sensitive to the initial disk mass. Variation of this parameter from 0.05 to 0.1 to 0.15 M$_\odot$ causes these two traps to produce planets of entirely different classes. In particular, we find that the two traps produce 1 AU Jupiters in the heaviest disks, warm Jupiters in the fiducial case, and super Earths in the lightest disks. The X-ray dead zone and heat transition traps, on the other hand, are insensitive to the particular disk mass used. While it remains true for these two traps that the planet formation timescale increases for smaller disk masses, a disk lifetime of 4 Myr still produced gas giants in both traps, even in the lightest disks considered. Moreover, the final locations of planets that result from the X-ray dead zone and heat transition occupy a small region on the mass-semimajor axis diagram. Both traps produce hot Jupiters in all cases considered, and the final masses and locations of the planets do not depend heavily on the initial disk mass used.

\subsection{Super Earth Abundances}

\begin{figure*} \begin{center}
\includegraphics[width = 2.3 in]{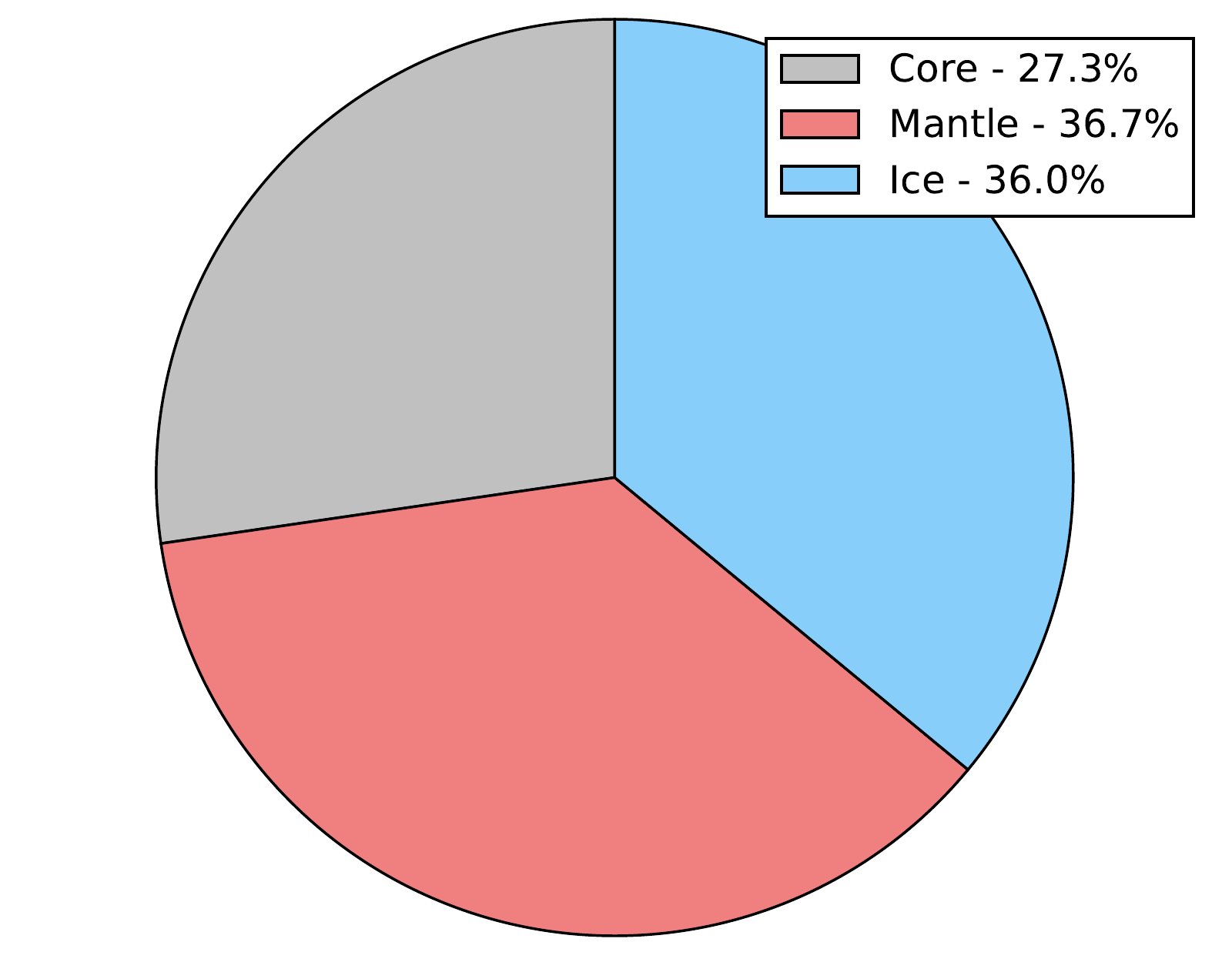} \includegraphics[width = 2.3in]{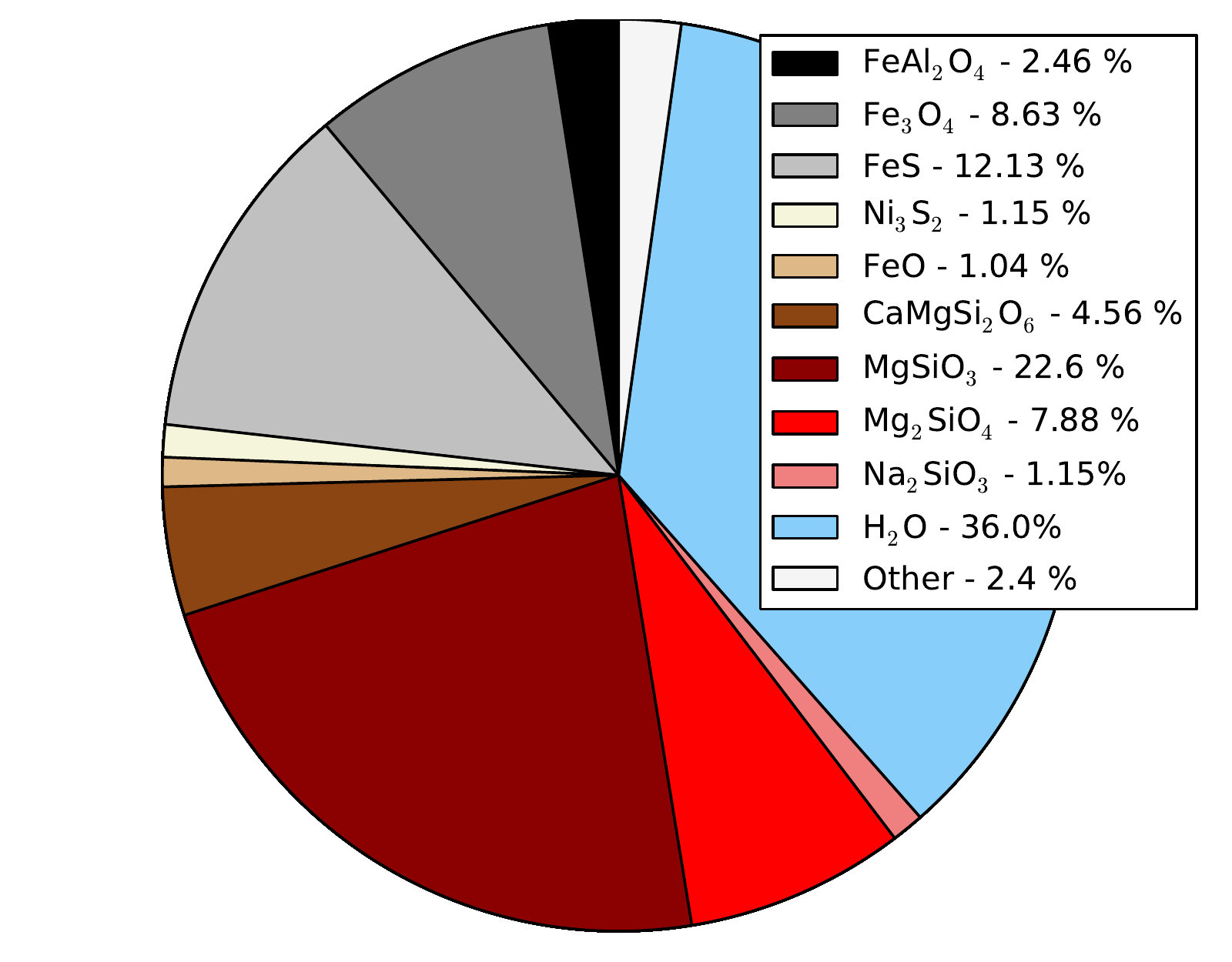} 
\includegraphics[width = 2.3 in]{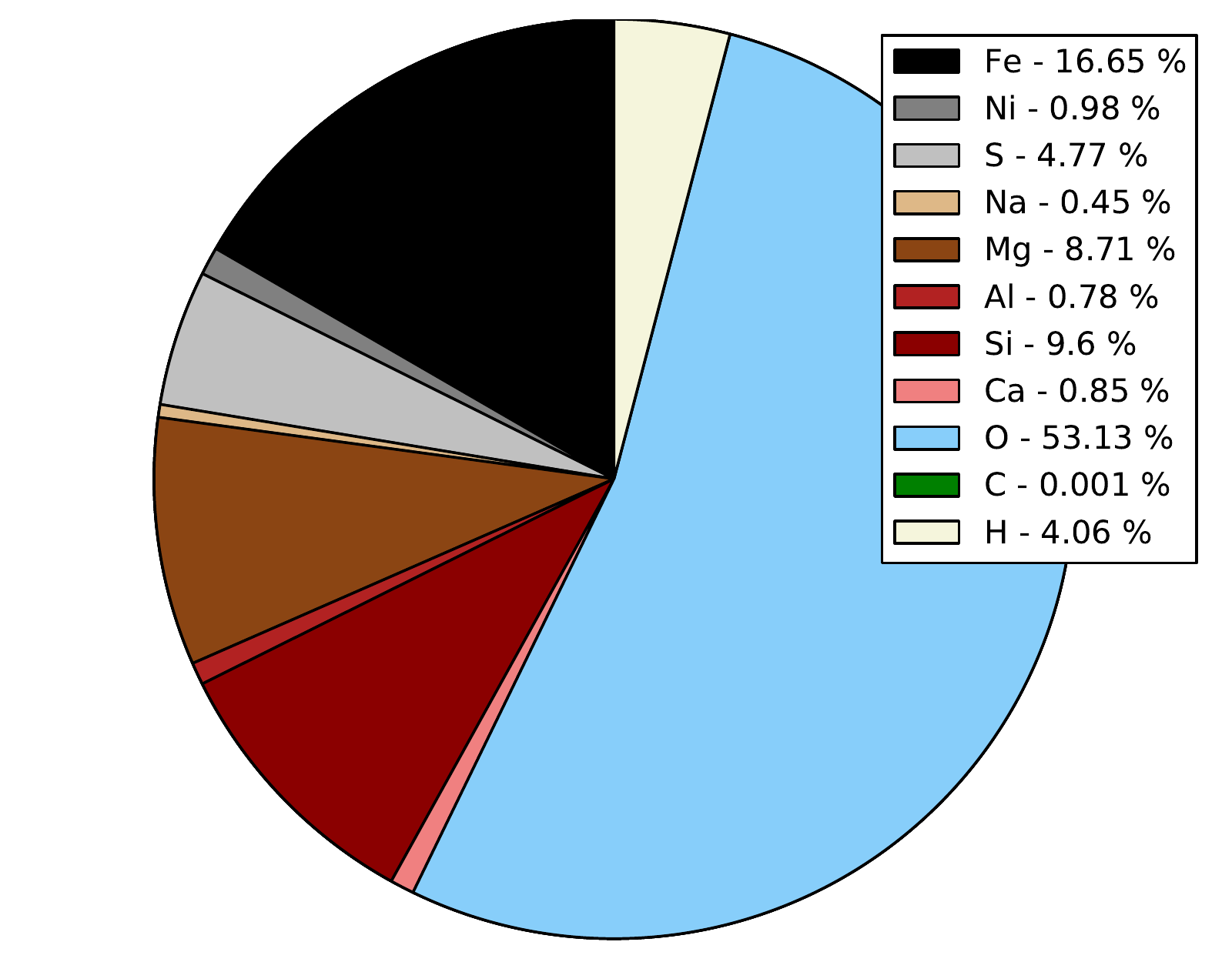} \end{center}
\caption{\textbf{Left}: The mass distribution among solid components for the super Earth formed in the heat transition in the $t_{LT} = 3$ Myr run (figure \ref{DecreasedLifetime}, top panel), with final mass 5.4 M$_\oplus$ and semimajor axis of roughly 0.2 AU. Since the planet accretes solids primarily exterior to the ice line, it has a substantial mass fraction in ice. \textbf{Middle}: Mass fractions in individual solids are plotted for the same planet. The first five materials in the legend are classified as core materials. The next four are considered mantle materials. Water ice shows the same abundances in both plots as it is the only ice considered in our chemical model. All solids that had a mass fraction of $< 1$ \% on the planet were binned as other on the pie chart. \textbf{Right}: Mass fractions in individual elements are shown for this planet.}
\label{3MyrHTPlanet} 
\end{figure*}

\begin{figure*}
\includegraphics[width = 2.3 in]{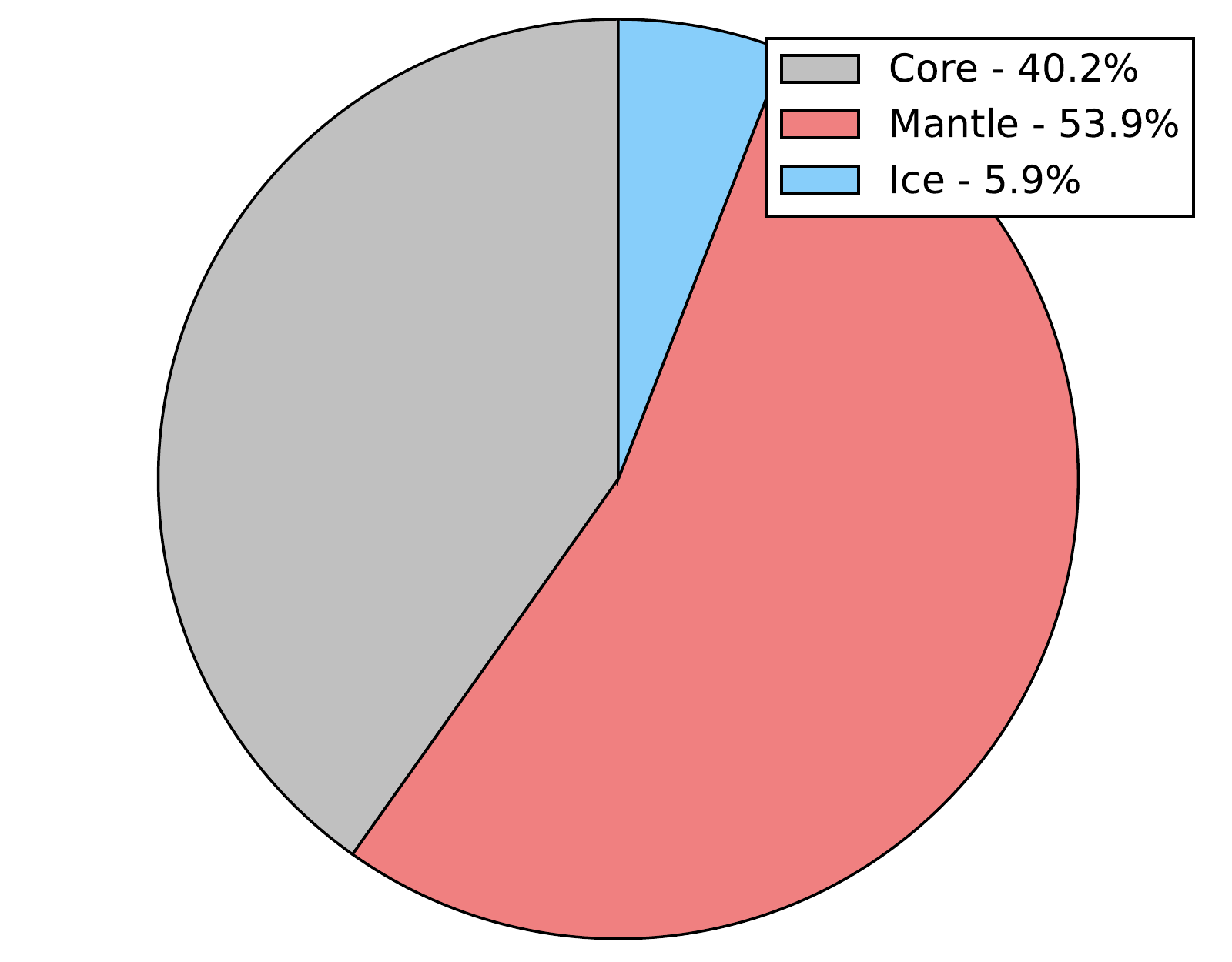} \includegraphics[width = 2.3in]{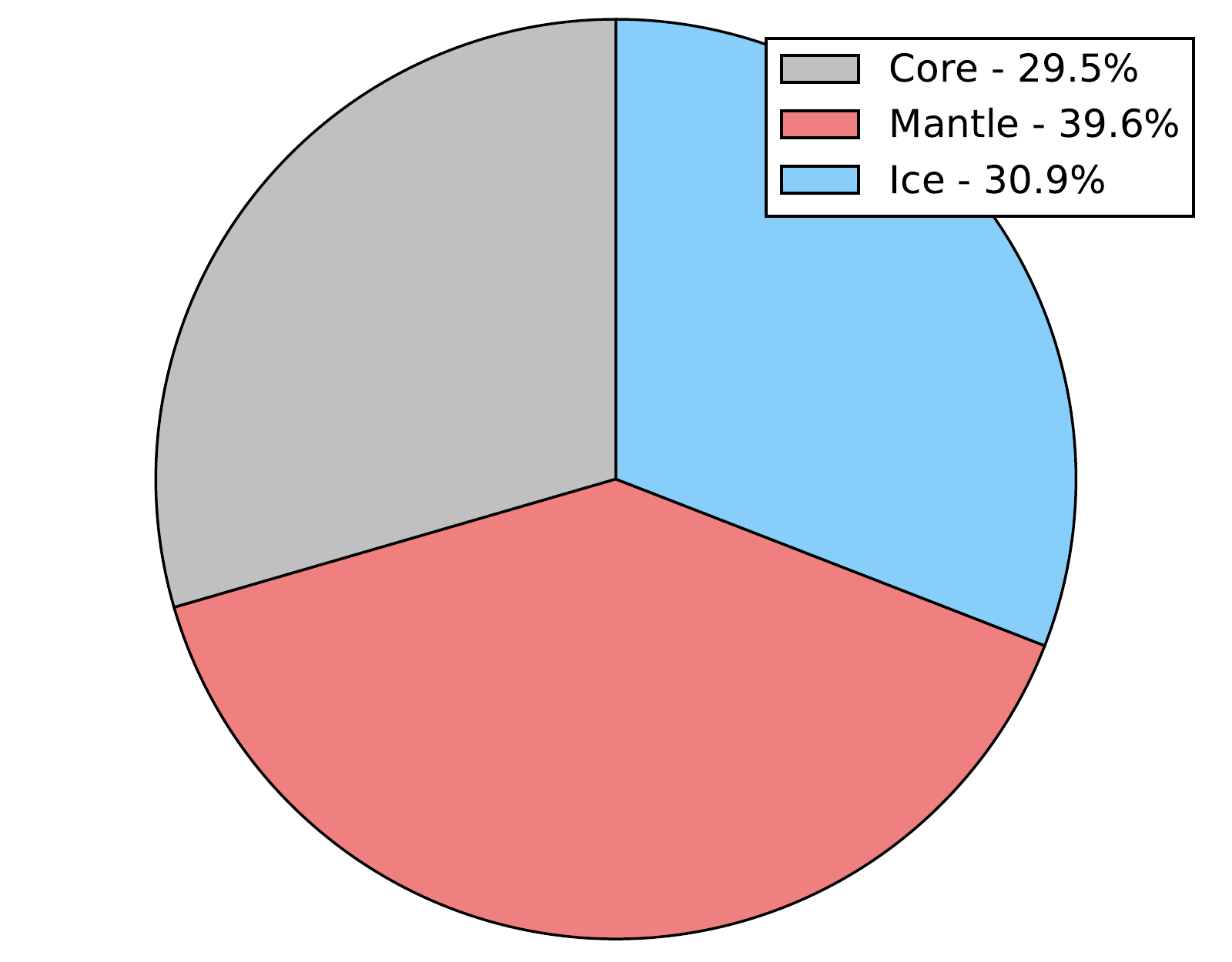}
\includegraphics[width = 2.3in]{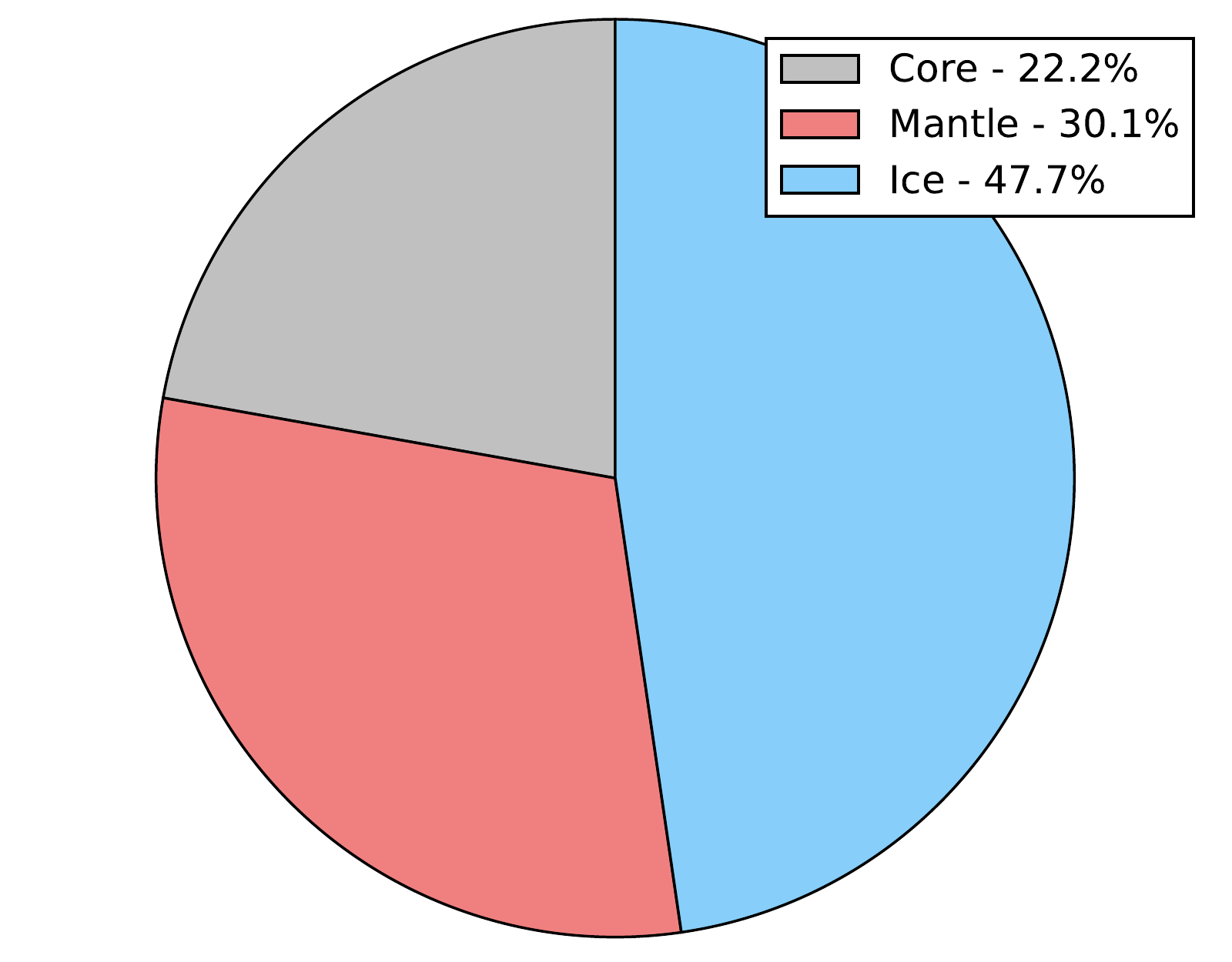} 
\caption{Solid mass abundances among solid components for the super Earths formed in the simulation with $t_{LT}$ = 2 Myr, shown in figure \ref{DecreasedLifetime} (bottom). The dead zone planet (left) is quite dry as it forms interior to the ice line, while the ice line (middle) and heat transition (right) planets have accreted a substantial mass in ice as they form in colder regions of the disk.}
\label{SuperEarths}
\end{figure*}

In figure \ref{3MyrHTPlanet}, we show the the solid composition of the super Earth produced in the heat transition in the $t_{LT}$ = 3 Myr disk (see figure \ref{DecreasedLifetime}, top panel, for the planet's formation track). The left panel shows how this planet's solid mass is distributed among core materials, mantle materials, and ice. Since the heat transition trap lies exterior to the ice line for the majority of the disk's lifetime, the super Earth spends the majority of its time accreting solids in an ice-rich environment. This leads to the planet accreting a substantial amount of ice (36\% of its mass) prior to the time of disk dispersal. Because the planet migrates throughout its formation, it samples disk chemistry over a range of disk radii. Therefore, its final composition does not correspond to that of any single radius in the disk.

Figure \ref{3MyrHTPlanet} also shows how the 3 Myr heat transition planet's mass is distributed among specific solids in the chemistry model. In addition to having over one third of it's solid mass in ice, there are several other core and mantle refractories that comprise a large fraction of the planet's solid mass. The planet's core material component is dominated by mass in troilite (FeS) and magnetite (Fe$_3$O$_4$), while the major silicates that have been accreted onto this planet are enstatite (MgSiO$_3$) and forsterite (Mg$_2$SiO$_4$). 

Lastly, the right panel of figure \ref{3MyrHTPlanet} shows the elemental abundances of this super Earth's solid component. We find the planet is very enriched in oxygen compared to Solar abundances. The large oxygen content mainly results from the abundant amount of ice the planet has accreted. Iron and sulphur comprise the majority of the planet's content in core materials, while magnesium and silicon make up its mass in mantle materials. We also find that the planet has a negligible amount of carbon in its solid component. This is consistent with \citet{Bond2010}, who found little carbon content among terrestrial planets forming in disks with C/O ratios similar to the Solar value of 0.54. 

\citet{Bond2010} found that substantial amounts of graphite can form along the midplane only in disks with C/O ratios over 2 times the Solar value, leading to an appreciable fraction of planets' solid masses being comprised of graphite. Our disk model, conversely, uses a Solar C/O ratio, leading to negligible amounts of graphite forming along the disk midplane. Because of this, our planet formation models result in very small C/O ratios in the solid components of super Earths.

By tracking solids accreted onto the three super Earths formed in the 2 Myr disk (see figure \ref{DecreasedLifetime}, lower panel, for formation tracks), we can compare solid abundances that arise from super Earth formation within the ice line, cosmic ray dead zone, and heat transition. 

In figure \ref{SuperEarths}, we show how each planet's solid mass is distributed among solid components at the end of their formation. The cosmic ray dead zone planet has the lowest ice content among the three planets shown (6 \% ice by mass) as it spends the majority of its time accreting solids interior to the ice line, acquiring most of its mass in refractory materials. It is only at late stages of its formation that the planet is situated close to the ice line, and is able to accrete a small amount of icy solids. By definition of the trap, the planet forming within the ice line is able to accrete a substantial amount of icy solids during its formation. At the end of its formation, the planet formed in this trap has roughly one third of its solid mass in ices. Lastly, as the planet formed in the heat transition lies exterior to the ice line, it is able to accrete a lot of icy solids, resulting in nearly half of its solid mass being ice. We do not show pie charts of abundances for the super Earths formed in the 0.05 M$_\oplus$ case, as the abundances are consistent with the super Earths formed in the cosmic ray dead zone and ice line shown in figure \ref{SuperEarths}.

\subsubsection{Migration Across the Ice Line: A Means of Achieving Time-Dependent Composition}

\begin{figure} 
\includegraphics[width = 3.5 in]{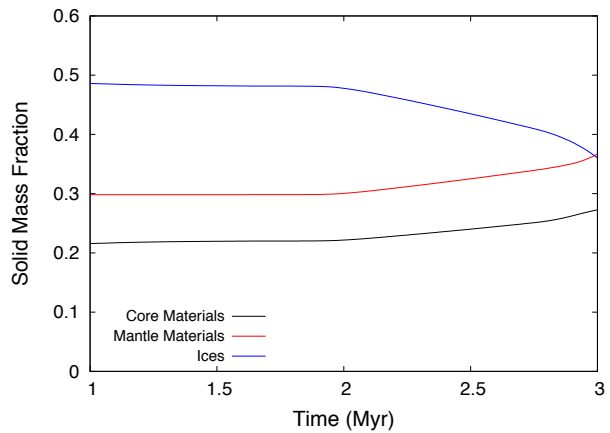}
\caption{The time dependent mass fraction in solid components for the planet forming in the heat transition. We connect the compositions at 2 and 3 Myr (shown in figures \ref{SuperEarths} bottom, and figure \ref{3MyrHTPlanet}, respectively) by tracking the mass fractions between these two times. Since the heat transition lies interior to the ice line after 2 Myr, the planet spends the last Myr of its formation accretion dry refractory materials.}
\label{CompositionTime}
\end{figure}

By comparing the composition of the heat transition planet at 2 Myr (figure \ref{SuperEarths}, right) and at 3 Myr (figure \ref{3MyrHTPlanet}), the effects of time dependent chemistry can be seen as the planet's composition changes over the last Myr of its formation. In particular, we see that after 2 Myr the planet has nearly half its solid mass in ice, and after 3 Myr it has decreased to roughly one third. In order to connect these two snapshots, we plot the continuous solid abundances of the planet during its formation in figure \ref{CompositionTime}. Compositional changes are expected in planets that encounter compositional gradients throughout the disk during their formation. The most recognizable compositional gradient in protoplanetary disks is the ice line, and this has a direct effect on the ice content in solids that a planet can accrete. The location of a trap (in this case, the heat transition) with respect to the ice line dictates the types of material available for formation. Interior to the ice line, there are few icy solids available for accretion, while outside they are in abundance. 

In the case of figure \ref{CompositionTime}, the heat transition trap intersects the ice line at roughly 2 Myr (see figure \ref{traps}), and the planet forming within the heat transition encounters a steep compositional gradient, causing a decrease in its ice abundance. Prior to the 2 Myr point, it accretes in an ice rich environment, building up nearly half its solid mass in ice. After the traps intersect, the planet transitions to an ice deficient environment interior to the ice line. Solid accretion in this dry region of the disk results in only refractories being accreted. This results in the ice abundance decreasing from 47.7 \% to 36 \% between 2 and 3 Myr. In turn, the mass abundances of core materials and mantle materials increase during this period.

\subsubsection{Comparing Cosmic Ray and X-ray Dead Zone Results}

While the X-ray dead zone does not form a super Earth in any of the simulations presented, the time marker in figures \ref{4Jupiters} and \ref{DecreasedLifetime} indicate that a disk that is photoevaporated after only 1 Myr will result in the X-ray dead zone producing a super Earth of mass $7.8$ M$_\oplus$ at 0.65 AU. While a 1 Myr disk lifetime is short compared to the fiducial value of 3 Myr discussed earlier, it is not entirely unreasonable as it lies within the 0.5-10 Myr range of estimated disk lifetimes as suggested by observations \citep{Hernandez2007, Mamajek2009}. Other theoretical models \citep{HP13} have used a $t_{LT} \sim$ 1 Myr as a lower limit to a range of disk lifetimes in core accretion scenarios.

\begin{figure}
\includegraphics[width = 3 in]{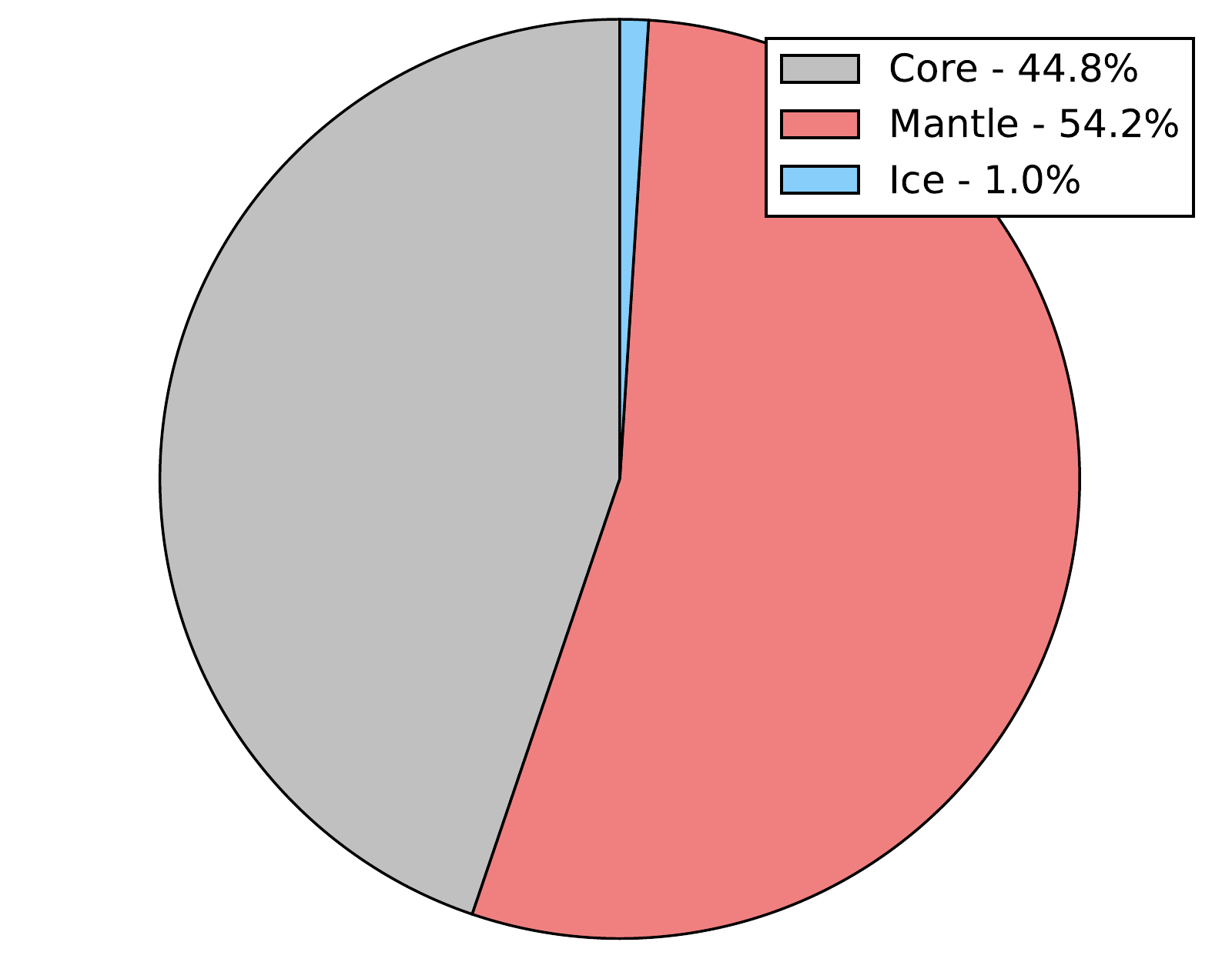}
\includegraphics[width = 3. in]{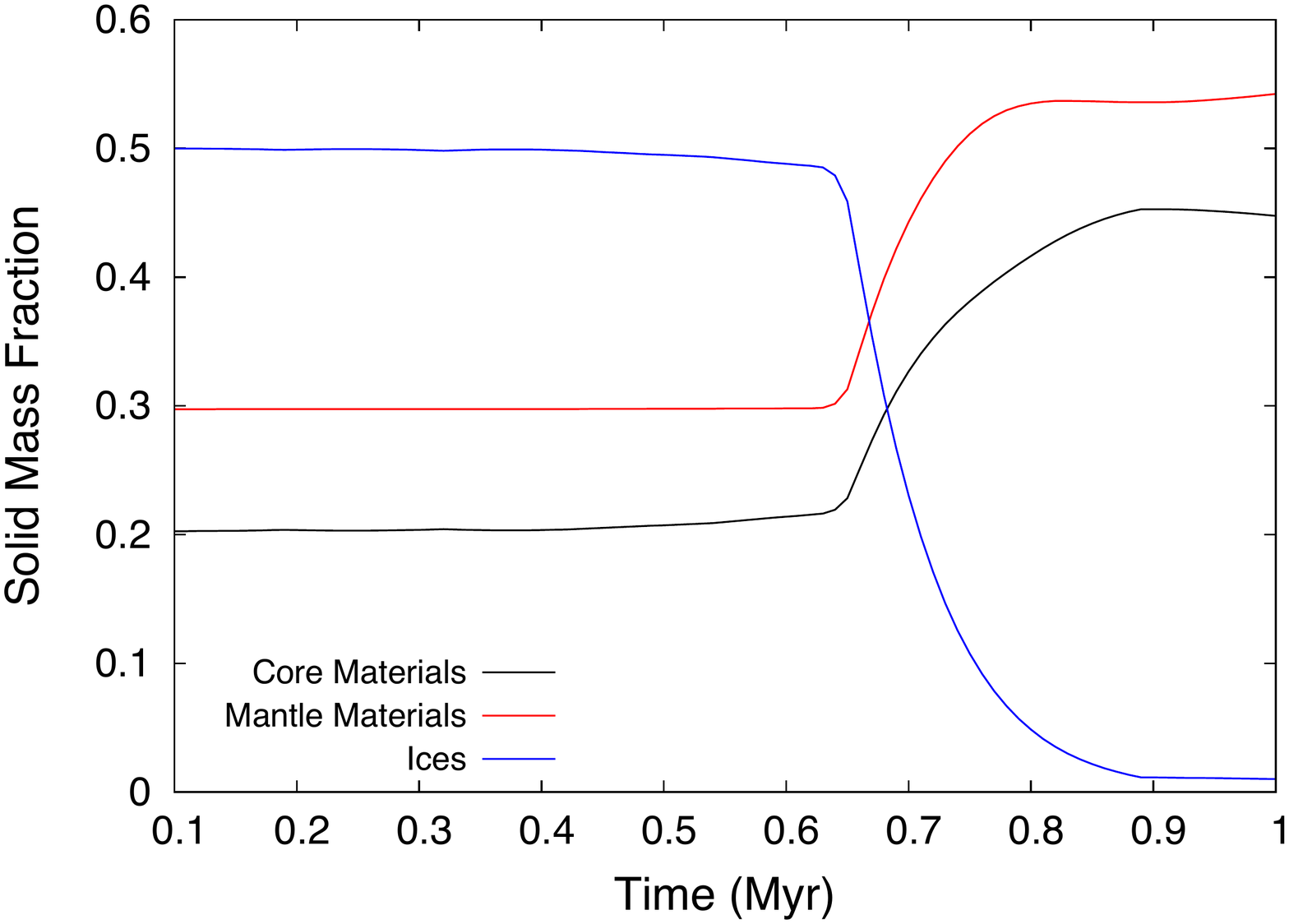}
\caption{\textbf{Top}: The solid abundances of the planet forming within the X-ray dead zone are shown 1 Myr into its formation. At this time, its mass is 7.8 M$_\oplus$, consistent with a super Earth. It is very dry at this time in its formation, suggesting it accreted most of its solids interior to the ice line. \textbf{Bottom}: By plotting the time dependent abundances of the three solid components on this planet, we see that the planet started out accreting in an icy environment (outside the ice line), prior to migrating inwards, and accreting only dry, rocky materials.}
\label{1MyrXRDZ}
\end{figure}

In Figure \ref{1MyrXRDZ}, upper panel, we show the mass abundance of the 1 Myr X-ray dead zone planet in solid components. We see that the planet is very dry at this point in its formation, with only 1 \% of its solid mass being in ice. The X-ray dead zone intersects the ice line early in the fiducial disk's lifetime, at roughly 0.7 Myr. As previously noted, a planet forming within a trap that sweeps past the ice line will have an evolving composition during its formation. We therefore expect the planet forming within the X-ray dead zone to have a time-dependent solid composition over the first Myr of its formation. 

In figure \ref{1MyrXRDZ} (lower panel), we show the time dependent mass abundance of the planet forming in the X-ray dead zone from the start of its formation (0.1 Myr) to the 1 Myr point where its mass is consistent with a super Earth. We find that the planet initially accretes icy solids, having a substantial mass abundance in ice of nearly 50 \% before decreasing drastically at 0.7 Myr to only 1 \%. At this time the X-ray dead zone sweeps past the ice line, leaving the planet to accrete from dry regions of the disk, resulting in the decreasing ice abundance on the planet. 

The small ice abundance of 1 \% on this planet suggests that it must have accreted much more solids inside the ice line than early in its formation when it was outside the ice line. We can confirm this by noting the solid accretion rate onto a planet is proportional to the planet's mass, with the scaling $\dot{M}_p \propto M_p^{2/3}$, as is shown in equations \ref{Solid_Accretion} and \ref{Solid_Accretion2}. This causes the planet forming in the X-ray dead zone to accrete solids faster in the later stages of its formation (interior to the ice line) than in early stages of its formation (outside the ice line). Thus, the total mass accreted in the dry regions of the disk between 0.7 Myr and 1 Myr greatly exceeds the mass accreted before 0.7 Myr in the icy regions of the disk. This causes the mass in ice that the planet was able to accrete early in its formation to comprise only a small fraction of 1 \% of the planet's total mass after 1 Myr.

\begin{table*}
  \centering
  \caption{Abundances in specific minerals for the five super Earths whose solid abundances are shown in this results section. In the top row, we note the trap the planets formed within as well as the lifetimes of their natal disks. The upper portion of the table gives mass abundances for the top three iron minerals and the top three silicate minerals, calculated with equation \ref{SolidAbundance}. The lower portion normalizes these mass abundances in terms of the planet's total mass in core materials (for the 3 iron minerals, using equation \ref{RelativeIron}) or mantle materials (for the 3 silicate minerals, using equation \ref{RelativeSilicate}).}
    \begin{tabular}{ccccccc}
    \toprule
        Planet: Disk t$_{LT}$ \& Trap & 3 Myr HT& 2 Myr HT & 2 Myr CRDZ & 2 Myr IL & 1 Myr XRDZ \\
    \midrule
    Figure &    \ref{3MyrHTPlanet}   &   \ref{SuperEarths} right    &    \ref{SuperEarths} left   &   \ref{SuperEarths} middle    & \ref{1MyrXRDZ}  \\
    \midrule
    \multicolumn{6}{c}{\% of planet's solid mass} \\
    \midrule
    Troilite (FeS)   & 12.13 & 9.88 & 17.66 & 12.83 & 19.01 \\
    Magnetite (Fe$_3$O$_4$) & 8.63 & 8.1 & 16.44 & 11.88 & 3.97 \\
    Hercynite (FeAl$_2$O$_4$) & 2.46 & 2.06 & 3.72 & 2.72 & 3.76 \\
    Enstatite (MgSiO$_3$) & 22.61 & 19.86& 35.71 & 26.2 & 24.14 \\
    Forsterite (Mg$_2$SiO$_4$) & 7.88 & 5.45 & 9.88 & 7.2 & 20.21 \\
    Diopside (CaMgSi$_2$O$_6$) & 4.56 & 3.71 & 6.74& 4.93& 7.16\\
    \midrule
    \multicolumn{6}{c}{Mass relative to planet's mass in core materials} \\
    \midrule
    Troilite (FeS)   & 0.44 & 0.45 & 0.44 & 0.44 & 0.42 \\
    Magnetite (Fe$_3$O$_4$) & 0.32 & 0.37 & 0.41 & 0.40 & 0.09 \\
    Hercynite (FeAl$_2$O$_4$) & 0.09 & 0.09 & 0.09 & 0.09 & 0.08 \\
    \midrule
    \multicolumn{6}{c}{Mass relative to planet's mass in mantle materials} \\
    \midrule
    Enstatite (MgSiO$_3$) & 0.62 & 0.66 & 0.66 & 0.66 & 0.45 \\
    Forsterite (Mg$_2$SiO$_4$) & 0.22 & 0.18 & 0.18 & 0.18 & 0.37 \\
    Diopside (CaMgSi$_2$O$_6$) & 0.12 & 0.12 & 0.13 & 0.12 & 0.13 \\
    \bottomrule
    \end{tabular}%
  \label{AbundanceTable}%
\end{table*}%

Contrasting the X-ray dead zone planet's composition after 1 Myr with the super Earth formed in the cosmic ray dead zone (Figure \ref{SuperEarths}, left panel), we see that the cosmic ray ionization model results in a super Earth with 6 \% of its mass in ices as opposed to 1 \% in the case of the planet formed in the X-ray ionized model. 

The differences can be explained by considering the difference in locations of the two traps. While the cosmic ray dead zone always lies interior to the ice line, it is always within 1 AU of it. While the planets forming in the cosmic ray dead zone are dry, they are still able to accrete a small amount of ice due to the trap's proximity to the ice line. Conversely, the X-ray dead zone lies drastically interior to the ice line, except for the first 0.7 Myr. This causes planets forming within this trap to accrete effectively no ice while forming in the dry regions of the disk.

\subsubsection{Super Earth Mineral Abundances}

In Table \ref{AbundanceTable}, we show abundances of the top three iron and silicate minerals for the five super Earth examples from our model discussed in this section. In the top portion of this table, we show solid abundances of each mineral that were calculated using equation \ref{SolidAbundance} on each planet. While the raw abundances show what minerals dominate the planet's solid mass, we note that these quantities are heavily dependent on each planet's mass in core minerals and mantle minerals. For example, a dry planet with little water will naturally have large abundances of core and mantle minerals, as is the case for the super Earths formed in the dead zone traps. Conversely, the planets that formed in the heat transition and ice line traps have systematically lower abundances of these minerals due to the high ice content on the planets.  

To remove the systematic variation in mineral abundances with each planet's total water content, we normalize the mineral abundances in the following manner. For each iron mineral, we compute its relative abundance with respect to the planet's total mass in core materials,
\begin{equation} \textrm{Abundance of}\;i\;\textrm{relative to core} = \frac{X_{\textrm{i, planet}}^{\textrm{solid}}}{X_{\textrm{core, planet}}^{\textrm{solid}}}\;. \label{RelativeIron} \end{equation}
A similar approach is taken to normalize the mantle minerals,
\begin{equation} \textrm{Abundance of}\;i\;\textrm{relative to mantle} = \frac{X_{\textrm{i, planet}}^{\textrm{solid}}}{X_{\textrm{mantle, planet}}^{\textrm{solid}}}\;. \label{RelativeSilicate} \end{equation}
By normalizing the abundances in this manner, we can discern how much a particular mineral contributes to the total mass of the planet's core (or mantle). 

In the lower portion of Table \ref{AbundanceTable}, abundances of the dominating iron and silicate minerals are shown relative to the planet's total mass in core minerals or mantle minerals. As an example, we see that troilite (FeS) contributes roughly 43-44 \% of each super Earth's mass in core materials, regardless of the trap each particular planet formed within. Therefore, when considering our 3-component chemistry model (core materials, mantle materials, \& ice), we can conclude (albeit with a small sample of five super Earths) that each planet's iron content is composed of roughly 43-44 \% in troilite. Aside from the planet formed in the X-ray dead zone, the other minerals follow a similar trend across the planets, as each mineral comprises a similar fraction of their corresponding planet's iron (or silicate) content. Unlike the case of troilite, however, there is a variation of $\sim$10\% in some cases in the abundances of a particular mineral on different planets.

Table \ref{AbundanceTable} shows that the super Earth formed in the X-ray dead zone has mineral abundances that differ greatly from the other four planets shown. This is due to the fact that the X-ray dead zone planet forms early in the disk's evolution, and accretes all its solids prior to 1 Myr. This, coupled with its orbital migration to the inner regions of the disk ($< 1$ AU) causes it to sample disk chemistry at higher temperatures than the other four planets shown. The X-ray DZ planet shows that we cannot simply assume a constant fraction of enstatite, for example, in the planet's silicate content. In this way, binning our solids into core and silicate components hides the ratios of the underlying minerals. While the three-component chemistry model does provide a simple description of the planet's composition giving the necessary density information for modelling its interior structure (as in \citet{Valencia2007}), it is important to realize that each planet's unique formation history results in different abundances of particular minerals that are hidden when quoting compositions in terms of the summed components.

\section{Discussion}

\subsection{Observational Constraints on Disk Chemistry}

\begin{table*}
  \centering
  \caption{Mass abundances of secondary gases on atmospheres of Jovian planets formed in the $t_{LT}$ = 4 Myr run (see figure \ref{4Jupiters}). Each column denotes the natal trap of a particular planet. The second row shows the planet's final masses. Planets that accreted their gas in cooler regions of the disk, such as the planet formed in the cosmic ray dead zone, have larger abundances of H$_2$O, CH$_4$, and NH$_3$. Conversely, gas accretion from hot regions of the disk, as is the case for the planet formed in the X-ray dead zone, results in larger abundances of CO, N$_2$, and SiO. }
    \begin{tabular}{ccccc}
    \toprule
         & Ice Line & Heat Transition & Cosmic Ray Dead Zone & X-ray Dead Zone \\
    \midrule
    M$_\textrm{P}$/M$_{\textrm{Jupiter}}$ &1.17 & 0.71& 1.47& 2.13 \\
    \midrule
 H$_2$O& 0.352 \% & 0.163 \% & 0.446 \%& 0.26 \%\\
 CO & 0.183 \% & 0.485 \% & 0 \%& 0.484 \%\\
 CH$_4$& 0.169 \% & 6.15$\times10^{-4}$ \%&0.279 \% & 1.54$\times10^{-4}$ \% \\
 N$_2$& 8.19$\times10^{-2}$ \%& 8.24$\times10^{-2}$ \% & 4.56$\times10^{-2}$ \% & 8.22$\times10^{-2}$ \%\\
NH$_3$ & 7.42$\times10^{-4}$ \%& 1.47$\times10^{-4}$ \%& 4.52$\times10^{-2}$ \%& 2.47$\times10^{-7}$ \%\\
H$_2$S & 1.51$\times10^{-2}$ \%& 3.79$\times10^{-2}$ \%& 6.62$\times10^{-8}$ \%& 2.59$\times10^{-2}$ \%\\
SiO & $0$ \%& 7.27$\times10^{-5}$ \%& $0$ \% & 0.103 \%\\
    \bottomrule
    \end{tabular}%
  \label{GasTable}%
\end{table*}%

One method of constraining our chemistry results is via the use of observed locations of condensation fronts. For example, in \citet{Zhang2013} the location of the ice line in TW Hya was shown to have an upper limit of 4.2 AU using the observed water vapour content throughout the disk. Additionally, they found that the water vapour content drops rapidly at the location of the ice line, over a short distance of 0.5 AU. The ice line location found in our work falls within their constrained regime, as do the sharp transitions between water vapour and ice profiles we find at the ice line in our model. 

In addition to the water ice line, the CO condensation front has been observed in disks at roughly 30 AU \citep{Qi2011}. Our equilibrium chemistry model does not predict the existence of carbon monoxide in solid or gas phase in the outer regions of the disk and cannot predict the location of the condensation front, which is a limitation of the model. We note, however, that the inclusion of a CO ice line in this model would not greatly affect the resulting super Earth compositions. This is because super Earth formation in all traps leads to solid accretion from regions of the disk within 30 AU, and in most cases within 10 AU. In addition to observations of condensation fronts, future ALMA observations of chemical signatures in disks can be used to further constrain our disk chemistry model. 

Compositions of exoplanet atmospheres can place additional constraints on our disk chemistry and planet formation models. By comparing atmospheric abundances of modelled planets with exoplanetary data, abundances of the underlying disks can be contrasted \citep*{Cridland2016}. While there are currently only several exoplanets with atmospheric abundance data, there are many future prospects whose compositional data may become available with the advent of JWST.  

While gas phase chemistry throughout the disk is not a main focus of this paper, we are still able to track the gases our modelled planets accrete throughout their formation as computed by equilibrium chemistry. Our results show that Jovian planets have roughly Solar abundances in their atmospheres, which are composed almost entirely of molecular hydrogen and helium. This is a direct result of the initial condition of Solar abundances being assumed for the disk chemistry calculation. Variations in secondary gas abundances do exist between different Jovian planets formed in our model, however, and we discuss these below.

In table \ref{GasTable}, we show the abundances of secondary gases for planets formed in the fiducial disk with a lifetime of 4 Myr (see figure \ref{4Jupiters}). Among these four Jovian planets arising from each of the traps in our model, we see variations among the secondary gases that result from each planet's unique formation history. The location of each particular planet at the time when it undergoes runaway growth plays a key role in these results, as the disk's composition at this radius will be reflected in the planet's atmosphere.

For example, the planet formed in the cosmic ray dead zone undergoes runaway growth in the coolest region of the disk compared to the other planets. As a result, it has the largest abundance of H$_2$O, CH$_4$, and NH$_3$. Conversely, planets that accrete gas in hot regions of the disk, as is the case for the planet formed in the X-ray dead zone, achieve the highest compositions in CO, N$_2$, and SiO. In particular, the planet formed in the X-ray dead zone accretes the largest amounts of gaseous SiO due to it accreting gas from the hottest regions of the disk (small radii prior to 2 Myr into the disk's evolution). These results show that abundances of certain gases can constrain a planet's formation history to have taken place within a particular region of the disk.

\subsection{Planet Formation Model}

\subsubsection{Planet-Planet Dynamics}

Throughout this work, we have noted that traps intersect throughout the disk's lifetime. Thus, planets forming in these traps would undergo a dynamical interaction. Our work is limited as we do not account for the dynamical interaction between multiple forming planets. Rather, our planets form in isolation. Detailed N-body simulations in \citet{HellaryNelson2012} show several interesting effects take place when dynamic interactions are accounted for while multiple planet cores are forming in a disk with an opacity transition. Handling the detailed dynamics between multiple forming planets in a disk with multiple traps is a prospect for future work, and in doing so we hope to see the effects of scattering and resonant traps in our model.

\subsubsection{Oligarchic Growth or Pebble Accretion?}

The first stage of our planet formation model assumes that planetary cores accrete km-sized solids via oligarchic growth, an N-body process. An alternative method of core growth has been proposed in recent models, such as in \citet{Ormel2010}, and \citet*{Bitsch2015}, and is referred to as pebble accretion. In this model, planetary cores grow by accreting from a sea of cm-sized pebbles coupled to the gas. These cm-sized pebbles are seeded by forming via the streaming instability \citep{Johansen2007}, and their constant production can result in large accretion rates onto the planetary core, even in MMSN disks. 

While our model uses oligarchic growth to handle solid accretion, it is possible that pebble accretion can be used as well, providing the important physical processes can be captured in a semi-analytic framework. If the two methods of solid core growth result in a 5-10 M$_\oplus$ core in less than 1 Myr, in principle they should not result in drastically different planets after gas accretion has been terminated. However, if the core growth timescales are drastically different, as the rapid core growth calculated using pebble accretion suggests it may be \citep{Bitsch2015}, the final locations of planets on the mass-semimajor axis diagram at the end of gas accretion may vary appreciably as a result. 

\subsubsection{Final Masses of Gas Giants}

During the final phase of gas accretion, our planets undergo runaway growth, and are limited only by the Kelvin-Helmholtz timescale. In the final stages of planet formation in our model, the Kelvin-Helmholtz timescale allows the planet's accretion to supersede the accretion rate throughout the disk. Additionally, we parameterize the planet's final mass in terms of its gap opening mass, as opening a gap is a key step to shutting off gas flow onto a planet. This approach is necessary to limit the planet's mass from diverging prior to the disk being photoevaporated. 

As discussed in section 2.4, other works, such as \citet{Machida2010, Dittkrist2014, Bitsch2015}, \& \citet{Mordasini2015} use an alternative approach to limit the accretion onto the planet. In these models, the gas accretion onto a planet is limited by the accretion rate throughout the disk. Using a time-dependent decreasing disk accretion rate can lead to the planets reaching a few Jupiter masses at the end of the disk's lifetime, avoiding an abrupt termination of gas accretion onto the planet. However, a disk-limited accretion model does require a parameterization of the fraction of the disk's accretion that gets accreted onto the planet, making the planets' final masses in this alternative approach dependent on model parameters pertaining to late stages of accretion. In both approaches, the model parameters are estimating the fraction of the available gas reservoir that gets accreted onto the planet.

Comparing these two methods, the Kelvin-Helmholtz timescale leads to planets moving upwards on the mass-semimajor axis diagram during stage III of their formation. The alternate model causes planets to move diagonally on the diagram, migrating inwards on a similar timescale as their accretion rate during this final stage of gas accretion. These two methods would lead to different final locations of planets at the end of their formation, and this will directly impact the mass-period relation the two models predict. In a future paper focusing on population synthesis, we will further compare the two methods in their ability to reproduce exoplanet data. 

\subsection{Extension to Planet Population Synthesis}

By varying the disk's initial mass and lifetime, we have shown that our planet formation model can produce planets occupying entirely different regions of the mass-period diagram. For example, we have found that super Earth formation is strongly tied to the disk lifetime. Short-lived disks ($t_{LT} \lesssim 2$ Myr) typically result in the planets having insufficient time to accrete more than $\sim$ 10 M$_\oplus$ of disk material, and are stranded at an early stage of their formation at the time of disk dispersal. Meanwhile, in sufficiently long-lived disks, we have shown in figure \ref{MassParameter} that the types of Jupiters produced are tied not only to the traps that the planets formed within, but also to the mass of their natal disks. Heavier disks (0.15 M$_\odot$) typically result in planets forming at large semi-major axes, and are more prone to forming 1 AU Jupiters. Conversely, lighter disks (0.05 M$_\odot$) tend to form planets closer to their host stars, and may also result in super Earth formation due to a longer planet formation timescale in low mass disks.

The mutual effects of the disk's initial mass and lifetime provides a means of populating all the regions of the mass-period diagram coinciding with the observed locations of exoplanets. In a future work (Alessi et al. 2016, in prep.), we will employ a continuous range of these parameters within observational constraints in order to determine how frequently different regions of the mass-period diagram are populated, which is a similar approach taken in \citet{HP13}. 

In this work we have shown that planet traps define the regions of a disk that a planet can accrete from. In this way, super Earth compositions are tied to the trap they formed within, with each planet's composition reflecting the composition in the disk at the locations where it accreted its solids. In taking a population approach in our future work, we expect to find ranges of super Earth compositions that arise from formation in different traps. Based on our results in this work, we anticipate that traps sweeping past the ice line will produce super Earths with a variety of compositions. Meanwhile, we expect traps that do not migrate significantly will produce super Earths with relatively uniform compositions, regardless of the mass of the disk they form within.

With the model that has been developed in this paper, we are able to provide the initial conditions, namely a planet's mass and solid composition, necessary for models of interior structure of super Earths, such as \citet{Valencia2007}. The variety of compositions predicted from super Earth formation in different planet traps can be extended with an interior model to predict a range of mean densities of these planets. For example, the dry and rocky planets will have higher densities than planets of the same mass with a substantial amount of ice. Interior models provide a link between planet compositions calculated with our model and planet's locations on the mass-radius diagram. In a similar approach to \citet{Mordasini2012b, Mordasini2012c}, combining an interior model with our future population approach will allow us to determine how our formed planets distribute themselves on the mass-radius diagram, allowing us to further compare with observations.

We note that the planets shown in table \ref{AbundanceTable} have accreted non-negligible amounts of gas during their formation, ranging from 1.6 \% by mass in the smallest case (2 Myr ice line) up to 27 \% (corresponding to the 2 Myr cosmic ray dead zone planet). As shown in \citet{Lopez2013} a small amount of atmosphere can greatly increase the planet's radius. In this case, the planet's internal composition cannot be deduced from observations of the planet's radius. Our model does not track the evolution of the planets post-formation, or specifically their atmospheres. It is therefore unclear whether or not these planets would retain their atmospheres over billion-year timescales, and it remains possible that the planets' internal compositions are discernible through radius observations.

\section{Conclusions}

We have made a major extension of the model by \citet{HP11, HP12, HP13} by including the effects of disk chemistry in order to model the ice line's location, the dust to gas ratio, and most importantly, to track accreted solids onto super Earths. Our major findings in this work are listed below:
\begin{itemize}
\item Super Earth formation is linked to the timing of disk dispersal. Our model has resulted in super Earth formation in disks with lifetimes $\lesssim$ 2 Myr. Additionally, our model has produced super Earths in light disks (initial mass 0.05 M$_\odot$).
\item Super Earths formed within the ice line and heat transition traps have substantial ice contents, ranging from 30 \% of their masses and up to nearly 50 \%. Conversely, both the X-ray and dead zone traps produce dry and rocky super Earths, with as little as 1 \% of their mass in ice.
\item Troilite and magnetite make up the majority ($\sim$ 70 \% - 80 \%) of the core materials in the super Earths formed in this paper. Meanwhile, enstatite and forsterite make up the majority ($\sim$ 75 \% - 85 \%) of these planets' mantle materials. 
\item The types of Jupiters formed in our model depend on the trap they formed within. In sufficiently long-lived disks ($t_{LT}$ = 4 Myr) the heat transition and X-ray dead zone result in hot Jupiters while the ice line and cosmic ray dead zone produce Jupiters at 1 AU.
\item Variations in secondary gas abundances exist among the Jovian planets formed with our model, and are sensitive to the disk temperature where the planets undergo runaway growth. Abundances of CO, N$_2$, and SiO result from gas accretion in hot regions of the disk, while accretion from colder regions of the disk results in higher abundances of H$_2$O, CH$_4$, and NH$_3$.
\item We find that planet formation in the X-ray dead zone and heat transition traps are insensitive to the masses of the disks they form within. Planets forming from these two traps have final locations varying at most over 0.1 AU using a 0.05 M$_\odot$ - 0.15M$_\odot$ initial disk mass range. 
\item The cosmic ray dead zone and ice line traps result in planets whose masses and final locations are sensitive to disk mass. Increasing the disk mass from 0.1 M$_\odot$ to 0.15 M$_\odot$ increases the planet's final locations by up to an AU, while using a small disk mass of 0.05 M$_\odot$ resulted in super Earths forming out of these two traps.
\end{itemize}
We will extend these models in a future planet population synthesis paper that takes into account the ranges of disk parameters that can shape that observed mass-period relation of exoplanets. 

\section*{Acknowledgments}

We thank an anonymous referee for their useful comments that improved the quality of this manuscript. The authors would also like to thank Yasuhiro Hasegawa for his useful insights and discussions regarding this work. R.E.P. also thanks the MPIA and the Institut f\"{u}r Theoretische Astrophysik (ITA) in the Zentrum f\"{u}r Astronomie Heidelberg for support during his sabbatical leave (2015/16) during the final stages of this project. M.A. acknowledges funding from an Ontario Graduate Scholarship (OGS) and from the National Sciences and Engineering Research Council (NSERC) through a CGS-M scholarship. R.E.P. is supported by an NSERC Discovery Grant. A.J.C. acknowledges funding through the NSERC Alexander Graham Bell CGS/PGS Doctoral Scholarship.

\bsp

\section*{Appendix}
\appendix

\section{Type-I Migration Regimes}

As is discussed in \citet{HellaryNelson2012} and \citet{Dittkrist2014}, there are several type-I migration sub-regimes governing planet migration prior to the forming planet opening a gap in the disk. Namely, they are the locally isothermal regime, the trapped regime whereby the corotation torque is unstaturated, and lastly the saturated corotation torque regime.

Throughout this work we only considered the trapped type-I migration regime and here we validate that approach. As shown below, when considering alternate sub-regimes of type-I migration using our model's parameters, we find that planets in our model are always in the trapped type-I migration regime until they open a gap in the disk. This validates our assumption of trapped type-I migration prior to planets reaching their gap-opening masses.

We follow the approach in \citet{Dittkrist2014} that compared four distinct timescales to discern which sub-regime a type-I migrating planet belongs to. An important length scale in this discussion is the width of the horseshoe region, $x_s$, which denotes the range of radii around a planet where disk material will undergo horseshoe orbits. The form of $x_s$, taken from \citet{Masset2003} is,
\begin{equation} x_s = 0.96 r_p \sqrt{\frac{q}{h_p}} \;,\end{equation}
where $q = M_p/M_*$, and $h_p=H_p/r_p$ is the disk's aspect ratio at the planet's location.

To distinguish between the locally isothermal regime (which can apply for planets $\lesssim 5 M_\oplus$) and the trapped regime, we first compare the u-turn timescale, $t_{\textrm{u-turn}}$, and the cooling timescale, $t_{\textrm{cool}}$. The u-turn timescale characterizes how long it takes for a gas parcel on a horseshoe orbit to undergo a u-turn in front of or behind the planet,
\begin{equation} t_{\textrm{u-turn}} = \frac{64x_sh_p^2}{9qr_p\Omega_p} \;.\end{equation}
The cooling timescale for the gas parcel undergoing a u-turn is \citep{Dittkrist2014},
\begin{equation} t_{\textrm{cool}} = \frac{l_{\textrm{cool}}\rho C_V}{8\sigma T^3}\left(8\rho \kappa l_{\textrm{cool}} + \frac{1}{\rho \kappa l_{\textrm{cool}}}\right) \;,\end{equation}
where $l_{\textrm{cool}}$ is the minimum of $H_p$ and $x_s$. Here $t_{\textrm{cool}} > t_{\textrm{u-turn}}$ implies that the planet is in the trapped regime (providing the corotation torque is not saturated, see discussion below), while $t_{\textrm{cool}} < t_{\textrm{u-turn}}$ implies the planet is in the locally isothermal regime.

To determine whether or not the corotation torque is saturated, we compare the viscous timescale \citep{Masset2003, Cridland2016},
\begin{equation} t_{\textrm{vis}} = \frac{x_s^2}{3\nu}\;,\end{equation}
to the libration timescale, characterizing the duration of a gas parcel's horseshoe orbit,
\begin{equation} t_{\textrm{lib}} = \frac{4 \pi r_p}{1.5\Omega_p x_s} \;.\end{equation}
In this case, $f_{\textrm{vis}}t_{\textrm{vis}} < t_{\textrm{lib}}$ implies that the corotation torque is unsaturated and the planet is trapped, while the converse case $f_{\textrm{vis}}t_{\textrm{vis}} > t_{\textrm{lib}}$ implies that the corotation torque is saturated. The parameter $f_{\textrm{vis}}$ is a factor of order unity introduced in \citet{Dittkrist2014} who considered a range of $f_{\textrm{vis}}$ values from 0.125 - 1.0, but found that $f_{\textrm{vis}}=0.55$ provided a best fit between their model and hydrodynamics simulations.

For all planets tracks presented, we find the u-turn timescale to be longer than the cooling timescale for planet masses less than $\sim 3-5 M_\oplus$ (depending on the specific trap used), meaning that the locally isothermal regime applies to low-mass planet cores in our model. We do not include the effects of this migration regime in this work, and force the low-mass planets to be trapped even when $t_{cool}<t_{u-turn}$. Additionally, we find the quantity $t_{\textrm{lib}}/f_{\textrm{vis}}t_{\textrm{vis}}$ is greater than one for all planets prior to them opening a gap, implying that corotation torque saturation does not apply for planets in the type-I migration regime in our model.

\begin{figure}
\includegraphics[width = 3.5 in]{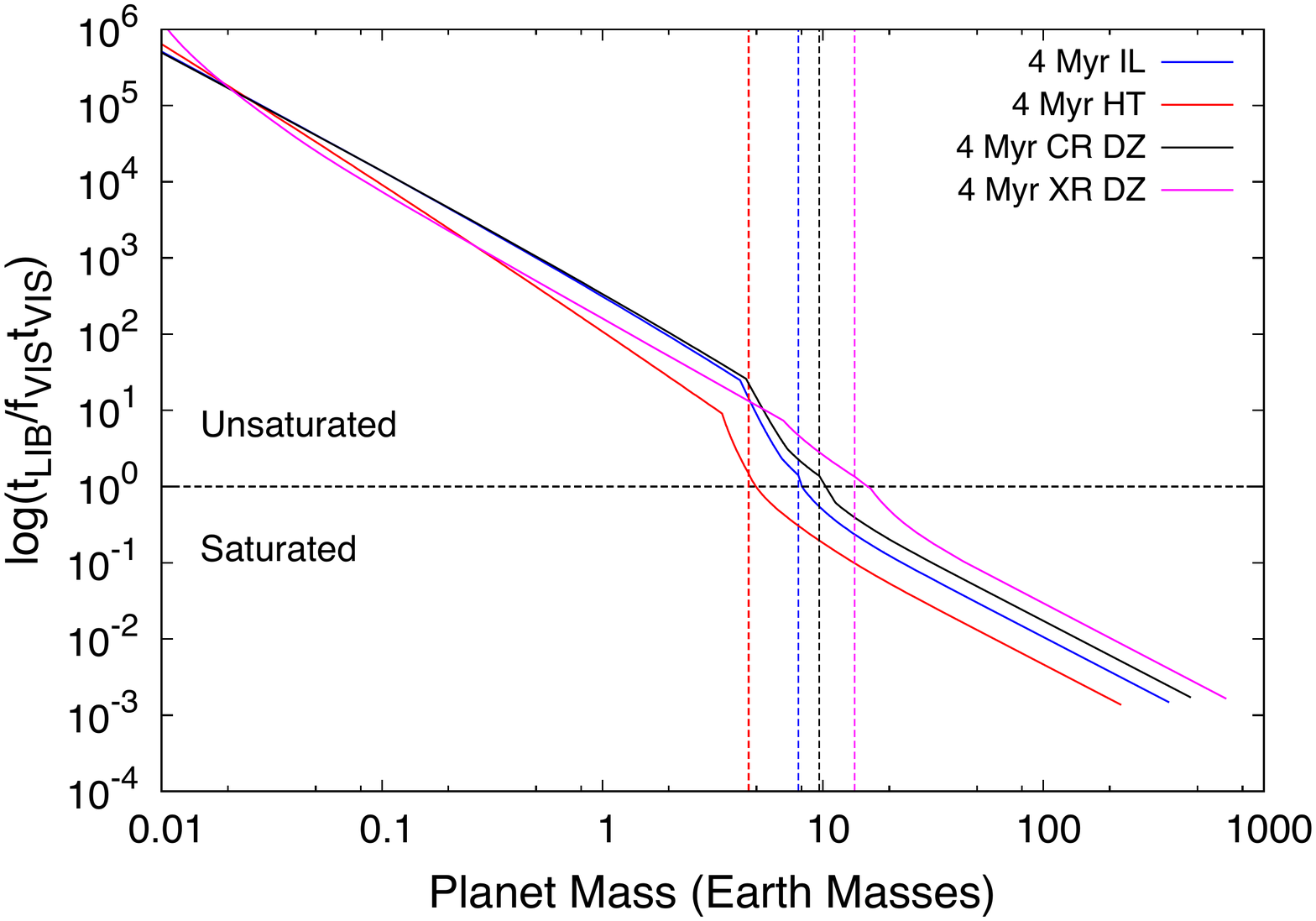}
\caption{We plot the ratio $t_{\textrm{lib}}/f_{\textrm{vis}}t_{\textrm{vis}}$ for the planets formed in the 4 Myr disk run. The vertical dashed lines represent the gap opening masses for each planet. The horizontal dashed line represents a ratio of one where the corotation torque saturates. In all cases, the planets reach their gap opening masses prior to the corotation torque saturating, meaning that they are in the trapped regime for the entirety of type-I migration.}
\label{Lib_Vis_Compare}
\end{figure}

In figure \ref{Lib_Vis_Compare}, we plot the ratio $t_{\textrm{lib}}/f_{\textrm{vis}}t_{\textrm{vis}}$ for the four planets formed in the 4 Myr run in figure \ref{4Jupiters}. Our results show that these planets have unsaturated corotation torques for all times until they open a gap in the disk. At a slightly higher mass, the corotation torques would saturate as $t_{\textrm{lib}}/f_{\textrm{vis}}t_{\textrm{vis}}<1$, but this argument no longer applies as the planets are in the type-II migration regime. We show the exact gap-opening masses and saturation masses, $M_{\textrm{sat}}$ (defined to be the planet mass when $t_{\textrm{lib}}/f_{\textrm{vis}}t_{\textrm{vis}}<1$), for each planet track in the 4 Myr disk run in table \ref{SaturateTable}. For planet formation in all four traps, the planets open gaps prior to their corotation torques saturating, implying that they are in the trapped regime for the entirety of type-I migration.

\begin{table}
\centering
\caption{Here we show the gap-opening masses and corotation torque saturation masses (defined where $t_{\textrm{lib}} = f_{\textrm{vis}} t_{\textrm{vis}}$) for each of the Jovian planets formed in the fiducial 4 Myr run (see figure \ref{4Jupiters}). We also include the disk aspect ratio at the time and orbital radius where each planet opens a gap. Our results show that each planet opens a gap prior to its corotation torque saturating, validating our assumption that planets are trapped during type-I migration.}
\begin{tabular}{ccccc}
\toprule
Trap & M$_{\textrm{gap}}$ (M$_\oplus$) & M$_{\textrm{sat}}$ (M$_\oplus$) & $h = H/r$\\
\midrule
Ice Line & 7.67 & 8.08 & 2.66 $\times 10^{-2}$\\
Heat Transition & 4.56 & 5 & 2.16 $\times 10^{-2}$\\
C.R. Dead Zone & 9.55 & 10.36 & 2.9 $\times 10^{-2}$ \\
X.R. Dead Zone & 13.83 & 16.44 & 3.36 $\times 10^{-2}$ \\
\bottomrule
\end{tabular}
\label{SaturateTable}
\end{table}

This timescale approach is an order-of-magnitude estimate of where the corotation torque saturates. Furthermore, the results we obtain depend on the values of parameters, for example $f_{\textrm{vis}}$. While the results in this section show that this approach predicts that the corotation torque will not saturate during type-I migration with the current choices of parameters, a higher setting of $f_{\textrm{vis}}$ will lead to the corotation torque saturating slightly before gap formation takes place. We predict, however, that the difference between M$_{\textrm{gap}}$ and M$_{\textrm{sat}}$ will be roughly a few tenths M$_\oplus$ for planet tracks in our model, making the saturated corotation torque phase short-lived. This will limit the effect that corotation torque saturation can have on planet tracks presented in this paper.

The planet tracks computed using our model are qualitatively similar to those presented in \citet{Dittkrist2014} during early phases of planet formation. We note, however, that their model includes the effects of multiple type-I migration sub-regimes in addition to the trapped phase. The other two sub-regimes (the locally isothermal regime and the saturated corotation torque regime) can cause the planet to undergo rapid inward migration or outward migration on timescales shorter than the disk's viscous timescale. This contrasts with the trapped inward migration on the viscous timescale planets in our model undergo.

\section{Tracking Planet Compositions}

The following algorithm is used to track planets' compositions throughout their formation. A time step of $\Delta t \sim 100 - 1000$ years is used during our planet formation runs.

\begin{description}
\item[1.] Initial conditions: 0.01 M$_\oplus$ planet core at a radius coinciding with a planet trap at 10$^5$ years. The solid mass abundances in the disk at the planet's location are scaled up to 0.01 M$_\oplus$ to obtain the planet's initial composition.
\item[2.] Update the planet's orbital radius, $r_p(t)$, due to the inward migration during $\Delta t$. This could be either due to the planet trap moving inwards, or, for high mass planets, to type II migration. Steps 3 through 8 are skipped if the planet is in stage IV of its formation (accretion has stopped).
\item[3.] Calculate the disk's temperature, pressure, and surface density at $r_p$ and the time into the disk's lifetime.
\item[4.] Calculate mass abundances of solids, $X_{i,\textrm{solid}}(r_p(t),t)$ and gases, $X_{i,\textrm{gas}}(r_p(t),t)$.
\item[5.] Calculate the accretion rate of solids, $\dot M_{p,\textrm{solid}}$, onto the planet using equations \ref{Solid_Accretion} and \ref{Solid_Accretion2}. The mass $\dot M_{p,\textrm{solid}} \Delta t$ is then added onto the planets current mass.
\item[6.] The amount of solid substances in the planet is updated using the local abundances of solids,
\begin{equation} \begin{aligned} M_{i,\textrm{planet}}&(t) = M_{i,\textrm{planet}}(t-\Delta t) 
\\ & + \dot{M}_{p,\textrm{solid}}\Delta t X_{i,\textrm{solid}}(r_p(t),t) \;.\end{aligned}\end{equation}
Steps 7 and 8 can be skipped if the planet in stage I ($M_p < M_{c,crit}$)
\item[7.] Calculate the accretion rate of gases, $\dot M_{p,\textrm{gas}}$ using equations \ref{Gas_Accretion} and \ref{Gas_Accretion2}. The mass $\dot M_{p,\textrm{gas}} \Delta t$ is then added onto the planet.
\item[8.] Update the amount of gaseous substances  within the planet using the local abundance of gases,
\begin{equation}\begin{aligned} M_{i,\textrm{planet}}&(t) = M_{i,\textrm{planet}}(t-\Delta t) 
\\ &+ \dot{M}_{p,\textrm{gas}}\Delta t X_{i,\textrm{gas}}(r_p(t),t) \end{aligned} \end{equation}
\item[9.] Repeat steps 2-8 throughout the planet's formation, until the time exceeds the disk's lifetime. At this point the gas will be dispersed, shutting off all accretion and planet-disk interactions (migration). 
\end{description}
We denote the solid mass abundance of material or component $i$ on a planet as the ratio between the component's mass on the planet, $M_{i \textrm{,planet}}$, and the planet's total mass in solids, $M_{\textrm{solid,planet}}$,
\begin{equation} X_{i\textrm{,planet}}^{\textrm{solid}} = 100 \frac{M_{i \textrm{,planet}}}{M_{\textrm{solid,planet}}} \, \% \;, \label{SolidAbundance} \end{equation}
where we use planet in the subscript to distinguish between mass abundances throughout the disk.

\label{lastpage}


\begin{thebibliography}{99}

\bibitem[\protect\citeauthoryear{Aikawa \& Herbst}{1999}]{Aikawa1999} Aikawa Y., Herbst E., 1999, ApJ, 526, 314
\bibitem[\protect\citeauthoryear{Alibert et al.}{2005}]{Alibert2005} Alibert Y., Mordasini C., Benz W., Winisdoerffer C., 2005, A\&A, 434, 343
\bibitem[\protect\citeauthoryear{Alibert et al.}{2013}]{Alibert2013} Alibert Y., Carron F., Fortier A., Pfyffer S., Benz W., Mordasini C., Swoboda D., 2013, A\&A, 558, A109
\bibitem[\protect\citeauthoryear{Bitsch, Lambrechts, \& Johansen}{Bitsch et al.}{2015}]{Bitsch2015} Bitsch B., Lambrechts M., Johansen A., 2015, A\&A, 582, A112
\bibitem[\protect\citeauthoryear{Blaes \& Balbus}{1994}]{Blaes1994} Blaes O.M., Balbus S.A., 1994, ApJ, 421, 163
\bibitem[\protect\citeauthoryear{Bond, O'Brien \& Lauretta}{Bond et al.}{2010}]{Bond2010} Bond J.C., O'Brien D.P., Lauretta D.S., 2010, ApJ, 715, 1050
\bibitem[\protect\citeauthoryear{Borucki et al.}{2011}]{Borucki2011} Borucki W.J., et al., 2011, ApJ, 728, 117
\bibitem[\protect\citeauthoryear{Cassan et al.}{2012}]{Cassan2012} Cassan A., et al., 2012, Nature, 481, 167
\bibitem[\protect\citeauthoryear{Ceccarelli et al.}{2014}]{Ceccarelli2014} Ceccarelli C., Dominik C., L\'{o}pez-Sepulcre A., Kama M., Padovani M., Caux E., Caselli P., 2014, ApJ, 790, L1
\bibitem[\protect\citeauthoryear{Chambers}{2009}]{Chambers2009} Chambers J.E., 2009, ApJ, 705, 1206
\bibitem[\protect\citeauthoryear{Chiang \& Laughlin}{2013}]{ChiangLaughlin2013} Chiang E., Laughlin G., 2013, MNRAS, 431, 3444
\bibitem[\protect\citeauthoryear{Cieza et al.}{2015}]{Cieza2015} Cieza F., Williams J., Kourkchi E., Andrews S., Casassus S., Graves S., Schreiber M., 2015, arXiv:1504.06040v2
\bibitem[\protect\citeauthoryear{Cleeves, Adams \& Bergin}{Cleeves et al.}{2013}]{Cleeves2013} Cleeves L.I., Adams F.C., Bergin E.A., 2013, ApJ, 772, 5
\bibitem[\protect\citeauthoryear{Cleeves, Bergin \& Adams}{Cleeves et al.}{2014}]{Cleeves2014} Cleeves L.I., Bergin E.A., Adams F.C., 2014, ApJ, 794, 123
\bibitem[\protect\citeauthoryear{Coleman \& Nelson}{2014}]{Coleman2014} Coleman G.A.L., Nelson R.P., 2014, MNRAS, 445, 479
\bibitem[\protect\citeauthoryear{Coleman \& Nelson}{2016}]{Coleman2016} Coleman G.A.L., Nelson R.P., 2016, arXiv:1604.05191v2 
\bibitem[\protect\citeauthoryear{Crida, Morbidelli \& Masset}{Crida et al.}{2006}]{Crida2006} Crida A., Morbidelli A., Masset F., 2006, Icar, 181, 587
\bibitem[\protect\citeauthoryear{Cridland, Pudritz \& Alessi}{Cridland et al.}{2016}]{Cridland2016} Cridland A.J., Pudritz R.E., Alessi M., 2016, arXiv:1605.09407v1
\bibitem[\protect\citeauthoryear{D'Alessio et al.}{1998}]{DAlessio1998} D'Alessio P., Cant\'{o} J., Calvet N., Lizano S., 1998, ApJ, 500, 411
\bibitem[\protect\citeauthoryear{D'Alessio et al.}{1999}]{DAlessio1999} D'Alessio P., Calvet N., Hartmann L., Lizano S., Cant\'{o} J., 1999, ApJ, 527, 893
\bibitem[\protect\citeauthoryear{D'Alessio, Calvet \& Hartmann}{D'Alessio et al.}{2001}]{DAlessio2001} D'Alessio P., Calvet N., Hartmann L., 2001, ApJ, 553, 321
\bibitem[\protect\citeauthoryear{Dittkrist et al.}{2014}]{Dittkrist2014} Dittkrist K.-M., Mordasini C., Klahr H., Alibert Y., Henning T., 2014, A\&A, 567, A121
\bibitem[\protect\citeauthoryear{Eisner et al.}{2005}]{Eisner2005} Eisner J.A., Hillenbrand L.A., White R.J., Akeson R.L., Sargent A.I., 2005, ApJ, 623, 952
\bibitem[\protect\citeauthoryear{Elser, Meyer \& Moore}{Elser et al.}{2012}]{Elser2012} Elser S., Meyer M.R., Moore B., 2012, Icar, 221, 859
\bibitem[\protect\citeauthoryear{Fleming, Stone \& Hawley}{Fleming et al.}{2000}]{Fleming2000} Fleming T.P., Stone J.M., Hawley J.F., 2000, ApJ, 530, 464
\bibitem[\protect\citeauthoryear{Fortney, Marley \& Barnes}{Fortney et al.}{2007}]{Fortney2007} Fortney J.J., Marley M.S., Barnes J.W., 2007, ApJ, 659, 1661
\bibitem[\protect\citeauthoryear{Fortney et al.}{2013}]{Fortney2013} Fortney J.J., Mordasini C., Nettelmann N., Kempton E.M.-R., Greene T.P., Zahnle K., 2013, ApJ, 775, 80
\bibitem[\protect\citeauthoryear{Gammie}{1996}]{Gammie1996} Gammie C.F., 1996, ApJ, 457, 355
\bibitem[\protect\citeauthoryear{Glassgold, Najita \& Igea}{Glassgold et al.}{1997}]{Glassgold1997} Glassgold A.E., Najita J., Igea J. 1997, ApJ, 480, 344
\bibitem[\protect\citeauthoryear{Goldreich \& Tremaine}{1980}]{GoldreichTremaine1980} Goldreich P., Tremaine S., 1980, ApJ, 241, 425
\bibitem[\protect\citeauthoryear{Gressel et al.}{2015}]{Gressel2015} Gressel O., Turner N.J., Nelson R.P., McNally C.P., 2015, ApJ, 801, 84
\bibitem[\protect\citeauthoryear{Gressel \& Pessah}{2015}]{Gressel2015b} Gressel O., Pessah M.E., 2015, ApJ, 810, 59
\bibitem[\protect\citeauthoryear{Hasegawa \& Pudritz}{2010}]{HP10} Hasegawa Y., Pudritz R.E., 2010, ApJ, 710, L167
\bibitem[\protect\citeauthoryear{Hasegawa \& Pudritz}{2011}]{HP11} Hasegawa Y., Pudritz R.E., 2011, MNRAS, 417, 1236
\bibitem[\protect\citeauthoryear{Hasegawa \& Pudritz}{2012}]{HP12} Hasegawa Y., Pudritz R.E., 2012, ApJ, 760, 117
\bibitem[\protect\citeauthoryear{Hasegawa \& Pudritz}{2013}]{HP13} Hasegawa Y., Pudritz R.E., 2013, ApJ, 778, 78
\bibitem[\protect\citeauthoryear{Hasegawa \& Pudritz}{2014}]{HP14} Hasegawa Y., Pudritz R.E., 2014, ApJ, 794, 25
\bibitem[\protect\citeauthoryear{Hellary \& Nelson}{2012}]{HellaryNelson2012} Hellary P., Nelson R.P., 2012, MNRAS, 419, 2737
\bibitem[\protect\citeauthoryear{Hern\'{a}ndez et al.}{2007}]{Hernandez2007} Hern\'{a}ndez J. et al., 2007, ApJ, 662, 1067
\bibitem[\protect\citeauthoryear{Howard et al.}{2013}]{Howard2013} Howard A.W. et al., 2013, Nature, 503, 381
\bibitem[\protect\citeauthoryear{Hubickyj, Bodenheimer \& Lissauer}{Hubickyj et al.}{2005}]{Hubickyj2005} Hubickyj O., Bodenheimer P., Lissauer J.J., 2005, Icar, 179, 415
\bibitem[\protect\citeauthoryear{Hueso \& Guillot}{2005}]{Hueso2005} Hueso R., Guillot T., 2005, A\&A, 442, 703
\bibitem[\protect\citeauthoryear{Ida \& Lin}{2004}]{IdaLin2004} Ida S., Lin D.N.C., 2004, ApJ, 604, 388
\bibitem[\protect\citeauthoryear{Ida \& Lin}{2008}]{IdaLin2008} Ida S., Lin D.N.C., 2008, ApJ, 673, 487
\bibitem[\protect\citeauthoryear{Ida \& Lin}{2010}]{IdaLin2010} Ida S., Lin D.N.C., 2010, ApJ, 719, 810
\bibitem[\protect\citeauthoryear{Ida, Lin \& Nagasawa}{Ida et al.}{2013}]{Ida2013} Ida S., Lin D.N.C., Nagasawa M., ApJ, 2013, 775, 42
\bibitem[\protect\citeauthoryear{Ikoma, Nakazawa \& Emori}{Ikoma et al.}{2000}]{Ikoma2000} Ikoma M., Nakazawa K., Emori H., 2000, ApJ, 537, 1013
\bibitem[\protect\citeauthoryear{Ivanov, Papaloizou \& Polnarev}{Ivanov et al.}{1999}]{Ivanov1999} Ivanov P.B., Papaloizou J.C.B., Polnarev A.G., 1999, MNRAS, 307, 79
\bibitem[\protect\citeauthoryear{Jang-Condell \& Sasselov}{2004}]{JangCondellSasselov2004} Jang-Condell H., Sasselov D.D., 2004, ApJ, 608, 497
\bibitem[\protect\citeauthoryear{Johansen et al.}{2007}]{Johansen2007} Johansen A., Oishi J.S., Mac Low M.-M., Klahr H., Henning T., Youdin A., 2007, Nature, 448, 1022
\bibitem[\protect\citeauthoryear{Kley}{1999}]{Kley1999} Kley W., 1999, MNRAS, 303, 696 
\bibitem[\protect\citeauthoryear{Kley \& Dirksen}{2006}]{KleyDirksen2006} Kley W., Dirksen G., 2006, A\&A, 447, 369
\bibitem[\protect\citeauthoryear{Kokubo \& Ida}{2002}]{KokuboIda2002} Kokubo E., Ida S., 2002, ApJ, 581, 666
\bibitem[\protect\citeauthoryear{Lissauer et al.}{2009}]{Lissauer2009} Lissauer J.J., Hubickyj O., D'Angelo G., Bodenheimer P., 2009, Icar, 199, 338
\bibitem[\protect\citeauthoryear{Lopez \& Fortney}{2013}]{Lopez2013} Lopez E.D., Fortney J.J., 2013, ApJ, 776, 2
\bibitem[\protect\citeauthoryear{Lubow, Seibert \& Artymowicz}{Lubow et al.}{1999}]{Lubow1999} Lubow S.H., Seibert M., Artymowicz P., 1999, ApJ, 526, 1001
\bibitem[\protect\citeauthoryear{Lubow \& D'Angelo}{2006}]{Lubow2006} Lubow S.H., D'Angelo G., 2006, ApJ, 641, 526
\bibitem[\protect\citeauthoryear{Lynden Bell \& Pringle}{1974}]{LBP1974} Lynden Bell D., Pringle J.E., 1974, MNRAS, 168, 603
\bibitem[\protect\citeauthoryear{Lyra et al.}{2009}]{Lyra2009} Lyra W., Johansen A., Zsom A., Klahr H., Piskunov N., 2009, A\&A, 497, 869
\bibitem[\protect\citeauthoryear{Lyra, Paardekooper \& Mac Low}{Lyra et al.}{2010}]{Lyra2010} Lyra W., Paardekooper S.-J., Mac Low M.-M., 2010, ApJ, 715, L68
\bibitem[\protect\citeauthoryear{Machida et al.}{2010}]{Machida2010} Machida M.N., Kokubo E., Inutsuka S.-I., Matsumoto T., 2010, MNRAS, 405, 1227
\bibitem[\protect\citeauthoryear{Mamajek}{2009}]{Mamajek2009} Mamajek E.E., 2009, in AIP Conf. Ser. 1158, Exoplanets and Disks: Their Formation and Diversity, ed. T. Usuda, M. Tamura, \& M. Ishii (Melville, NY: AIP), 3
\bibitem[\protect\citeauthoryear{Marcus et al.}{2015}]{Marcus2015} Marcus P.S., Pei S., Jiang C.-H., Barranco J.A., Hassanzadeh P., Lecoanet D., 2015, ApJ, 808, 87
\bibitem[\protect\citeauthoryear{Masset}{2001}]{Masset2001} Masset F.S., 2001, ApJ, 558, 453
\bibitem[\protect\citeauthoryear{Masset}{2002}]{Masset2002} Masset F.S., 2002, A\&A, 387, 605
\bibitem[\protect\citeauthoryear{Masset \& Papaloizou}{2003}]{Masset2003} Masset F.S., Papaloizou J.C.B., 2003, ApJ, 588, 494
\bibitem[\protect\citeauthoryear{Masset et al.}{2006}]{Masset2006} Masset F.S., Morbidelli A., Crida A., Ferreira J., 2006, ApJ, 642, 478
\bibitem[\protect\citeauthoryear{Matsumura \& Pudritz}{2003}]{MP2003} Matsumura S., Pudritz R.E., 2003, ApJ, 598, 645
\bibitem[\protect\citeauthoryear{Matsumura \& Pudtitz}{2005}]{MP2005} Matsumura S., Pudritz R.E., 2005, ApJ, 618, L137
\bibitem[\protect\citeauthoryear{Matsumura \& Pudritz}{2006}]{MP2006} Matsumura S., Pudritz R.E., 2006, MNRAS, 365, 572
\bibitem[\protect\citeauthoryear{Matsumura, Pudritz \& Thommes}{Matsumura et al.}{2007}]{MP2007} Matsumura S., Pudritz R.E., Thommes E.W., 2007, ApJ, 660, 1609
\bibitem[\protect\citeauthoryear{Mayor et al.}{2011}] {Mayor2011} Mayor M., et al., 2011, arXiv:1109.2497v1
\bibitem[\protect\citeauthoryear{Menou \& Goodman}{2004}]{MenouGoodman2004} Menou K., Goodman J., 2004, ApJ, 606, 520
\bibitem[\protect\citeauthoryear{Min et al.}{2011}]{Min2011} Min M., Dullemond C.P., Kama M., Dominik C., 2011, Icar, 212, 416
\bibitem[\protect\citeauthoryear{Mohanty, Ercolano \& Turner}{Mohanty et al.}{2013}]{Mohanty2013} Mohanty S., Ercolano B., Turner N.J., 2013, ApJ, 764, 65
\bibitem[\protect\citeauthoryear{Molli\`{e}re et al.}{2015}]{Molliere2015} Molli\`{e}re P., van Boekel R., Dullemond C., Henning T., Mordasini C., 2015, ApJ, 813, 1
\bibitem[\protect\citeauthoryear{Mordasini et al.}{2012a}]{Mordasini2012a} Mordasini C., Alibert Y., Benz W., Klahr H., Henning T., 2012, A\&A, 541, A97
\bibitem[\protect\citeauthoryear{Mordasini et al.}{2012b}]{Mordasini2012b} Mordasini C., Alibert Y., Klahr H., Henning T., 2012, A\&A, 547, A111
\bibitem[\protect\citeauthoryear{Mordasini et al.}{2012c}]{Mordasini2012c} Mordasini C., Alibert Y., Georgy C., Dittkrist K.-M., Klahr H., Henning T., 2012, A\&A, 547, A112
\bibitem[\protect\citeauthoryear{Mordasini et al.}{2014}]{Mordasini2014} Mordasini C., Klahr H., Alibert Y., Miller N., Henning T., 2014, A\&A, 566, A141 
\bibitem[\protect\citeauthoryear{Mordasini et al.}{2015}]{Mordasini2015} Mordasini C., Molli\`{e}re P., Dittkrist K.-M., Jin S., Alibert Y., 2015, International Journal of Astrobiology, 14, 201
\bibitem[\protect\citeauthoryear{Moriarty, Madhusudhan \& Fischer}{Moriarty et al.}{2014}]{Moriarty2014} Moriarty J., Madhusudhan N., Fischer D., 2014, ApJ, 787, 81
\bibitem[\protect\citeauthoryear{Morton et al.}{2016}]{Morton2016} Morton T.D. et al., 2016, arXiv:1605.02825
\bibitem[\protect\citeauthoryear{\"{O}berg et al.}{2011}]{Oberg2011} \"{O}berg K.I., Adwin Boogert A.C., Pontoppidan K.M., Van Den Broek S., Van Dishoeck E.F., Bottinelli S., Blake G.A., Evans II N.J., 2011, ApJ, 740, 109
\bibitem[\protect\citeauthoryear{Oppenheimer \& Dalgarno}{1974}]{Oppenheimer1974} Oppenheimer M., Dalgarno A., 1974, ApJ, 192, 29
\bibitem[\protect\citeauthoryear{Ormel \& Klahr}{2010}]{Ormel2010} Ormel C. W., Klahr H. H., 2010, A\&A, 520, A43
\bibitem[\protect\citeauthoryear{Owen, Ercolano \& Clarke}{Owen et al.}{2011}]{Owen2011}Owen J.E., Ercolano B., Clarke C.J., 2011, MNRAS, 412, 13
\bibitem[\protect\citeauthoryear{Padovani et al.}{2015}]{Padovani2015} Padovani M., Hennebelle P., Marcowith A., Ferri\`{e}re K., 2015, arXiv:1509.06416v1
\bibitem[\protect\citeauthoryear{Pascucci \& Sterzik}{2009}]{Pascucci2009} Pascucci I., Sterzik M., 2009, ApJ, 702, 724
\bibitem[\protect\citeauthoryear{Pasek et al.}{2005}]{Pasek2005}Pasek M., Milsom J., Ciesla F., Lauretta D., Sharp C., Lunine J., 2005, Icar, 175, 1
\bibitem[\protect\citeauthoryear{Pignatale et al.}{2011}]{Pignatale2011} Pignatale F.C., Maddison S.T., Taquet V., Brooks G., Liffman K., 2011, MNRAS, 414, 2386
\bibitem[\protect\citeauthoryear{Pollack et al.}{1996}]{Pollack1996} Pollack J.B., Hubickyj O., Bodenheimer P., Lissauer J.J., Podolak M., Greenzweig Y., 1996, Icar, 124, 62
\bibitem[\protect\citeauthoryear{Pontoppidan et al.}{2014}]{Pontoppidan2014} Pontoppidan K.M., Salyk C., Bergin E.A., Brittain S., Marty B., Mousis O., \"{O}berg K.I., 2014, Protostars and Planets VI, pp. 363-386.
\bibitem[\protect\citeauthoryear{Qi et al.}{2011}]{Qi2011} Qi C., D'Alessio P., \"{O}berg K.I., Wilner D.J., Hughes A.M., 2011, ApJ, 740, 84
\bibitem[\protect\citeauthoryear{Rowe et al.}{2014}]{Rowe2014} Rowe J.F. et al., 2014, ApJ, 785, 45
\bibitem[\protect\citeauthoryear{Sano et al.}{2000}]{Sano2000} Sano T., Miyama S.M., Umebayashi T., Nakano T., 2000, ApJ, 543, 486
\bibitem[\protect\citeauthoryear{Shakura \& Sunyaev}{1973}]{SS1973} Shakura N.I., Sunyaev R.A., 1973, A\&A, 24, 337
\bibitem[\protect\citeauthoryear{Siess, Dufour \& Forestini}{Siess et al.}{2000}]{Siess2000} Siess L., Dufour E., Forestini M., 2000, A\&A, 358, 593
\bibitem[\protect\citeauthoryear{Simon et al.}{2013}]{Simon2013} Simon J.B., Bai X.-N., Stone J.M., Armitage P.J., Beckwith K., 2013, ApJ, 764, 66
\bibitem[\protect\citeauthoryear{Spitzer \& Tomasko}{1968}]{Spitzer1968} Spitzer L.J., Tomasko M.G., 1968, ApJ, 152, 971
\bibitem[\protect\citeauthoryear{Stepinski}{1998}]{Stepinski1998} Stepinski T.F., 1998, Icar, 132, 100
\bibitem[\protect\citeauthoryear{Toppani et al.}{2006}]{Toppani2006} Toppani A., Libourel G., Robert F., Ghanbaja J., 2006, Geochimica et Cosmochimica Acta, 70, 5035
\bibitem[\protect\citeauthoryear{Umebayashi \& Nakano}{1981}]{Umebayashi1981} Umebayashi T., Nakano T., 1981, PASJ, 33, 617
\bibitem[\protect\citeauthoryear{Valencia, Sasselov \& O'Connell}{Valencia et al.}{2007}]{Valencia2007} Valencia D., Sasselov D.D., O'Connell R.J., 2007, ApJ, 665, 1413
\bibitem[\protect\citeauthoryear{Visser \& Bergin}{2012}]{VisserBergin2012} Visser R., Bergin E.A., 2012, ApJ, 754, L18
\bibitem[\protect\citeauthoryear{White, Johnson \& Dantzig}{White et al.}{1958}]{White1958} White W.B., Johnson S.M., Dantzig G.B., 1958, J. Chem. Phys., 28, 751
\bibitem[\protect\citeauthoryear{Zhang et al.}{2013}]{Zhang2013} Zhang K., Pontoppidan K.M., Salyk C., Blake G.A., 2013, ApJ, 766, 82

\end{thebibliography}
\end{document}